\documentclass[iop]{emulateapj_table}

\usepackage{graphicx}
\usepackage{subfigure}

\usepackage[colorlinks=true,
		linkcolor=red,
		linkbordercolor={1 0 0},
		anchorcolor=magenta,
		citecolor=blue,
		citebordercolor={0 1 0},
		filecolor=magenta,
		filebordercolor={0 .5 .5},
		menucolor=magenta,
		menubordercolor={1 0 0},
		runcolor=magenta,
		urlcolor=magenta,
		urlbordercolor={0 1 1},
		frenchlinks=false]{hyperref}

\usepackage{multirow}
\usepackage{nicefrac}

\usepackage{longtable}
\usepackage{lscape}

\slugcomment{Accepted for publication in ApJ}
\shorttitle{SFR by galaxy components}\shortauthors{Catal\'an-Torrecilla et al. 2017}

\newcommand{\noprint}[1]{}

\begin{document}


%\title{Star formation in the local Universe from the CALIFA sample. II. SFR by components and quenching mechanisms in massive spiral galaxies}

\title{Star formation in the local Universe from the CALIFA sample. II. \\
Activation and quenching mechanisms in bulges, bars, and disks.}

%% Use \author, \affil, plus the \and command to format author and affiliation 
%% information.  If done correctly the peer review system will be able to
%% automatically put the author and affiliation information from the manuscript
%% and save the corresponding author the trouble of entering it by hand.
%%
%% The \affil should be used to document primary affiliations and the
%% \altaffil should be used for secondary affiliations, titles, or ema

%% Authors with the same affiliation can be grouped in a single

\author{C. Catal\'an-Torrecilla\altaffilmark{1}, A. Gil de Paz\altaffilmark{1}, A. Castillo-Morales\altaffilmark{1}, J. M\'endez-Abreu\altaffilmark{2}, J. Falc\'on-Barroso\altaffilmark{3,4}, S. Bekeraite\altaffilmark{5}, L. Costantin\altaffilmark{6}, A. de Lorenzo-C\'aceres\altaffilmark{2}, E. Florido\altaffilmark{7,8}, R. Garc\'ia-Benito\altaffilmark{9}, B. Husemann\altaffilmark{10}, J. Iglesias-P\'aramo\altaffilmark{9}, R. C. Kennicutt\altaffilmark{11}, D. Mast\altaffilmark{12}, S. Pascual\altaffilmark{1}, T. Ruiz-Lara\altaffilmark{7,8,13,14}, L. S\'anchez-Menguiano\altaffilmark{7,9}, S.F. S\'anchez\altaffilmark{15}, C. J. Walcher\altaffilmark{5}, J. Bland-Hawthorn\altaffilmark{16}, S. Duarte Puertas\altaffilmark{9}, R. A. Marino\altaffilmark{17}, J. Masegosa\altaffilmark{9}, P. S\'anchez-Bl\'azquez\altaffilmark{18,19} and CALIFA Collaboration}

\affil{\altaffilmark{1}Departamento de Astrof\'isica y CC. de la Atm\'osfera, Universidad Complutense de Madrid, E-28040 Madrid, Spain, \email{ccatalan@ucm.es}}
\affil{\altaffilmark{2} School of Physics and Astronomy, University of St. Andrews, SUPA, North Haugh, KY16 9SS, St. Andrews, UK}
\affil{\altaffilmark{3} Instituto de Astrof\'isica de Canarias, Calle V\'ia L\'actea s/n, E-38205 La Laguna, Tenerife, Spain}
\affil{\altaffilmark{4} Departamento de Astrof\'isica, Universidad de La Laguna, E-38200 La Laguna, Tenerife, Spain}
\affil{\altaffilmark{5} Leibniz-Institut fur Astrophysik Potsdam (AIP), An der Sternwarte 16, 14482, Potsdam, Germany}
\affil{\altaffilmark{6} Dipartimento di Fisica e Astronomia 'G. Galilei', Universita di Padova, vicolo dell'Osservatorio 3, I-35122 Padova, Italy}
\affil{\altaffilmark{7} Departamento de F\'isica Te\'orica y del Cosmos, Universidad de Granada, Campus de Fuentenueva, E-18071 Granada, Spain}
\affil{\altaffilmark{8} Instituto Carlos I de F\'isica Te\'orica y computacional, Universidad de Granada, E-18071 Granada, Spain}
\affil{\altaffilmark{9} Instituto de Astrof\'isica de Andaluc\'ia (CSIC), Glorieta de la Astronom\'ia s/n, Aptdo. 3004, E-18080 Granada, Spain}
\affil{\altaffilmark{10} European Southern Observatory, Karl-Schwarzschild-Str. 2, 85748, Garching b. Munchen, Germany }
\affil{\altaffilmark{11}Institute of Astronomy, University of Cambridge, Madingley Road, Cambridge, CB3 0HA, UK}
\affil{\altaffilmark{12} Instituto de Cosmologia, Relatividade e Astrof\'{\i}sica - ICRA, Centro Brasileiro de Pesquisas F\'{\i}sicas, Rua Dr.Xavier Sigaud 150, CEP 22290-180, Rio de Janeiro, RJ, Brazil}
\affil{\altaffilmark{13} Instituto de Astrof\'isica de Canarias, Calle V\'ia L\'actea s/n, E-38205 La Laguna, Tenerife, Spain}
\affil{\altaffilmark{14} Departamento de Astrof\'isica, Universidad de La Laguna, E-38200 La Laguna, Tenerife, Spain}
\affil{\altaffilmark{15} Instituto de Astronom\'\i a, Universidad Nacional Auton\'oma de M\'exico, A.P. 70-264, 04510 M\'exico, D.F.,  Mexico}
\affil{\altaffilmark{16} Sydney Institute for Astronomy, School of Physics, University of Sydney, NSW, Australia}
\affil{\altaffilmark{17} Institute for Astronomy, Department of Physics, ETH Z\"urich, Switzerland$.$}
\affil{\altaffilmark{18} Departamento de F\'isica Te\'orica, Universidad Aut\'onoma de Madrid, Cantoblanco, 28049, Spain}
\affil{\altaffilmark{19} Instituto de Astrof\'isica, Pontifica Universidad de Chile, Av. Vicuna Mackenna 4860, 782-0436 Macul, Santiago, Chile}

%% Use the \and command so offset the last author.


\begin{abstract}

We estimate the current extinction-corrected H$\alpha$ star formation rate (SFR) of the different morphological components that shape galaxies (bulges, bars, and disks). We use a multi-component photometric decomposition based on SDSS imaging to CALIFA Integral Field Spectroscopy datacubes for a sample of 219 galaxies. This analysis reveals an enhancement of the central SFR and specific SFR (sSFR $=$ SFR/$M_{\star}$) in barred galaxies. Along the Main Sequence, we find more massive galaxies in total have undergone efficient suppression (quenching) of their star formation, in agreement with many studies. We discover that more massive disks have had their star formation quenched as well. We evaluate which mechanisms might be responsible for this quenching process. The presence of type-2 AGNs plays a role at damping the sSFR in bulges and less efficiently in disks. Also, the decrease in the sSFR of the disk component becomes more noticeable for stellar masses around 10$^{10.5}$ M$_{\odot}$; for bulges, it is already present at $\sim$10$^{9.5}$ M$_{\odot}$. The analysis of the line-of-sight stellar velocity dispersions ($\sigma$) for the bulge component and of the corresponding Faber-Jackson relation shows that AGNs tend to have slightly higher $\sigma$ values than star-forming galaxies for the same mass. Finally, the impact of environment is evaluated by means of the projected galaxy density, $\Sigma$$_{5}$. We find that the SFR of both bulges and disks decreases in intermediate-to-high density environments. This work reflects the potential of combining IFS data with 2D multi-component decompositions to shed light on the processes that regulate the SFR.
 
\end{abstract}


\section{Introduction} \label{intro}

Among the multiple open issues on galaxy formation and evolution, arguably the most fundamental are related to the evolution of the baryonic component and, more specifically, on the relative role of the different mechanisms that can trigger and quench star formation (SF) in galaxies. 

For the processes that can activate and regulate SF, these may vary depending on the location within the galaxy. Secular internal evolution \citep{Kormendy_2004} and the accretion of gas \citep{Dekel_2009,Sanchez_almeida_2014} are likely dominant in galaxy disks, with the latter process being progressively more important as we move outwards in the disks. In the case of the central regions, {\it in-situ} SF is strongly affected by the amount of gas inflow that is driven to the center due to the presence of bars \citep{Sakamoto_1999,Sheth_2005} or by galaxy mergers \citep{Barnes_1991}.

With respect to the quenching of {\it in-situ} SF in galaxies, these are also expected to differ depending on whether we are talking about the formation of stars associated to bulges, bars or disks. Some of the mechanisms that have been proposed to be responsible for the star formation shutdown are related with the gas consumption, such as the termination of gas supply, i.\,e., strangulation \citep{Kawata_2008,Peng_2015}, or ram-pressure stripping \citep{Book_2010,Steinhauser_2016}. The previous mechanisms that transform galaxies are related to the influence of the environment in regulating the SFR in galaxies \citep{Hashimoto_1998,Koyama_2013}. Galaxy harassment \citep{Moore_1996,Moore_1998,Bialas_2015} or morphological quenching \citep{Martig_2009} are also important.

The role of active galactic nuclei (AGN) at enhancing \citep{Silk_2005,Silk_2013} or suppressing the star formation in the host galaxy \citep{Oppenheimer_2010, Page_2012, Shimizu_2015, Hopkins_2016, Carniani_2016}, the effect of SNe-driven winds \citep{Stringer_2012,Bower_2012} and the feedback from massive stars \citep{Dalla_Vecchia_2008,Hopkins_2012} have important implications for the evolution of galaxies as well.

Different mechanisms act on different spatial scales and are sensitive to the presence of specific structural components (spiral arms, bars, etc). That is why having high spatial resolution is crucial to solve the problem. Besides, it is also important to quantify how these mechanisms compete not only as a function of different galaxy properties but also as a function of redshift. One of the most fundamental parameters that characterizes galaxies is its Star Formation Rate (SFR). A better understanding of the distribution of the SFR in the different stellar structures that shaped galaxies in the local Universe will shed some light on their formation and evolution processes. The advance of Integral Field Spectroscopy (IFS) techniques gives us the opportunity to accurately measure the SFR at the different components that are forming the galaxies such as unresolved nuclear sources, bulges, bars, and disks. We can also explore the capacity of forming new stars with respect to the stellar mass in each of these stellar structures. This is a determining path if we want to know the different contributions of each component to the integrated value of the SFR in each galaxy. The Calar Alto Legacy Integral Field Area (CALIFA) survey \citep{Sanchez_2012} provides us with excellent data to answer these questions in a spatially resolved way. Some early attempts based on radial profiles of the SFR as a function of galaxy morphology suggests that galaxies are quenched inside-out, and that this process is faster in the central, bulge-dominated part than in the disks \citep{Gonzalez_Delgado_2016}. Here, we do a more precise analysis by isolating the galaxies in their basic stellar structures. We combine for the first time in a large sample of galaxies the two dimensional (2D) photometric decomposition of the CALIFA galaxies \citep{mendez_abreu_2016} with IFS data to measure the SFR in the different morphological components of galaxies. 

This paper is organized as follows: in Section~\ref{sample}, we describe the CALIFA reference sample used in this article; in Section~\ref{analysis}, we describe the analysis and methodology applied to the data, including the concept of ``smooth-aperture", the 2D photometric decomposition in bulges, bars, and disks and the derivation of the corresponding IFS-based SFRs. Our results are discussed in Section~\ref{results}. Finally, in Section \ref{conclusions}, we summarize the main conclusions of this work. Throughout our paper we use a cosmology defined by H$_{0}$ $=$ 70 km\,s$^{-1}$\,Mpc$^{-1}$, $\Omega$$_{\Lambda}$ $=$ 0.7 and a flat universe.

\section{CALIFA Sample} \label{sample}

The galaxies used in this work are part of the Calar Alto Legacy Integral Field Area (CALIFA) Survey \citep{Sanchez_2012}. Data were obtained with the Potsdam Multi-Aperture Spectrophotometer \citep[PMAS,][]{Roth_2005} in the PPak mode \citep{Kelz_2006} mounted on the 3.5m telescope at the Calar Alto Observatory. As a brief summary, galaxies have spectroscopic redshifts in the range 0.005 $<$ z $<$ 0.03 and angular isophotal diameter in the range 45'' $<$ D25 $<$ 80'' in the SDSS {\it r}-band. The properties of the CALIFA mother sample are fully described in \citet{Walcher_2014}.

The observations span the whole optical wavelength range in two overlapping setups. The V500 grating covers the range 3745-7500\,\AA\ at a spectral resolution of R $\sim$ 850 while the V1200 grating is restricted to 3650-4840\,\AA\ but with a higher resolution (R $\sim$ 1650). As our aim is to calculate extinction-corrected H$\alpha$ luminosities in each stellar galaxy component is desirable to have both H$\beta$ and H$\alpha$ emission lines in the same observing range. This is the reason why we use the V500 setup thorough this work. The V1200 data are restricted to the analysis of the line-of-sight velocity dispersions (Section \ref{kinematics}).

This paper makes use of 545 CALIFA galaxies that have been observed and processed with the V500 grating, are part of the Data Release 3 (DR3) \citep{Sanchez_2016} and belong to the CALIFA mother sample. This criterion should guarantee that we maintain the limits where the mother sample is representative of the general galaxy population: 9.7 and 11.4 in log(M$_{\star}$/M$_{\odot}$), $-$19.0 and $-$23.1 in {\it r}-band absolute magnitude and 1.7 and 11.5 kpc in half-light radius \citep{Walcher_2014}. As we are interested on the SFR properties of these galaxies in their different components, our sample is further constrained to those galaxies that are eligible for the 2D photometric decomposition. Galaxies meeting any of the following criteria were excluded: (1) if they are forming a pair, an interacting system or they have a heavily distorted morphology and (2) if they are highly inclined galaxies as the projection effects can affect the results (typically {\it i} $>$ $70^{\circ}$). More details about the sample selection are given in \citet{mendez_abreu_2016}. A total of 204 galaxies were excluded in this way. We also reject 122 galaxies that do not show detectable H$\alpha$ emission based on a signal-to-noise (S/N) criteria including also galaxies classified as elliptical in the 2D decomposition analysis. We impose a minimum of S$/$N $>$ 5 for the detection of both H$\beta$ and H$\alpha$ emission lines in each photometric structure of the galaxies (more details are given in Section \ref{fluxes}). This leads to the final sample of 219 CALIFA galaxies for this work.

\section{Analysis} \label{analysis}

In this Section, we describe the method applied to obtain a extinction-corrected H$\alpha$ SFR value for each galaxy morphological component (nuclear point source, bulge, bar, and disk). This method relies on the combination of 2D decomposition of multi-band photometry on IFU spectral datacubes. Consequently, our galaxy components are defined based exclusively on the fitting to the photometry. Our objective is to determine how these components will grow in stellar mass due to {\it in-situ} star formation which ultimately dominates the total mass growth in the Local Universe. We aim to identify the mechanism(s) that either trigger or quench star formation in each of these regions, and, therefore, in galaxies as a whole, but going beyond the use of simple ill-defined spectro-photometric apertures on the datacubes.

\subsection{Assigning SFR values to morphological components defined based on purely photometric criteria}\label{limitations}

Different approaches can be used to perform a spatially-resolved analysis of the SFR in galaxies, including individual pixels, full 2D maps, radial profiles or, as in this paper, multi-component decomposition. The analyses based on 2D maps or individual pixels have difficulty in combining information from different galaxies and are also limited in our case by the coarse spatial resolution of the CALIFA datacubes. A simplified approach would have been to identify the transition radius between the bulge and disk components in one-dimensional (1D) surface brightness profiles and use this radius (and the galaxy ellipticity and position angle at that radius) to define spectro-photometric apertures for those two components. However, early tests already showed that this 1D approach does not allow to properly isolate the emission coming from the bulge and the disk, neither to deal with objects where a clear bar is present. In fact, some studies \citep{Aguerri_2005, Weinzirl_2009,Meert_2015} have demonstrated the importance of including the bar to obtain the precise parameters for the bulge component. 

An alternative would have been to use the distinct kinematic features of bulge, disk and bar main-sequence stars to separate components. However, it is quite likely that the stars currently being formed do not follow the same balance but might all form in a kinematically cold component and are being heated up afterwards. In addition, for this particular aim, the CALIFA spectral resolution is at the limit of what is needed to perform such multi-component kinematical fitting. Future high-efficiency IFS facilities working at R $>$ 5000 such as MEGARA \citep[at GTC;][]{Gil_de_paz_2016} or WEAVE \citep[at WHT;][]{Dalton_2014} will help in that regard. 

To overcome the previous limitations, we introduce here the concept of ``smooth-aperture" as the optimal option in our case. Instead of imposing a fixed aperture to define and isolate the different galaxy components (which would have to be defined based on terms of corresponding scale lengths), we allow the light associated with each spaxel in the datacube to have a contribution coming from different components (bulge, bar, disk). The starting stellar structures parameters are recovered using a multi-component decomposition as it has been widely proved to be one of the best methods for that purpose \citep{de_Souza_2004, Gadotti_2009,Weinzirl_2009, Salo_2015,Meert_2015,Meert_2016} and these parameters are obtained based exclusively on photometric criteria (see Section~\ref{2d_photometric}).

Bulge and disk SFR values are assigned as those measured in regions where the stellar content is dominated by stars that follow either a bulge or disk light profile. In particular, the central regions of galaxies could be classified as either classical bulges or pseudobulges. The latter ones are thought to have a complex star formation history where young stellar populations could be present. Nevertheless, we do not aim to probe the mechanism by which the stars were formed in each component but to provide a measurement of its current SFR focusing on the stars that are associated to their spatial distribution at present. Consequently, these are also the regions where the SFR is expected to later contribute to the growth of the stellar mass of these components.

\subsection{2D Photometric decomposition analysis} \label{2d_photometric}

We use the structural parameters derived for the CALIFA galaxies in \citet{mendez_abreu_2016}. These values were obtained by applying the 2D photometric decomposition code GASP2D \citep{mendez_abreu_2008, mendez_abreu_2014} over the {\it g}-, {\it r}- and {\it i}-band images from the Sloan Digital Sky Survey Data Release 7 \citep[SDSS-DR7,][]{Abazajian_2009}. The use of SDSS images is justified in terms of their higher spatial resolution in comparison with CALIFA making the method more precise. GASP2D makes use of the widely used Levenberg-Marquardt algorithm (i.\,e. damped least squares method) to fit the 2D surface brightness distributions of galaxies. This code allows the simultaneous fitting of different galaxy structures such as nuclear point sources, bulges, bars, and disks (including breaks). The reader is referred to \citet{mendez_abreu_2016} for more details about the methodology of the fitting algorithm.

For the purpose of this work, we use the parameters derived using the SDSS {\it g}-band images. This band is the one that provides the best compromise between image depth and being able to fit analytic functions to the light distribution of the youngest possible stellar populations. We have discarded the use of other bands for the following reasons: (a) trying to fit these 2D analytic components to the UV bands leads to catastrophic failures in all but the very early type systems, (b) the {\it u}-band is significantly less deep than {\it g} in SDSS (and it is subject to the same problems than the UV, especially in late-type spirals), (c) redder bands would progressively trace older stellar populations. It is worth emphasizing here that the main objective of the use of the {\it g}-band data is to provide relative weights for the different components in those regions where they compete in surface brightness (inner disk, outer regions of the bar, etc.). However, the actual SFR is dominated by the amount of extinction-corrected H$\alpha$ luminosity provided by the CALIFA datacubes to which these weights are applied.

\begin{figure*}
\includegraphics[trim={0cm 0cm 0cm 0cm},width=\hsize]{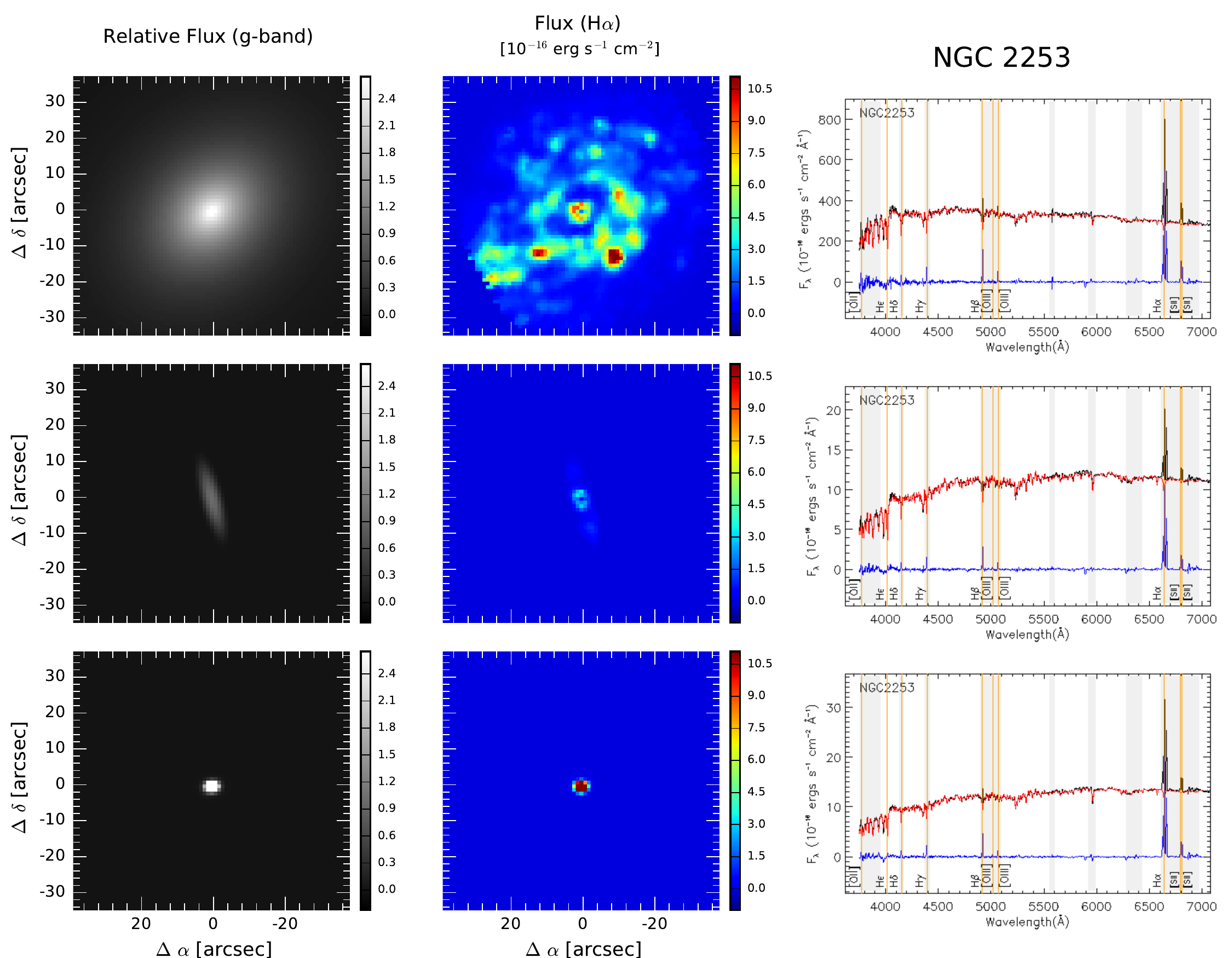}
\caption{Left panels: 2D {\it g}-band models derived from the multi-component photometric decomposition. Disk, bar, and bulge components are shown from top to bottom, respectively. Units for the flux are given relative to the central surface brightness of the bar component (if the bar is not present the central surface brightness of the disk component is used instead). Central surface brightness and the rest of the parameters needed to create these {\it g}-band models are provided in \citet{mendez_abreu_2016}. Middle panels: Distribution of the continuum-subtracted H$\alpha$ luminosity in the different stellar galaxy components. To create these H$\alpha$ maps, the original CALIFA datacubes have been multiplied by the corresponding weight maps in each morphological component, i.\,e., there is a weighted-datacube for each galaxy component, and then, analyzed spaxel by spaxel. Right panels: the integrated spectrum extracted for the weighted-datacube for each galaxy structure (bulge, bar, and disk) is shown in black. Gray-colored vertical ranges correspond to the emission lines and sky lines masked out during the fitting procedure. The red spectrum corresponds to the best fit for the underlying stellar population. The emission-line spectrum originated by the ionized gas is shown in blue. The latest is the one used to measure the H$\alpha$ and H$\beta$ fluxes associated to each component. The complete figure set (219 images) showing the models, H$\alpha$ maps and their corresponding spectra for each of the galaxies used in this work is available in the electronic edition of the journal. \label{fig:2d_model}}
\end{figure*}

The process followed to create the datacubes for each of the stellar components in each galaxy is the following. Firstly, we create the photometric characterization of the multiple stellar structures (nuclear point source, bulge, bar or disk), i.\,e., their best-fitting 2D {\it g}-band models as illustrated in the left panels of Figure~\ref{fig:2d_model}. Then, we create weight maps for each stellar component. The weight maps are defined as the ratio between the light in each galaxy structure (nuclear point source, bulge, bar or disk) and the total luminosity of the galaxy, as given by the SDSS {\it g}-band best-fitting models. These weight maps are computed for each individual CALIFA spaxel. Finally, the original CALIFA datacube of the galaxy is multiplied by these weight maps. This means that a 3D datacube is now created for each of the photometric structures. 

Figure~\ref{fig:2d_model} illustrates the process. The complete figure set (219 images) is available in the online version of the journal. Once we have created the final weighted-datacube for each component, we can obtain the corresponding distribution of the continuum-subtracted H$\alpha$ luminosity. Middle panels of Figure~\ref{fig:2d_model} show the continuum-subtracted H$\alpha$ luminosity for the disk, the bar and the bulge (from top to bottom). We emphasize that these H$\alpha$ maps are given as a visual tool to prove the goodness of the method but the actual H$\alpha$ luminosity is computed using the corresponding spectrum per component as explained in the next paragraph. We have identified 15 galaxies (7\,$\%$ of the bulge components) in the online figure set that show a clear contamination coming from the internal parts of the disks. These objects are marked with a nuclear 3-arcsec green aperture (see Section~\ref{agn}).

Finally, we obtain the integrated spectrum for each galaxy structure (right panels of Figure~\ref{fig:2d_model}) and for it the corresponding H$\alpha$ flux to derive the SFR. The analysis of the spectra extracted from the CALIFA datacubes is explained with detail in the following Section \ref{fluxes} and it is similar to that described in \citet{catalan_torrecilla_2015}. We emphasize here that the spectra obtained for each component might not be optimal for the study of intermediate-to-old stellar populations in each of these regions since these (more evolved) populations do show distinct kinematical properties in bulges, bars, and disks that could be used instead \citep{Johnston_2017,Tabor_2017}. Indeed, these properties are the ones that ultimately define what bulges, bars and disks truly are.

%%%%%%%%%%%%%%%%%%%%%%%%%%%%%%%%%%%%%%%%%%%%%%%%%%%%%%%%%%%%%%%%%%%%%%
%%%%%%%%%%%%%%%%%%%% FIGURE SET (goes here) %%%%%%%%%%%%%%%%%%%%%%%%%%
%%%%%%%%%%%%%%%%%%%%%%%%%%%%%%%%%%%%%%%%%%%%%%%%%%%%%%%%%%%%%%%%%%%%%%

\subsection{CALIFA: Extinction-corrected H$\alpha$ luminosities, continuum subtraction and line-flux measurements} \label{fluxes}

Once we have the final datacube for each component, we obtain the integrated spectra (see below) and the corresponding H$\alpha$ fluxes. 

For each component (nuclear point source, bulge, bar, and disk), we spatially integrate their corresponding datacube to generate an integrated spectrum using an elliptical aperture with a major axis radius of 36 arcsec. The use of 36 arcsec apertures is justified in terms of assuring a homogenous way for computing the aperture effects that are mentioned at the end of this section. The minor-to-major axis ratio of the elliptical aperture is given by the isophotal major and minor axis in the {\it g}-band from SDSS-DR7 as well as the isophotal position angle (PA). Before extracting the integrated spectra, a spatial masking over the datacubes is performed to avoid light coming from spaxels contaminated by field stars or background objects.

The complete description of the methodology applied to obtain the H$\alpha$ and H$\beta$ fluxes is explained in detail in \citet{catalan_torrecilla_2015}. For the sake of completeness, we briefly describe here the main steps. Once we have the integrated spectrum of each component, we carefully remove the stellar continuum using a linear combination of two single stellar population (SSP) evolutionary synthesis models of \cite{Vazdekis_2010} based on the MILES stellar library \citep{Sanchez_blazquez_2006}. Two set of models with a Kroupa IMF \citep{Kroupa_2001} are combined. One set contains models (considered as a young stellar population) with ages of 0.10, 0.50 and 0.79 Gyr. A second set (considered as an old stellar population) involves ages of 2.00, 6.31 and 14.13 Gyr. For each age we considered five different metallicities with [M/H] values equal to 0.00, 0.20, -0.40, -0.71 and -1.31 dex offset from the solar value. The basic steps applied to obtain the H$\alpha$ and H$\beta$ fluxes are the following: (1) to shift the SSP templates to match the systemic velocity of the integrated spectrum, (2) to convolve each stellar population model with a Gaussian profile so the absorption features could be broadened to match those of the integrated spectrum, (3) to redden the spectrum using a k($\lambda$) $=$ R$_{V}$ ($\lambda$/5500\,\AA)$^{-0.7}$ power law, where R$_{V}$ $=$ 5.9, as given by \cite{charlot_fall_2000} (4) to determine the best linear combination of SSPs by a $\chi$$^{2}$ minimization. Finally, H$\alpha$ and H$\beta$ fluxes are obtained from fitting Gaussians to the pure emission line spectra. The fluxes uncertainties are estimated from a random redistribution of the residuals after the Gaussian fittings mentioned before. The procedure, that consists of adding this new residual spectrum to the pure emission-line spectrum and perform afterwards the Gaussian fittings, is repeated 1000 times. The standard deviation of the computed fluxes is taken as the error in the H$\alpha$ and H$\beta$ fluxes.

An important parameter to take into account is the amount of dust attenuation for our measured H$\alpha$ luminosities. In particular, we use Balmer decrements with a Galactic extinction curve and a foreground screen dust geometry approximation to estimate the attenuation. Although there is not a considerable number of edge-on galaxies in this work due to the selection criteria imposed for the 2D photometric decomposition, we refer the reader to the extensive analysis in \citet{catalan_torrecilla_2015} where we test that the use of the Balmer Decrement in foreground dust screen approximation does not have an important impact on the SFR derived for these galaxies.

As some galaxies could extend beyond the PPak Field of View (FoV) we have applied aperture corrections to our extinction-corrected H$\alpha$ measurements. Among all the morphological components analyzed, the light coming from disk is the only one that might extend beyond the FoV. As a consequence, we have applied these aperture corrections to the spectrum of the disk only. We have derived dust-corrected H$\alpha$ growth curves using elliptical integrated apertures centered at the center of mass of the galaxy with radii increasing by steps of 3 arcsec up to a maximum radius of 36 arcsec \citep[a similar methodology is used in][]{Gil_de_paz_2007}. The last aperture corresponds to the 36 arcsec aperture that is the one used previously to compute the integrate disk spectra. This method allows to estimate the aperture effects in all the disks that create our sample in a uniform way. Then, we calculate the gradients of the extinction-corrected H$\alpha$ growth curves as the ratio between the flux in each aperture and the corresponding radial interval. This gradient decreases and becomes nearly zero when it approaches the maximum radius as the flux tend to be constant in the last apertures. Finally, if we plot the flux as a function of these gradients, the intercept of this relation gives us the value of the aperture correction. The mean and the median values for the aperture correction multiplicative factors in our sample are 1.19 and 1.08, respectively. The extinction-corrected H$\alpha$ SFR measurements for each galaxy component are given in Table~\ref{table}.

\subsection{CALIFA: Stellar masses} \label{masses}

Stellar mass is a key parameter on the process of formation and evolution of galaxies. For this study, we rely on the CALIFA total stellar masses that were calculated by \citet{Walcher_2014} using \citet{Bruzual_Charlot_2003} stellar population models with a \citet{Chabrier_2003} stellar IMF to construct UV to NIR SEDs. In particular, FUV \citep[GALEX,][]{Martin_2005}, {\it u}, {\it g}, {\it r}, {\it i}, {\it z} \citep[SDSS-DR7,][]{Abazajian_2009} and J, H, K \citep[2MASS Extended Source Catalog,][]{Jarrett_2000} photometric data were used.

We are interested in determining the stellar masses not only for the galaxies as a whole but also for their different structural components. For that reason, we apply the recipe below that allows deriving the mass in each component using the {\it i}-band mass-to-light relation of each component, (M$_{*}$/L)$_{comp,i}$, the galaxy total stellar mass, M$_{*,total}$, and the bulge-to-total (B/T), bar-to-total (Bar/T) and disk-to-total (D/T) luminosity ratios in the {\it i}-band. The luminosity ratios are derived as by-products of the 2D photometric decomposition for our galaxies in {\it i} and {\it g}-bands (see Section \ref{2d_photometric} for more details). We use the {\it i}-band values as they will better reproduce the stellar mass distribution than the {\it g}-band. Thus, we obtain:

\begin{equation}
\frac{M_{*,comp}}{M_{*,total}} = \frac{\left(\frac{M_{*}}{L}\right)_{comp,i}}{\left(\frac{M_{*}}{L}\right)_{total,i}} \cdot  \frac{L_{comp,i}}{L_{total,i}}
\label{comp_total}
\end{equation}

We make use of the color-dependent M$_{*}$/L$_{i}$ ratio given by equation 7 in \citet{Taylor_2011} where the authors also assume a \citet{Chabrier_2003} IMF. The authors proposed the following empirical relation between M$_{*}$/L$_{i}$ and ({\it g}$-${\it i}) color:

\begin{equation}
log M_{*}/L_{i} = -0.68 + 0.70({\it g} - {\it i})
\label{ml_formula}
\end{equation}

In our case, the ({\it g}$-${\it i}) colors correspond to ({\it g}$-${\it i})$_{disk}$, ({\it g}$-${\it i})$_{bar}$ or ({\it g}$-${\it i})$_{bulge}$. The following expression is used to obtain the ({\it g}$-${\it i}) colors for each galaxy component:

\begin{equation}
({\it g} - {\it i})_{comp} = ({\it g} - {\it i})_{total} - 2.5 \cdot log \frac{\left(\frac{L_{comp}}{L_{total}}\right)_{g}}{\left(\frac{L_{comp}}{L_{total}}\right)_{i}} 
\label{color_formula}
\end{equation}

The ({\it g}$-${\it i})$_{total}$ color measurements came from the analysis of the growth curve magnitudes performed in \citet{Walcher_2014}.

To verify the goodness of our stellar mass values per component, we have checked that the sum of the stellar masses for the different components obtained via the previous equations reproduces the total stellar mass derived from SED fitting for each galaxy. Both methods yield similar results for 82\,$\%$ of the galaxies, with the difference between the sum of the stellar components and the SED stellar mass being less than $\pm$ 15\,$\%$. For the remaining 18\,$\%$ of the galaxies, a larger difference arises due to significant variations in the \nicefrac{(B/T)$_{i}$}{(B/T)$_{g}$} ratio. The latter case has a mean value of 2.26 for the  \nicefrac{(B/T)$_{i}$}{(B/T)$_{g}$} ratio in contrast to a value of 1.30 for the cases in which the sum of the derived stellar masses of the components and the SED total stellar mass are similar. The former case is consequence of the non-linearity between the luminosity ratios in both bands and the mass-luminosity relation in equation \ref{ml_formula}. 

As a final remark, we note here that the H$\alpha$ extinction-corrected SFR tracer used along this work and the stellar population models applied for the continuum subtraction (Section \ref{fluxes}) are both based on a \citet{Kroupa_2001} IMF. For the sake of consistency, we rescale the stellar masses derived in this section to the \citet{Kroupa_2001} IMF applying the factor 1.08 as obtained in \citet{Madau_Dickinson_2014}. This value is almost independent of the stellar population age and has a very weak dependence on metallicity. The stellar masses derived for each galaxy component are provided in Table~\ref{table}.

\subsection{AGN optical classification} \label{agn}

AGN feedback is one of the mechanisms proposed to explain the quenching of the star formation in classical bulges and in massive galaxies, as it has been put forward to explain the differences between models and observations mainly at the high-end of the galaxy luminosity function \citep{Silk_Mamon_2012}. Therefore, it is critical to determine which of our CALIFA galaxies host an AGN. We apply a classical emission-line diagnostics to classify the objects into star-forming or type-2 AGN. For that purpose, we use the [OIII]/H$\beta$ vs$.$ [NII]/H$\alpha$ diagram introduced by \citet{Baldwin_1981} with the demarcation lines of \citet{Kauffmann_2003} and \citet{Kewley_2001}. We extract the spectrum centered within 3 arcsec of the nucleus and we imposed a S/N $>$ 4 for the previous emission lines. We obtain that 74 out of 219 galaxies are Seyfert/LINER. From now on, we refer to Seyfert/LINER as type-2 AGN objects. We highlight that galaxies that have type-1 AGN signatures are excluded from the sample completely. In the Unified Model, the emission from the AGN in Seyfert 1 galaxies outshine that due to recently formed stars as the Broad Line Region (BLR) is directly observable, while in Seyfert 2 the BLR is highly obscured and the line emission from the AGN competes with that due to star formation. Alternatively, several studies have pointed out a different scenario where Seyfert 1 and Seyfert 2 might be indeed different classes of objects, suggesting that Seyferts 2 intrinsically lack the BLR \citep{Tran_2001,Tran_2011}. Since the spatial resolution in our data is not enough to disentangle whether the central contribution is coming totally from the AGN or it has some contamination from (or even dominated by) star formation, for the type-2 AGN objects we have decided to include them in our sample and to distinctly mark them as type-2 AGN when necessary. Table~\ref{table} provides information about the galaxies classified as AGN in our sample.

Although LINERs have been traditionally associated with low-luminosity active galactic nuclei \citep[LLAGN,][]{Ho_1993,Terashima_2000}, some authors have recently claimed the importance of differentiating between galaxies hosting a weakly active nuclei and galaxies that could be ionized by hot low-mass evolved stars \citep[a recent discussion about the nature of LINER galaxies is provided by][]{Singh_2013}. In that regard, \citet{stasinska_2008} and \citet{cid_fernandes_2010,cid_fernandes_2011} have proposed to use the observed H$\alpha$ equivalent widths (EW$_{\mathrm{H}\alpha}$) versus the [NII]/H$\alpha$ ratio in the so-called WHAN diagram in which the division between weak AGNs and galaxies that are ionized by their hot low-mass evolved stars is fixed at 3\,\AA. We restrict the estimation of the EW$_{\mathrm{H}\alpha}$ to the center of our galaxies, i.\,e., the 3 central arcsec, instead of using the total integrated spectrum as we want to know whether or not the AGN is the dominant photoionization mechanism in the nuclear regions. The 3\,\AA \,criterion admittedly overestimates the number of galaxies classified as ``retired galaxies" as diluted bona fide AGNs could be also included in this category \citep{cid_fernandes_2011}. For that reason, we analyze the trend for the 6-arcsec and 9-arcsec apertures in these objects. Radial EW$_{\mathrm{H}\alpha}$ profiles using CALIFA data have been previously probed to be optimal for the study of the nuclear and extranuclear nebular emission of the warm ionized gas \citep{Gomes_2016}. We find that there are two distinct types. On one hand, some galaxies show an increase in the EW$_{\mathrm{H}\alpha}$ and a reduction in the [NII]/H$\alpha$ ratio at larger apertures reflecting the presence of a star-forming component. Even more, the integrated spectrum shows values of the EW$_{\mathrm{H}\alpha}$ larger than 3\,\AA. On the other hand, there are galaxies for which the EW$_{\mathrm{H}\alpha}$ decreases while the [NII]/H$\alpha$ ratio maintains a roughly constant value when using larger apertures. There are two possibilities for this case: (a) the evolved stars that are responsible for the photoionization of these regions exhibit a gradient which might explain the radial variation in EW$_{\mathrm{H}\alpha}$ and/or (b) there is actually an AGN in the central region and the older populations in their surroundings create a decline in the EW$_{\mathrm{H}\alpha}$ measurements. Whether one or both of these possibilities is the responsible mechanism is beyond the scope of this paper. There are still, however, a fraction of 39.2\% of the galaxies initially classified as AGN (33.5\% of the sample) where an homogeneous population of evolved stars could generate, according to the predictions of \citet{cid_fernandes_2011}, the EW$_{\mathrm{H}\alpha}$ values and distribution observed (at least at the spatial resolution of CALIFA). Thus, galaxies that have a Seyfert/LINER central spectrum are referred as type-2 AGN even though a fraction of these could be actually powered by a source distinct from a truly AGN.

\section{Results} \label{results}

Along this work we use extinction-corrected H$\alpha$ (H$\alpha$$_{corr}$) as our SFR reference indicator following the recipe given by \citet{Kennicutt_Evans_2012}. From now on we will use H$\alpha$ instead of H$\alpha$$_{corr}$ to shorten the term along the text although we emphasize that all the H$\alpha$ SFR measurements used here are extinction-corrected.

We have previously investigated the goodness of H$\alpha$ as a SFR tracer for a representative sample of 272 CALIFA galaxies \citep[for more details see][]{catalan_torrecilla_2015}. For that purpose, we compared extinction-corrected H$\alpha$ integrated measurements with single band (FUV$_{corr}$, 22 $\mu$m and TIR) and hybrid (22 $\mu$m + H$\alpha$$_{obs}$, TIR + H$\alpha$$_{obs}$, 22 $\mu$m + FUV$_{obs}$, TIR + FUV$_{obs}$ ) tracers. The latter shows an excellent agreement with dispersions around 0.18 dex.  We also find that only 1\,$\%$ of our objects host highly-obscured SF. Bearing in mind the above considerations, we can safely conclude that the use of extinction-corrected H$\alpha$ is appropriate for our sample. Whether or not this calibration can be applied to other samples in the Local Universe or to higher redshifts depends strongly on the expected fraction of galaxies and SFR that could be locked into completely-obscured star-forming sites and also on the percentage of nuclear line emission in Sy2 coming from either SF or AGN (or even ionization from evolved stars).

In this Section we show the correlations found between the SFR in the different morphological components of the galaxies and other physical properties such as stellar mass, morphological type, the presence of an AGN, environment and stellar velocity dispersion. Among other aspects, we investigate the so-called ``Main Sequence" of galaxies using not only integrated values but also the values in each galaxy morphological component (i.\,e., nuclear point sources, bulges, bars, and disks).

\subsection{SFR ratios by components: SFR central enhancement due to the presence of bars} \label{sfr_ratios}

In this section, we explore the connection between the central SFR(H$\alpha$) with other parameters such as the morphological type and the B/T in the {\it g}-band.  Galaxy morphologies were inferred by a combination of independent visual classifications carried out by members of the CALIFA collaboration as described in \citet{Walcher_2014} while B/T values in the {\it g}-band came from the analysis of the 2D decomposition (Section~\ref{2d_photometric}). 

\begin{figure}[h!]
\includegraphics[width=87mm]{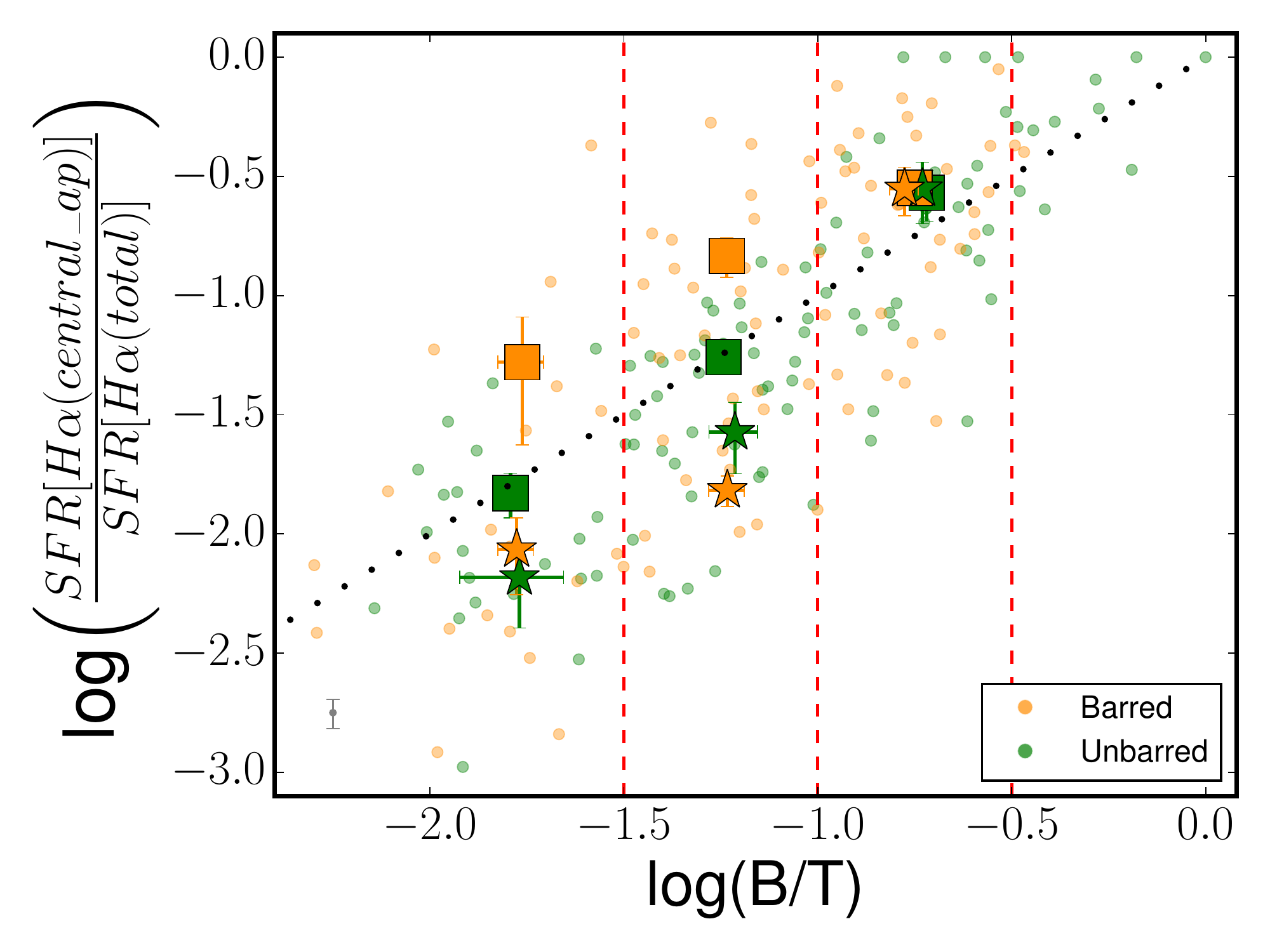}
\caption{Variation of the SFR[H$\alpha$(central)]/SFR[H$\alpha$(total)] ratio with the B/T parameter obtained for the {\it g}-band. The central SFR is referred to the amount of SFR in the smooth-aperture of the bulge component for unbarred  (green points) and barred galaxies (orange points). Squares correspond to purely star-forming galaxies while stars are referred to type-2 AGN objects. The previous symbols correspond to the logarithm of the mean value for the SFR[H$\alpha$(central)]/SFR[H$\alpha$(total)] ratio and the errors represent the standard deviation of the mean. \label{fig: sfr_central_bt_final}}
\end{figure}

\begin{figure}[h!]
\includegraphics[width=87mm]{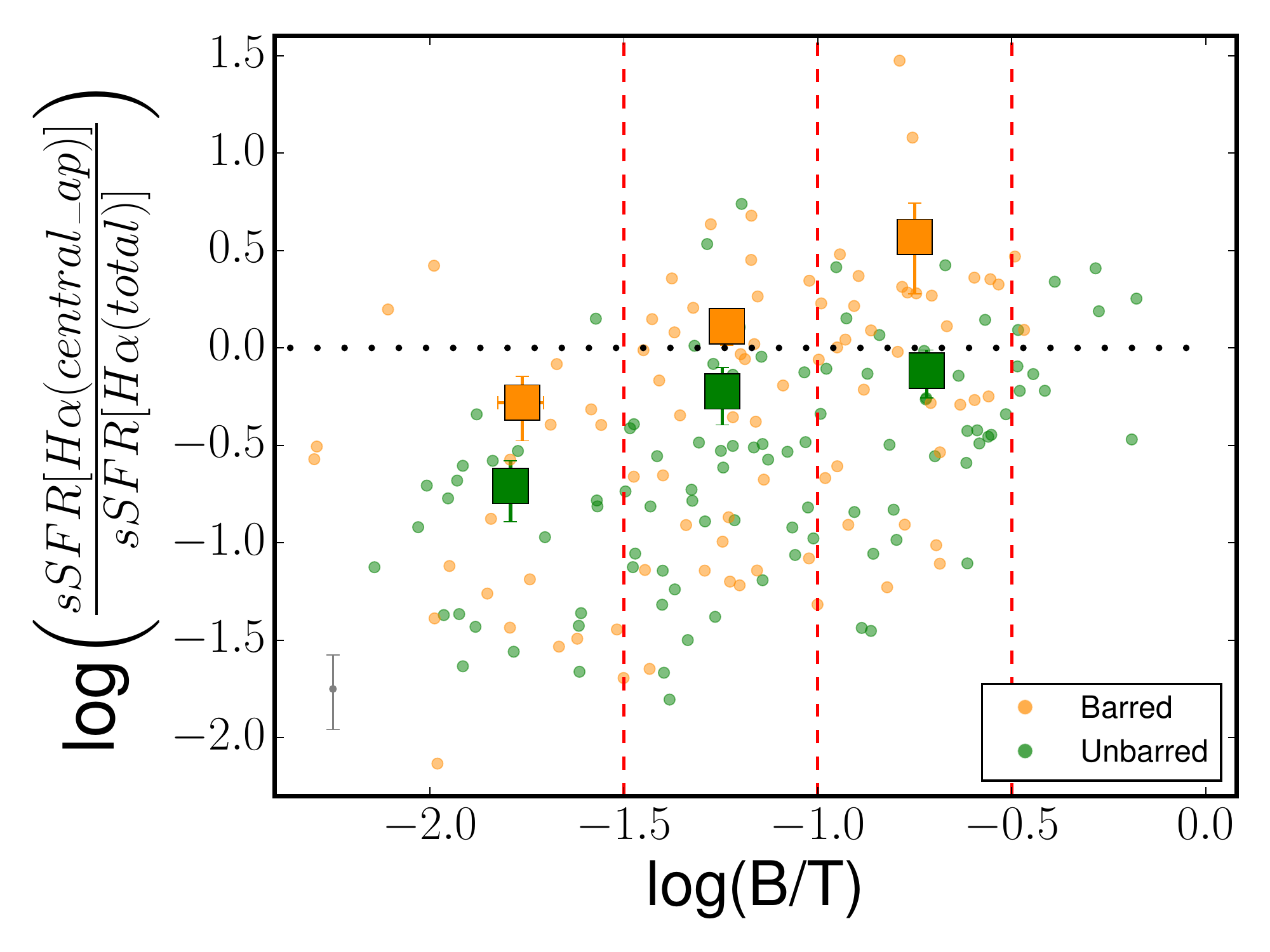}
\caption{Variation of the sSFR[H$\alpha$(central)]/sSFR[H$\alpha$(total)] ratio with the B/T parameter in {\it g}-band. Symbols and color coding are the same as in Figure~\ref{fig: sfr_central_bt_final}. \label{fig: ssfr_central_bt}}
\end{figure}

The analysis is performed only for galaxies that do show H$\alpha$ emission in the central regions. For the discussion below, central regions refer to the amount of SFR found in the aperture associated with the bulge component. To investigate whether the impact of the bar could trigger the star formation in the centers of galaxies, galaxies are classified into two main types, barred (orange points) and unbarred (green points) in Figure~\ref{fig: sfr_central_bt_final} and~\ref{fig: ssfr_central_bt}, respectively.

The bulges in our sample could be either classical bulges or pseudobulges \citep[for an extensive review see][]{Kormendy_2004}. Although with limitations, one can broadly discriminate between classical and pseudobulges using the S\'ersic index n$_b$ \citep[see][]{Fisher_2008, Fisher_2016}, where classical bulges are characterized by n$_{b}$ values greater than 2 while pseudobulges have values lower than 2. Using the n$_{b}$ parameters derived from the 2D photometric decomposition, 72\,$\%$ of our bulges would be classified as pseudobulges while the remaining 28\,$\%$ would appear as classical bulges. 

Attending to the previous criterion, the percentages for pseudobulges in our late-type galaxies is as follows: 74\,$\%$ for Sb, 91\,$\%$ for Sc, and, 80\,$\%$ for Sd (this is due to low number of galaxies in this bin, where 4 of the 5 galaxies are classified as pseudobulges). The median value of the SFR(central$_{ap}$)/SFR(total$_{ap}$) \footnote{The {\it ap} subscript indicates that these are smooth-aperture SFR measurements as explained in Section~\ref{limitations}. This subscript appears in the corresponding Figures but we have not included it along the text for simplicity.} is higher for the Sb/c barred galaxies in comparison with unbarred galaxies. This result points out that SFR in the central parts of these galaxies may be enhanced by the presence of a bar. Nevertheless, this trend is not found for other morphological types, perhaps due to much lower-number statistics in those types.

%% ANTES: The variation of the SFR[H$\alpha$(central)]/SFR[H$\alpha$(total)] ratio as a function of the morphological type is shown in Figure~\ref{fig: sfr_central_morpho}. It can be seen that the median values of the fraction of SFR contained in bulges is higher for the Sb/c barred galaxies (orange points) in comparison with unbarred galaxies (green points). This result points out that SFR in the central parts of these galaxies may be enhanced by the presence of a bar. Nevertheless, this trend is not found for other morphological types, perhaps due to much lower-number statistics in those types. 

As the majority of our galaxies are concentrated in the bin of Sb/c objects making the dynamic range of our morphological classification smaller, we also explore the behavior of the SFR[H$\alpha$(central)]/SFR[H$\alpha$(total)] ratio with the B/T parameter (Figure~\ref{fig: sfr_central_bt_final}). As commented before, B/T is obtained from the 2D decomposition analysis and does not depend on a visual classification. 

In order to quantify whether or not the presence of the bar is affecting the SFR in the bulge component, we split the sample in three bins: log(B/T) $<$ $-$1.5, $-$1.5 $<$ log(B/T) $<$ $-$1.0 and $-$1.0 $<$ log(B/T) $<$ $-0.5$. Big squares represent the logarithm of the mean value for the SFR[H$\alpha$(central)]/SFR[H$\alpha$(total)] ratios in each bin for purely star-forming galaxies while big stars refer to galaxies that have been classified as type-2 AGN. The 1:1 (dotted) line corresponds to the locus of galaxies that having only bulge and disk components would show the same extinction-corrected H$\alpha$-to-optical (g-band) luminosity ratio among these two components. The main result from this figure is that star-forming galaxies present higher mean SFR central values for barred galaxies (orange squares) than for unbarred ones (green squares). This effect is specially important for the cases of B/T smaller than 0.1. The enhancement of the central SFR due to the presence of bars has been pointed out by several authors using observational data \citep{de_Jong_1984,Devereux_1987,Ellison_2011,Wang_2012,Florido_2015} and also in recent dynamical simulations such as in \citet{Carles_2016}. As a result, a rejuvenation of the stellar populations in the center of barred galaxies has been also claimed by \citet{Fisher_2006} and \citet{Coelho_2011} among others. 

We can also analyze the connection between the presence of bars and AGN activity. We find that the optical bar fractions are similar for star-forming objects and type-2 AGN host galaxies, 43.9\,$\%$ and 52.2\,$\%$, respectively. This result is in accordance with previous works \citep{Mulchaey_1997,Hao_2009}. Nevertheless, galaxies hosting a type-2 AGN show less difference between the mean central SFR values for barred (orange stars) and unbarred (green stars) galaxies in comparison with purely star-forming objects. If bars and AGNs are simultaneously present, the effect of the bar in triggering the central SFR is reduced. Finally, it is also clear that type-2 AGNs are quenching the central SFR in their host galaxies, at least for small values of the B/T parameter.

To better understand the increase in the SFR in the central parts of star-forming barred galaxies we examine the behavior of the sSFR (sSFR $=$ SFR/$M_{\star}$) in fixed bins of B/T values. Figure~\ref{fig: ssfr_central_bt} shows that barred galaxies tend to have higher mean values of the sSFR in their central regions compared to unbarred galaxies. The horizontal dotted line here represents the location of galaxies that having only bulge and disk components would have the same sSFR in these two components. From this plot it is clear that low B/T galaxies with only bulge and disk components have higher disk sSFR values than their bulges. This is possibly related to blue optical-to-infrared colors and the presence of significant intermediated-aged stellar populations in their disks.

From this section, we can conclude that there is a clear relation between the SFR and sSFR in the central part of the galaxies and the presence of bars. Star-forming barred galaxies show higher values of their central SFR and sSFR than unbarred galaxies. This trend is present when we analyze the variation of the SFR with the B/T ratio while it is not as clear with the morphological type, probably, due to the low statistics for early-types and Sd/m galaxies. Besides, morphological type is also related with other aspects such as the definition of the spiral arms or the surface brightness. In contrast, the B/T is a more robust parameter to quantify the variation of the SFR in the central part of the galaxies as it is related with the bulge prominence. This finding supports the idea of bars driving gas efficiently toward the central regions of galaxies causing an enhancement of the SFR and the importance of the internal secular processes for the evolution of galaxies. On the contrary, in type-2 AGN we do not find a significant difference in the central SFR between barred and unbarred galaxies. Thus, nuclear activity should play a role at quenching the central SFR.

%\begin{figure}[h!]
%\includegraphics[width=87mm]{sfr_central_sfr_total_morpho_median_final_only_bulges.pdf}
%\caption{Distribution of the SFR[H$\alpha$(central)]/SFR[H$\alpha$(total)] ratio with the morphological type. Green points correspond to unbarred galaxies. Orange and blue points represent barred galaxies. In the former case the contribution of the SFR in the bar component is not added while in the latter case this contribution is included. The gray squares show the median value for each bin of measurements and the errors correspond to the standard error of the median, i./,e., 1.253$\times$$\sigma$/sqrt(N), where $\sigma$ is the interval that includes 68\,$\%$ of the data points around the median. \label{fig: sfr_central_morpho}}
%\end{figure}

%%%%%%%%%%%%%%%%%%%%%%%%%%%%%%%%%%%%%%%%%%%%%%%%%%%%%%%%%
%%%%%%%%%%%%%%%%%%%%%%%%%%%%%%%%%%%%%%%%%%%%%%%%%%%%%%%%%
\subsection{Main Sequence} \label{main_sequence}

\begin{figure*}
\includegraphics[width=85mm]{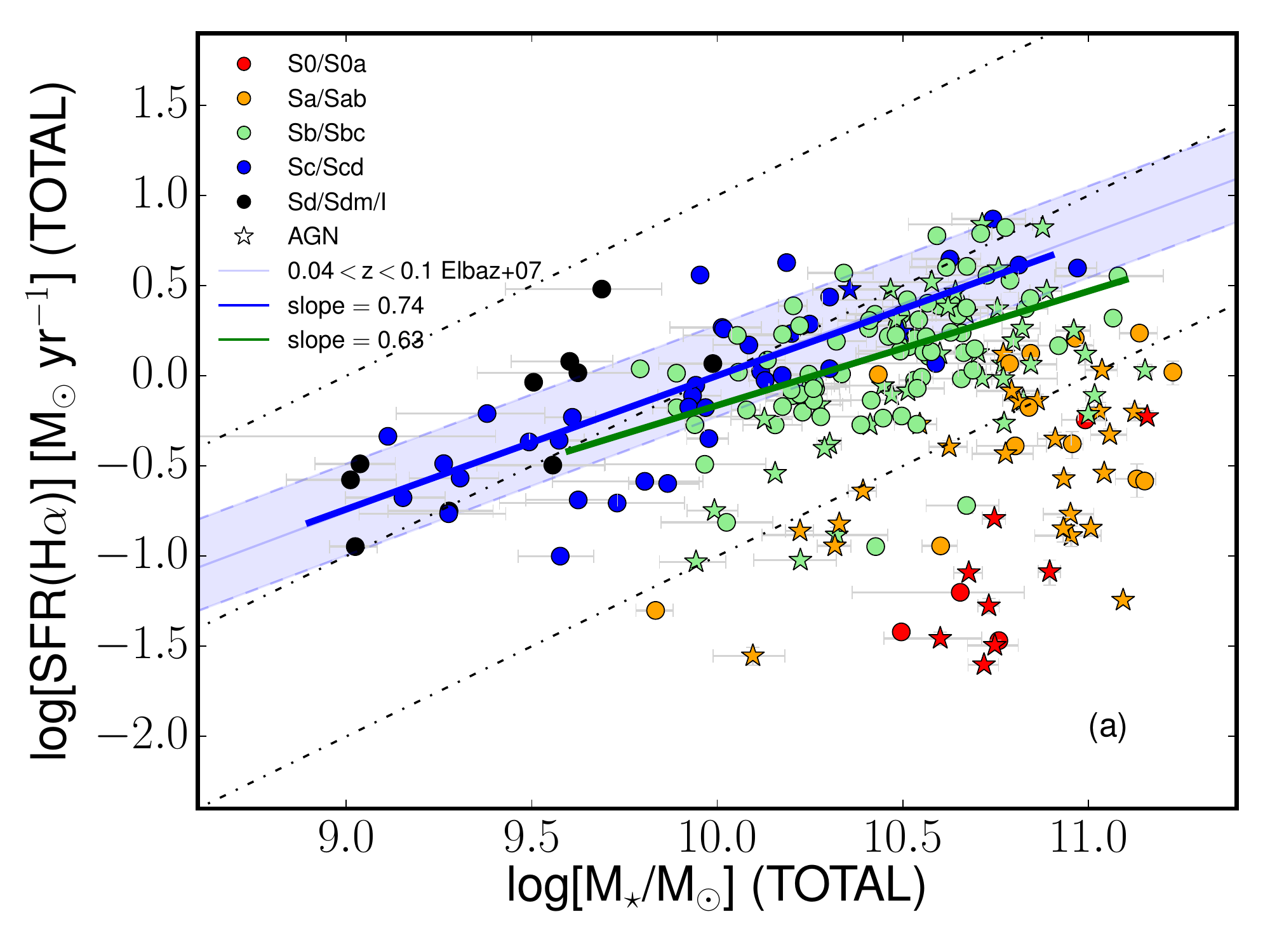} 
\includegraphics[width=85mm]{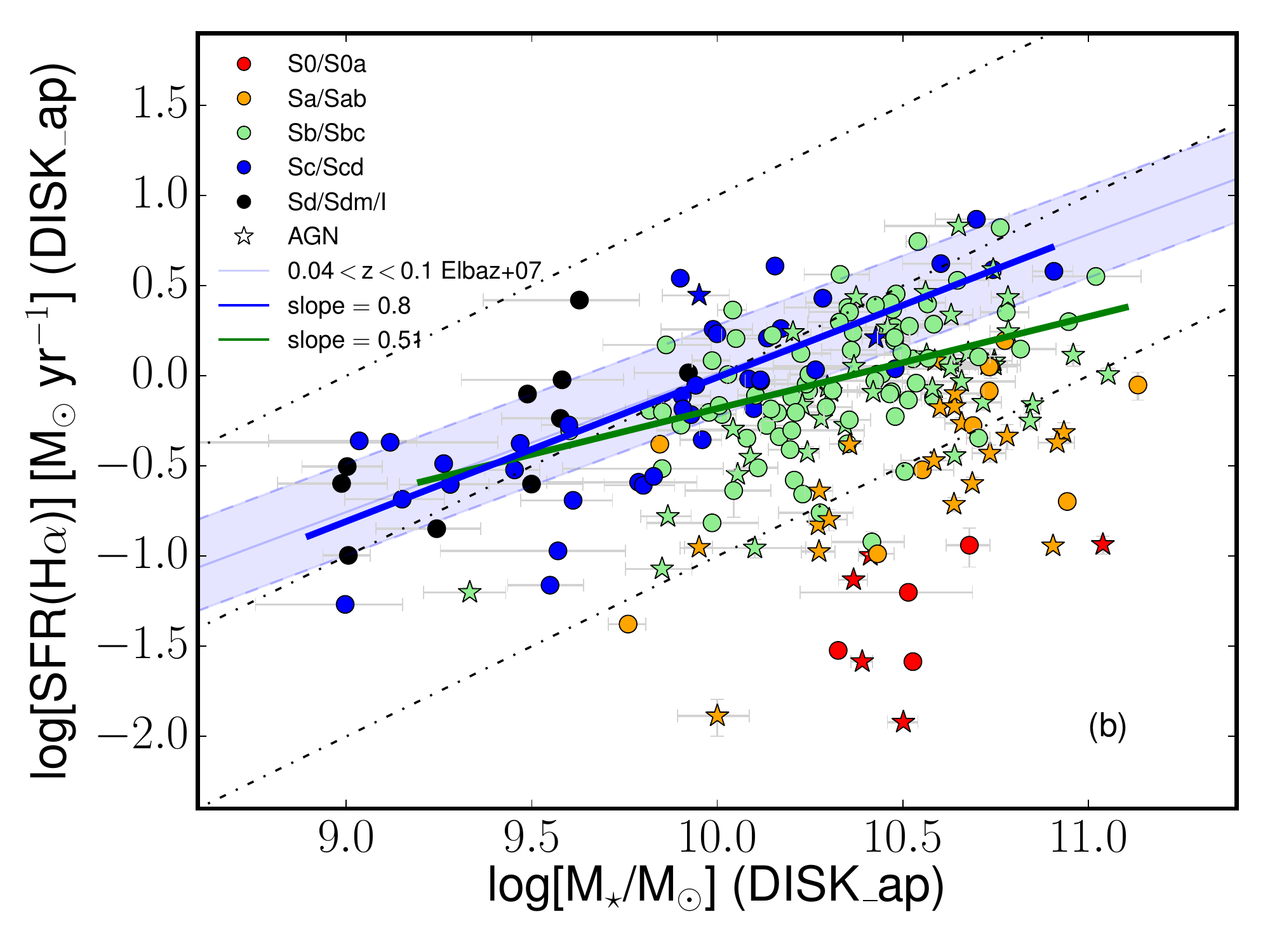} \\
\includegraphics[width=85mm]{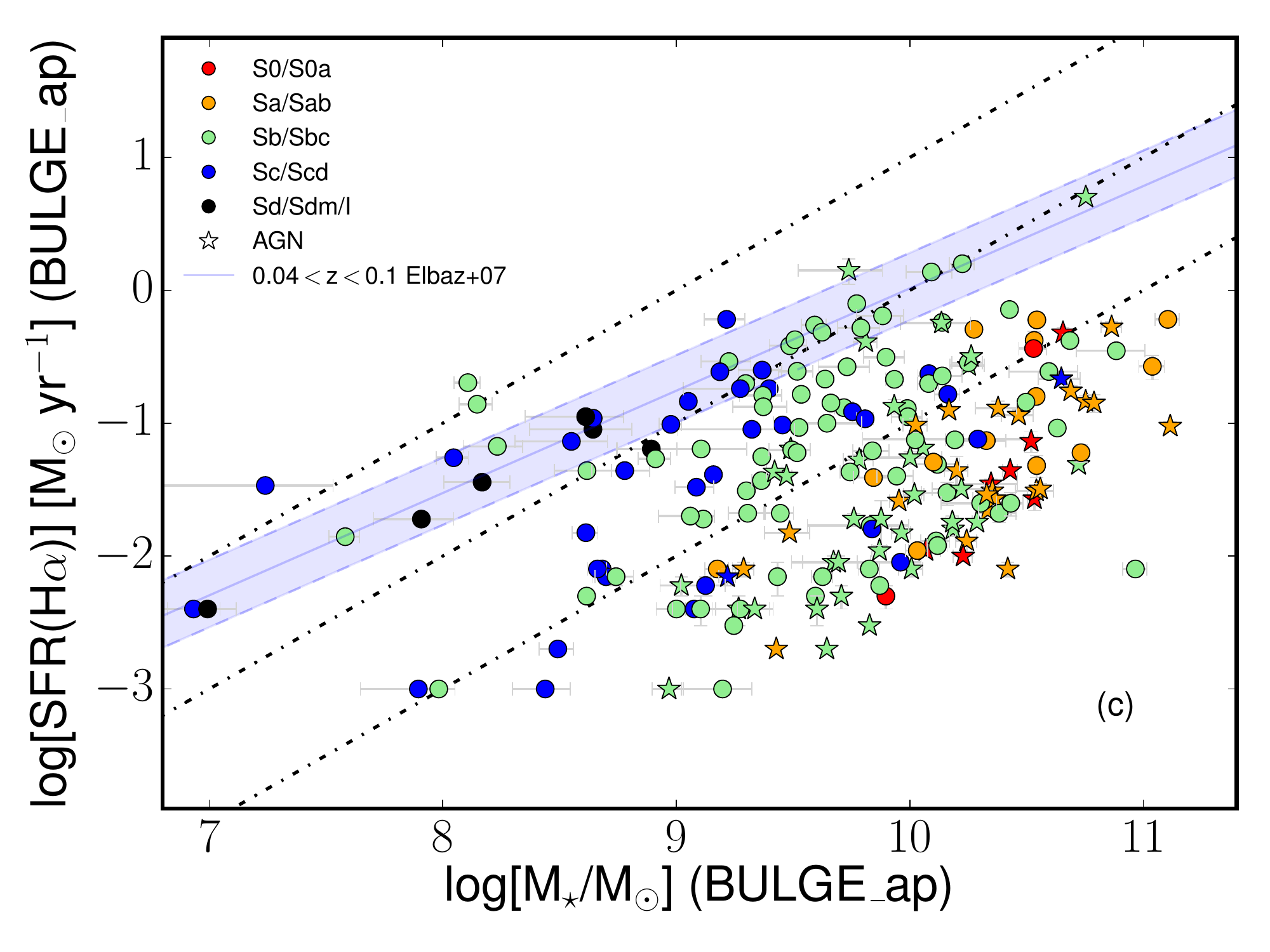} 
\includegraphics[width=85mm]{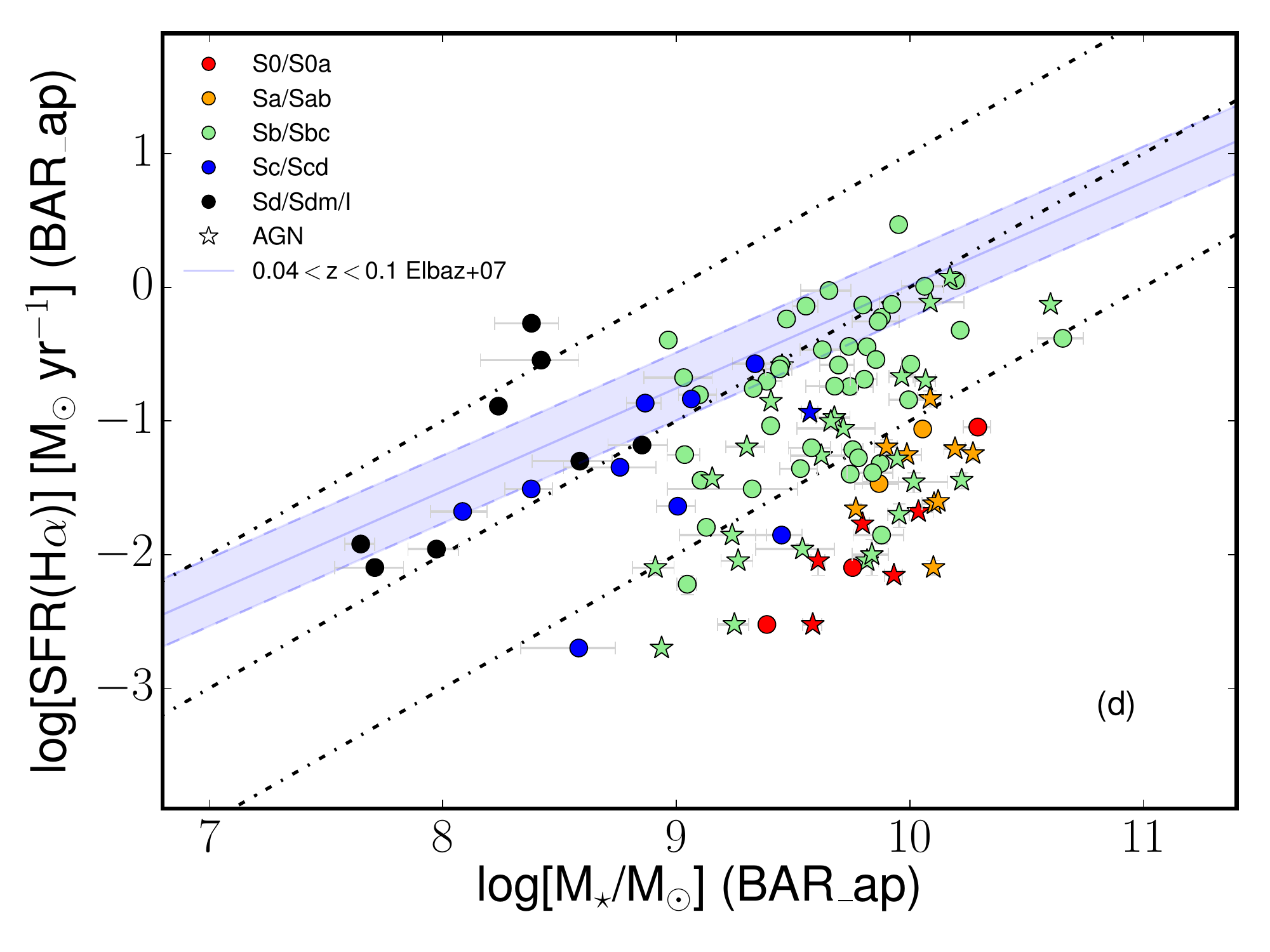} \\
\caption{Top left panel (a): SFR(H$\alpha$)-M$_{\star}$ plane for integrated values of the galaxies in our sample. The blue shaded area in this diagram shows the position of the Main Sequence (MS) using the fit by \citet{Elbaz_2007}. Star-forming galaxies are represented by circles and AGNs by stars symbols. The green line shows the fitting to Sb/Sbc star-forming objects while the blue line is the fitting to Sc/Scd/Sd/Sdm star-forming galaxies. The slope of the previous fittings appeared in the legend. The color-coding is used to distinguish the different morphological types of the galaxies. The dot-dashed lines indicate lines of constant sSFR (10$^{-9}$, 10$^{-10}$ and 10$^{-11}$ yr$^{-1}$ from top to bottom, respectively). Top right panel (b): SFR(H$\alpha$)-M$_{\star}$ plane for the disk component. Bottom left panel (c): SFR(H$\alpha$)-M$_{\star}$ plane for the bulge component. Bottom right panel (d): SFR(H$\alpha$)-M$_{\star}$ plane for the bar component. Lines, symbols and colors in panels (b), (c), and (d) are the same as in the panel (a). The {\it ap} subscript in the labels indicates that these are (smooth-)aperture measurements.}\label{fig: main_sequence_component}
\end{figure*}

The correlation observed between SFR and stellar mass (M$_{*}$) often referred to as the galaxy ``Main Sequence" (MS) has been extensively studied in the local Universe and at high redshift \citep{Noeske_2007, Daddi_2007, Elbaz_2007, Elbaz_2011, Wuyts_2011, Whitaker_2012, Whitaker_2014, Whitaker_2015, Speagle_2014, Magnelli_2014, Renzini_Peng_2015, catalan_torrecilla_2015, Lee_2015, Cano_diaz_2016, Duarte_puertas_2016}.

Top left panel in Figure~\ref{fig: main_sequence_component} shows the MS for the galaxies in our sample classified according to their morphological type. For the sake of clarity, we include here the fitting by \citet{Elbaz_2007} that shows the region of the diagram where local star-forming galaxies are placed. Most of the late-type galaxies in our sample are located in this region. On the other hand, S0/S0a, Sa/Sab and some Sb/Sbc galaxies are comparatively less efficient at forming stars at the present time, meaning that for the same stellar mass they are placed outside the MS as shown in this diagram. 

Some of the previously mentioned studies claimed that there is a turn over of the MS for stellar masses  M$_{*} > $10$^{10}$ M$_{\odot}$. We analyze whether or not this particular trend is also present in our sample. Instead of imposing a stellar mass cut, we divide the sample into two groups: i) Sb/Sbc objects and ii) Sc/Scd together with Sd/Sdm galaxies. Nevertheless, this morphological type cut-off is quite similar to the one used for stellar mass as the majority of Sb/Sbc galaxies tend to have stellar masses larger than 10$^{10}$ M$_{\odot}$ while most of the Sc/Scd and Sd/Sdm objects have masses below 10$^{10}$ M$_{\odot}$. Moreover, the fact that massive late-type spirals are clearly on the MS while early-type ones of the same mass are significantly offset does advice on the use of other criteria besides mass to perform the analysis of the MS. The fittings for both cases are shown in the top left panel of Figure~\ref{fig: main_sequence_component}  (green and blue lines, respectively). Star-forming galaxies in Figure~\ref{fig: main_sequence_component} are represented by circles while AGN objects appeared as stars symbols. The fittings are only done for star-forming galaxies. There is an offset between them in the sense that Sb/Sbc galaxies tend to have lower SFR values for the same stellar mass. It is also important the change in the slope (0.74 $\pm$ 0.09 for Sc/Scd/Sd/Sdm, 0.63 $\pm$ 0.12 for Sb/Sbc) that goes in the direction of an extra flattening in the case of the Sb/Sbc objects \footnote{The relation given by \citet{Elbaz_2007} is already tilted relative to the lines of constant sSFR}. As our sample does not contain highly inclined disks (due to the criteria imposed for the 2D decomposition, Section~\ref{2d_photometric}) we avoid effects that might be associated with an underestimate of the SFR which would affect the slope and width of the MS \citep[see][]{Morselli_2016}. Therefore, we are in agreement with the authors that find a turn over of the MS and we confirm this result for our sample. We go beyond this as we find that is not only mass driven but also related with the galaxy morphological type.

Although the analysis of the MS for integrated properties of galaxies is extremely valuable, we highlight the necessity of studying if the MS is also present when galaxies are separated in their stellar structures (bulges, bars, and disks). In fact, there is a key question that still remain unsolved, do disks of galaxies that are quenched as a whole (i.\,e. are found away from the MS) populate the MS?

To shed some light on this issue we analyze the ``Disks MS", that is, the relation between the SFR in the disk component and the stellar mass of the disk (top right panel of Figure~\ref{fig: main_sequence_component}). As we have done for the case of integrated values, we focus our attention on intermediate-to-late-type galaxies. We find that the global trend for the MS is also reproduced for the case of the disks. Moreover, the fittings to the different morphological types, Sb/Sbc and Sc/Scd-Sd/Sdm, show a similar behavior when compared with the integrated values. There is an offset between both fits and the slope is also steeper for Sc/Scd-Sd/Sdm galaxies (0.80 $\pm$ 0.10) compared to Sb/Sbc (0.51 $\pm$ 0.14). Then, we conclude that the current-to-past SFR has decreased in more massive disks and in earlier-type spirals relative to less massive and later-type systems. Not only entire galaxies but also disks in more massive systems have been more efficiently quenched. We note here that the dynamical range for stellar masses is quite similar for disks and for integrated galaxies, so when we refer to more massive systems, in general, we are referring to more massive disks as well. We find in the same figure that many of the disks, mainly S0/S0a and Sa/Sab, are still away from the MS on their own. 

The position of the bulges in the SFR(bulge)-M$_{\star}$(bulge) plane is shown in the bottom left panel of Figure~\ref{fig: main_sequence_component} while for the case of the bars the SFR(bar)-M$_{\star}$(bar) plane is shown in the bottom right panel of the same Figure. Bulges and bars are clearly much less efficient than disks in terms of their SFR even less if we take into account that in type-2 AGN some of the SFR associated to the central components might not be related to recent SF. 

\begin{figure}[h!]
\includegraphics[width=85mm]{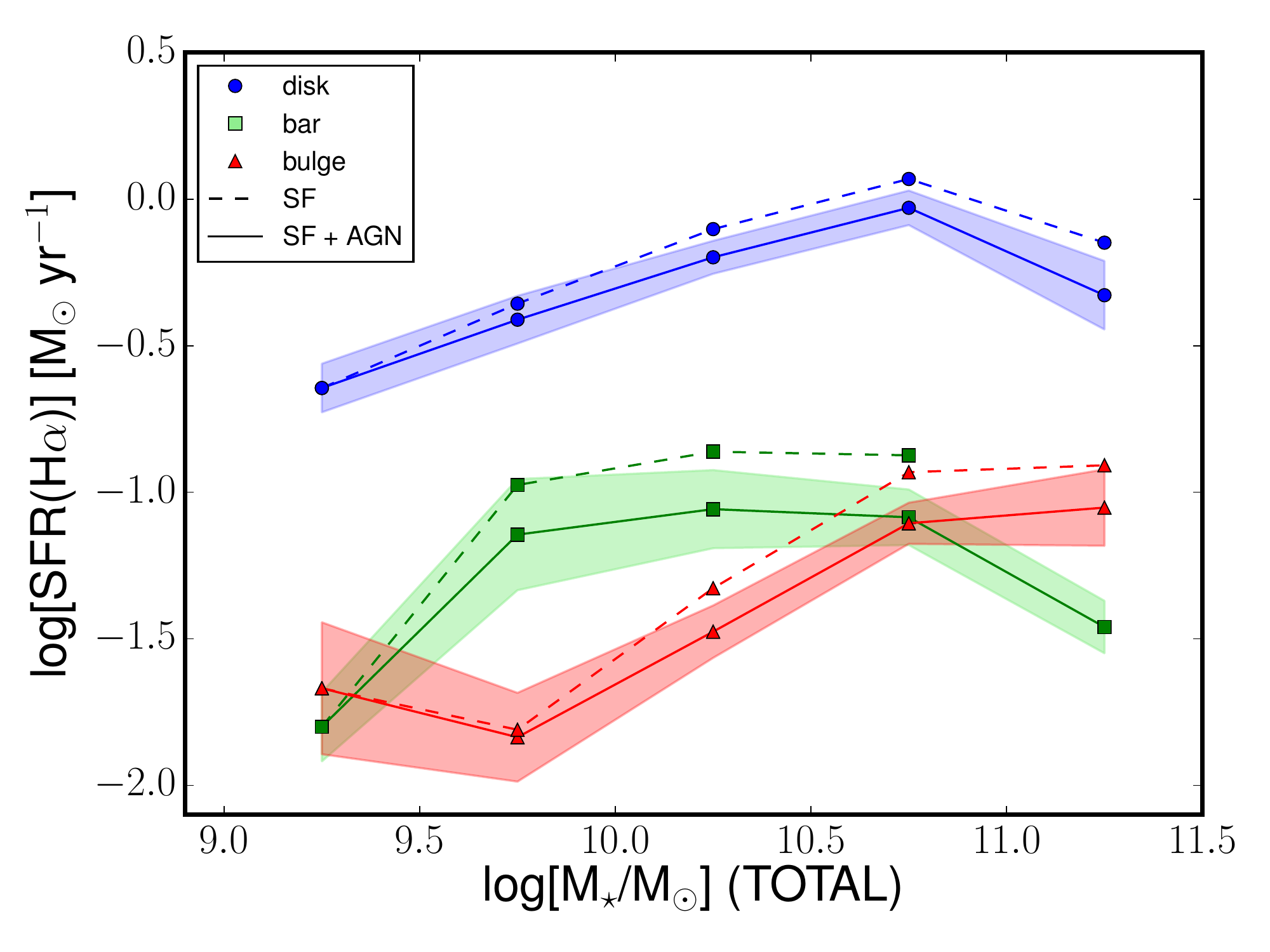}
\includegraphics[width=85mm]{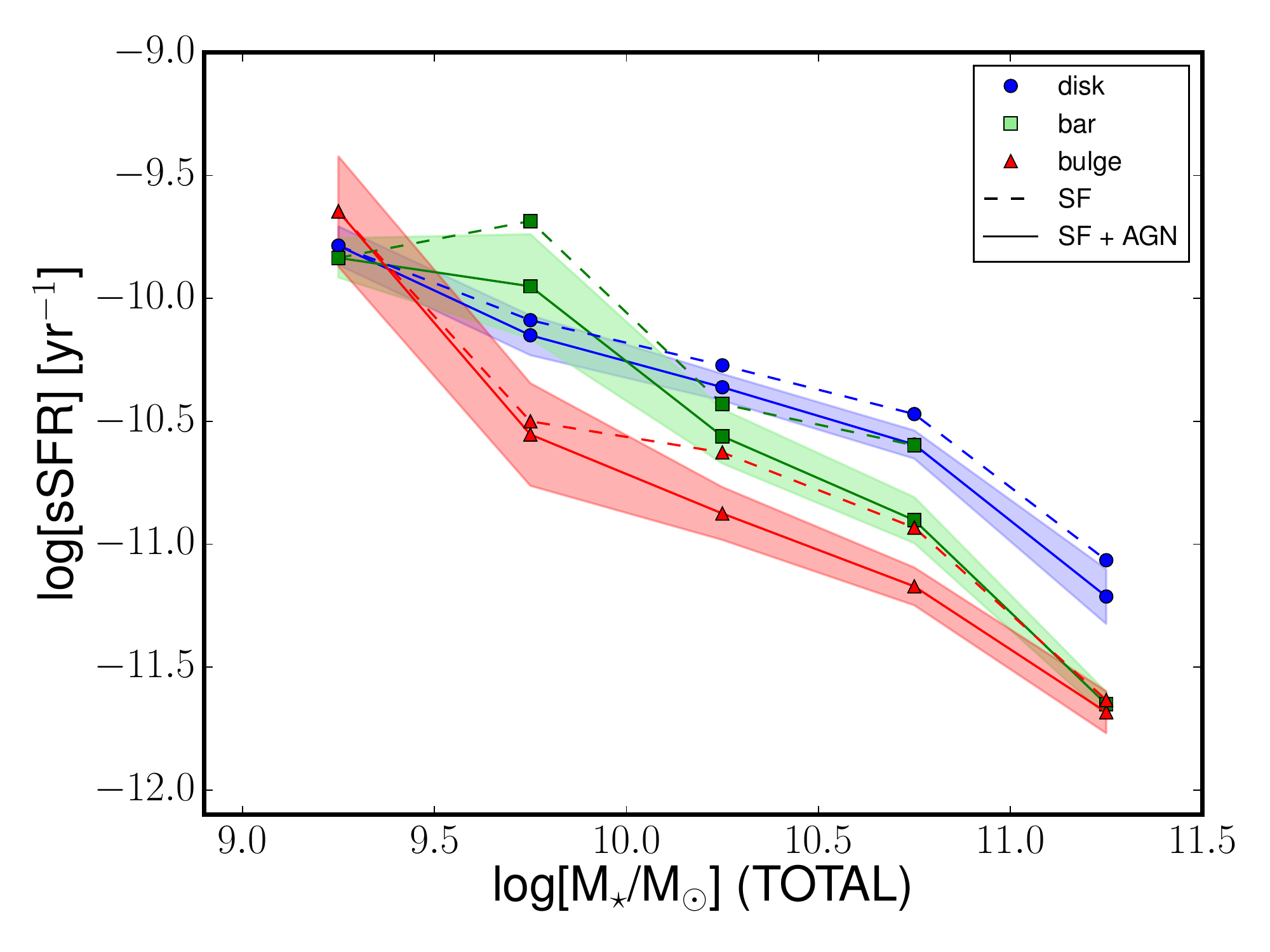} 
\caption{Top panel: Variation of the H$\alpha$-based SFR for the different morphological components of the galaxies (bulge, bar, and disk) with the total stellar mass of the galaxy. The trends for bulges, bars, and disks are shown in red triangles, green squares and blue circles, respectively. Filled contours represent the 1$\sigma$ dispersion around the mean value expressed as a red, green and blue solid line for the bulge, bar, and disk, respectively. Dashed lines show the trends just for the star-forming galaxies.  Bottom panel: Same description as the top panel this time for the sSFR.\label{fig:sfr_ssfr_mass}}
\end{figure}

Until now we have shown the SFR trend of each galaxy component with their corresponding stellar mass (bulges, bars, and disks). Now, we focus on the analysis of the SFR of each component with the total galaxy stellar mass instead. Top panel in Figure~\ref{fig:sfr_ssfr_mass} shows the trends for the variation of the SFR in the bulge, bar, and disk component in bins of 0.5 dex in stellar mass. We have combined at the same time all the morphological types for each component (which will obviously increase the dispersion as early-type spiral have lower values of their SFR specially for the disk component). It can be seen from this figure that most of the actual SFR in galaxies is located in the disk component as it is expected while bars and bulges show a smaller contribution for a fixed stellar mass. As seen previously for the disks, not only with morphological type but also with stellar mass there is a clear decrease in the SFR for more massive disk galaxies (i.\,e., more massive systems in general due to the similar range in total and disk stellar masses).

To conclude, we have demonstrated in this section that more massive star-forming disks and earlier-type spiral disks show a higher level of quenching. Previous studies have shown that more massive star-forming galaxies (understanding galaxies as entire systems) tend to be less efficient at forming new stars. Here, the important fact is that we treat disks as separate components of the galaxies.

%%%%%%%%%%%%%%%%%%%%%%%%%%%%%%%%%%%%%%%%%%%%%%%%%%%%%%%%%
%%%%%%%%%%%%%%%%%%%%%%%%%%%%%%%%%%%%%%%%%%%%%%%%%%%%%%%%%

%%%%%%%%%%%%%%%%%%%%%%%%%%%%%%%%%%%%%%%%%%%%%%%%%%%%%%%%%
%%%%%%%%%%%%%%%%%%%%%%%%%%%%%%%%%%%%%%%%%%%%%%%%%%%%%%%%%
\subsection{sSFR-M$_{\star}$ relation for bulges and disks: a clue for the quenching of massive systems} \label{specific_sfr}

\begin{figure*}
\includegraphics[width=85mm]{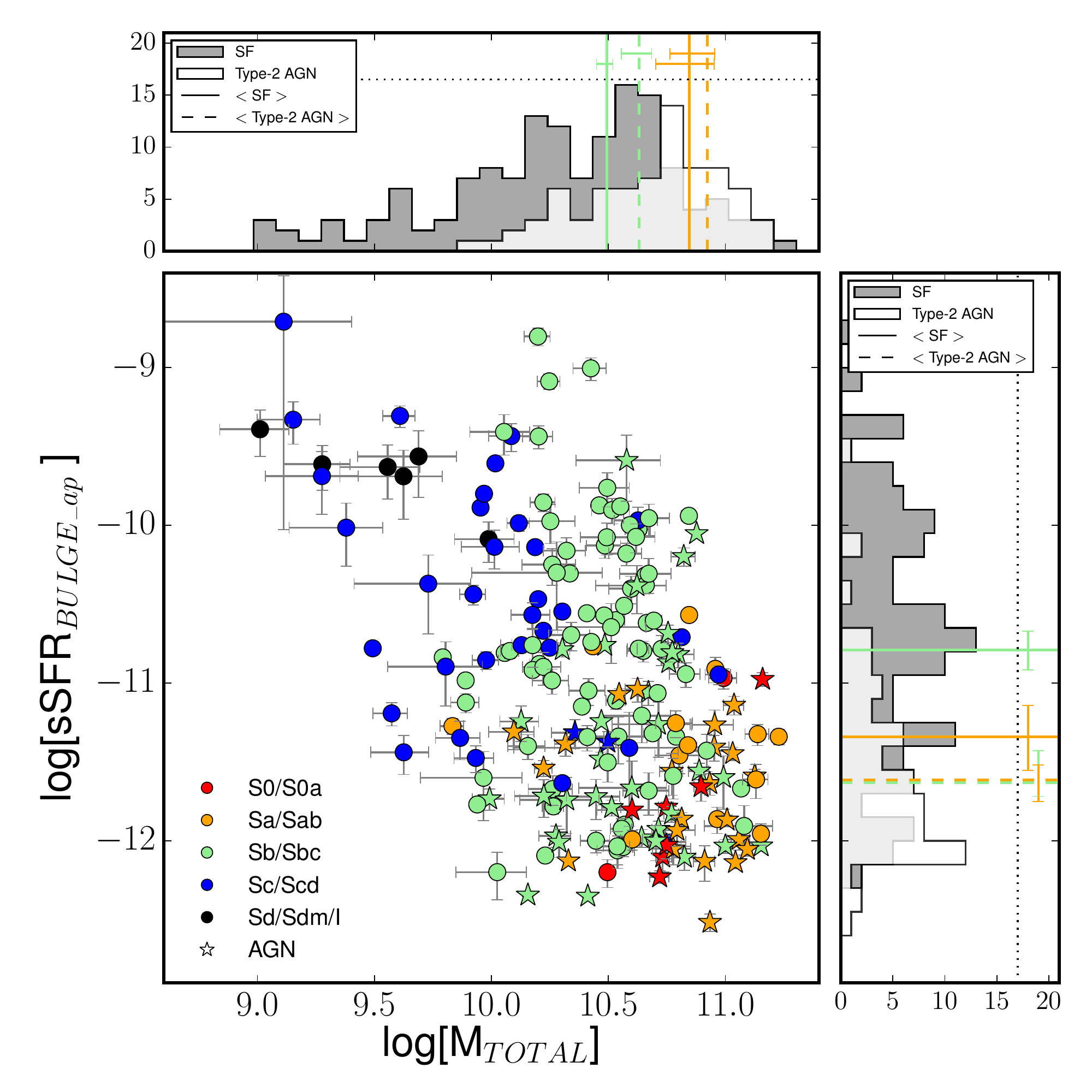} 
\includegraphics[width=85mm]{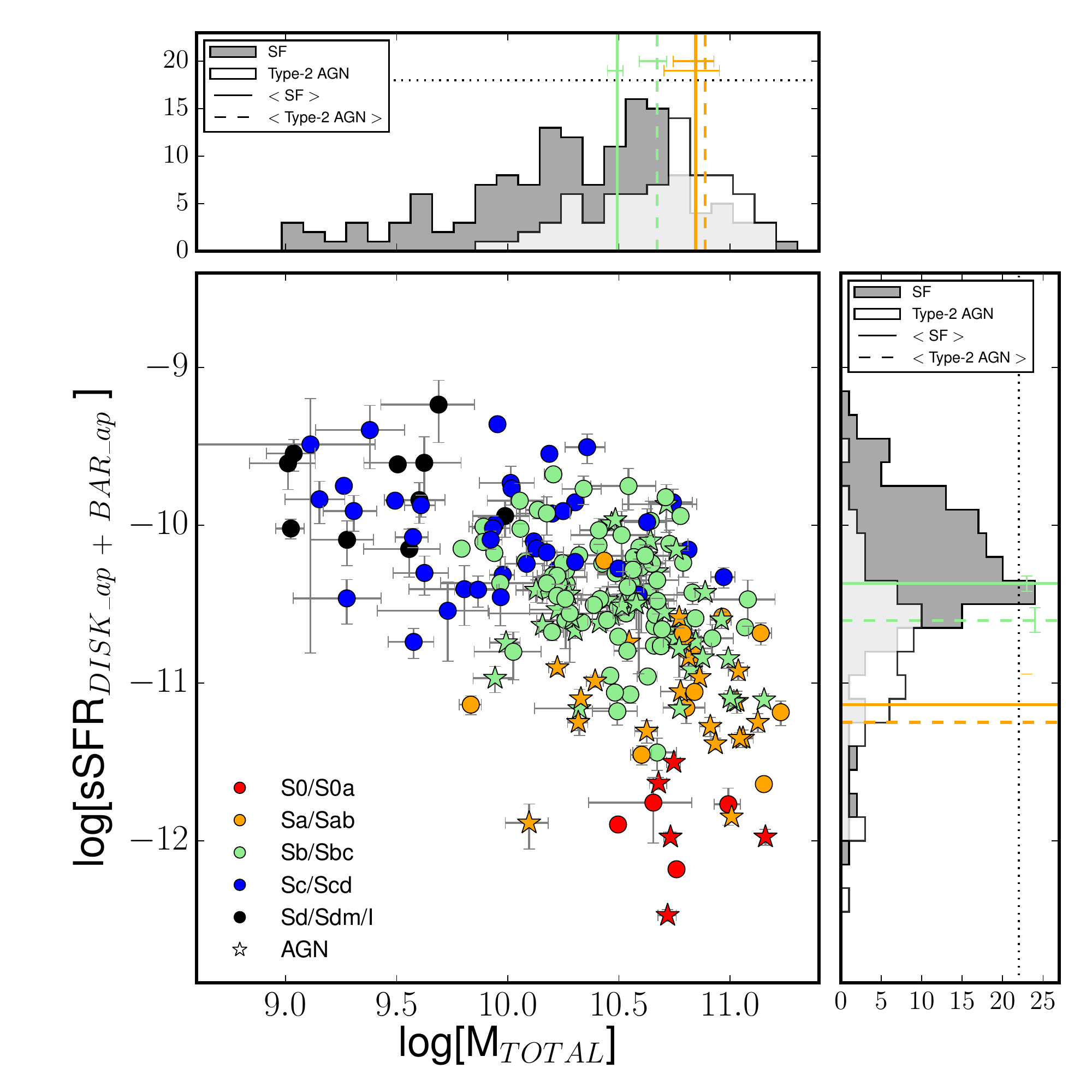}
\caption{Left panel: sSFR for the aperture of the bulge component as a function of the total stellar mass. The circles refer to the star-forming galaxies in our sample while the stars indicate the AGNs objects. The color of each symbol indicate the corresponding morphological type of the galaxy. The top histogram shows the distribution of the bulges with the total stellar mass of the corresponding galaxy that could be classified as star-forming or type-2 AGN. Vertical solid (dashed) lines represent the median value of SF (AGN) galaxies while the errors bars are computed using the standard error of the median, i./,e., 1.253$\times$$\sigma$/sqrt(N) where $\sigma$ is the interval that contains 68$\%$ of the points. Sb/Sbc (Sa/Sab) types appeared in light-green (orange). The right histogram shows the distribution of the galaxies as a function of the sSFR in the bulge component. Vertical lines represent median values and their errors. Right panel: same description as the left panel this time for the aperture of the disk + bar component. \label{fig: specific_sfr}}
\end{figure*}

In the previous Section, we have explored in detail the Main Sequence and the fact that the same relation that applies to star-forming galaxies as a whole is also valid for the disk component of the same galaxies. More surprisingly, however, galaxies that are offset from the main sequence have disks that are also forming stars at present at a lower rate than in the past compared to MS galaxies (i.\,e. they do not fall on the MS), so the position of galaxies relative to the MS is not only due to a larger contribution of the bulge component but also to a decrease in the recent SFR of their disks normalized to their mass (see below). This is true even for well-defined Sb/Sbc spirals. In other words, there is what we have called the ``Disk Main Sequence" and galaxies that are away the MS have disks that are also away from the MS. To associate this finding with the capacity for galaxies to form stars at present time (compared with that in the past), we analyze here the specific SFR of both bulges and disks as a function of the galaxy total stellar mass. Ultimately, we aim to answer the following question, what are the mechanisms responsible for the quenching of the most luminous and massive galaxies and their disks?

\citet{Abramson_2014} proposed that sSFR(disk) is approximately constant with stellar mass for M$_{\star}$ $>$ 10$^{10}$ M$_{\odot}$ and B/T $<$ 0.6. The authors assume sSFR(disk) $=$ SFR(total)/M$_{\star}$(disk) and that nuclear and bulge regions might have small contributions to the SF. If this was the case, the growth of bulges may be the potential cause to create the flattening in the MS for the higher stellar masses. Nevertheless, we argue here that bulges also contribute to the SFR in those galaxies with higher values of their stellar masses (as shown in the previous Section). Figure~\ref{fig: specific_sfr} shows the relation of the sSFR(bulge) and sSFR(disk + bar) with the total stellar mass for the galaxies in our sample. From this figure, we conclude that sSFR(disk + bar) is not constant with stellar mass meaning that disks are not equally active at forming stars in terms of their sSFR. Besides, from left panel in Figure~\ref{fig: specific_sfr} it can be seen that the sSFR(bulge) spans a wide range (more than 2 dex) of values and that the most active bulges present a not negligible value of their sSFR. As commented in Section~\ref{sfr_ratios}, attending to the n$_{b}$ parameter 72\,$\%$ of our bulges would be classified as pseudobulges while the remaining 28\,$\%$ would appear as classical bulges. \citet{Fisher_2016} established that bulges should be forming stars actively for sSFR $>$ 10$^{-11}$ yr$^{-1}$ (typically pseudobulges) while they might be either pseudobulges or classical bulges for lower values of the sSFR. We find a median value of 1.7 $\times$ 10$^{-11}$ (8.9 $\times$ 10$^{-12}$) yr$^{-1}$ for pseudobulges (classical bulges). Determining whether or not sSFR provide an accurate separation between bulges or pseudobulges is beyond the scope of this paper and would require of high-resolution imaging of the nuclear regions, which is not available for the vast majority of the galaxies in our sample.

Other potential mechanism to quench the star formation of the more massive galaxies could be the presence of an AGN. Although many studies include only galaxies that are strictly star-forming, we also include here type-2 AGN to study their relative position in the sSFR-stellar mass plane. The power of IFS data will certainly help us to resolve whether or not the presence of AGN contribute to the quenching of the massive galaxies. We recently reported in \citet{catalan_torrecilla_2015} (Figure 19) that AGN might have an impact at suppressing the total SFR in their host galaxies. Other works corroborate the idea of the suppression of the star formation by AGNs in the host galaxies \citep{Shimizu_2015,Leslie_2016}. In this Section, we investigate the role of AGN in the quenching of the SFR, not only in global terms but also in their bulges and disks separately. This is particularly important considering that, as shown above, galaxies that are away from the MS host disks that have their star formation depressed/suppressed, so AGN quenching should thus work at galactic-wide scales. The alternative is that AGN quenching is not the dominant mechanism but it is  coeval with another mechanism(s) that has an impact on the star formation at those scales. One possibility is the removal of a fraction of the high-angular momentum gas of the disks due to interactions towards the nucleus (leading to an AGN) becoming unavailable for star formation in the disk component. 

To investigate this possibility, we examine the sSFR-stellar mass plane shown in Figure~\ref{fig: specific_sfr} for bulges and disks, separately. Some interesting results emerge from these plots. First, type-2 AGN are not homogeneously distributed in the plane. They tend to be in high mass end. Indeed, type-2 AGN are mostly found in galaxies with stellar mass values in the range between [10$^{10}$ - 10$^{11.5}$] M$_{\odot}$ (white histograms on the top of both panels in Figure~\ref{fig: specific_sfr}). We also find that there is a clear decrease in the sSFR values when a type-2 AGN is present. Bulges of AGN hosts show a median sSFR(bulge) that is 0.89 dex below that of star-forming galaxies when the difference in median stellar mass is $+$0.32 dex. For the case of the disks there is a $-$0.52 dex difference in median sSFR(disk) and $+$0.41 dex in median mass. Nevertheless, it is important to quantify whether this effect is still present in terms of the same morphological type or not. Thus, due to the lack of type-2 AGN in most of our late-type galaxies, in agreement with previous works \citep{Moles_1995}, we restrict the following analysis to Sa/Sab and Sb/Sbc objects. Bulges of Sa/Sab (Sb/Sbc) show a median sSFR that is 0.27 (0.84) dex below that of star-forming galaxies while the difference in the median value of the stellar mass is 0.08 (0.14) dex. For the case of the disks, Sa/Sab (Sb/Sbc) galaxies exhibit a difference in the median values of sSFR for star-forming and AGNs of 0.11 (0.23) dex while the difference in stellar masses is 0.04 (0.16) dex (solid and dashed vertical lines in the top and right histograms of right panel in Figure~\ref{fig: specific_sfr}). If the bars are excluded the median values of sSFR for star-forming and AGNs are 0.13 (0.20) dex for Sa/Sab (Sb/Sbc). The previous results suggest a possible damping of the SFR in both components (bulges and disks) due to the presence of  AGNs. We prefer the term damping here as compared to quenching. It is not clear whether this decrease in the sSFR is enough (neither if it lasts enough) to make these galaxies evolve towards and remain in the red sequence, something for which galaxy evolution models require of a strong quenching of the star formation in massive galaxies at high redshift \citep[][and references therein]{Weinberger_2017}. Also, we find that bulges show a constant decline of the sSFR across the entire stellar mass range. On the contrary, the decrease in the disk component is more dramatic when galaxies reach a certain stellar mass, typically around 10$^{10.5}$ M$_{\odot}$. Finally, a significant trend with the morphological type is also found. Late-type galaxies have higher values of their sSFR for both components, bulges and disks. 

To clarify the previous trends, bottom panel in Figure~\ref{fig:sfr_ssfr_mass} shows the variation of the sSFR in the different morphological components (bulge, bar, and disk) in bins of 0.5 dex in total stellar mass. As done previously in the case of the SFR (top panel in the same figure), all the morphological types for each component are combined at the same time (spreading the dispersion as early-type spiral present lower sSFR values). Again, it is clear from this figure that the disk component is significantly more effective than the bulge at forming new stars, specially for M$_{\star}$ $>$ 10$^{9.5}$ M$_{\odot}$, and the stepper decline for the bulges at the lower stellar mass bin.

From the results in this section, we conclude that the presence of an AGN might be linked with some level of the damping of the SFR in both the bulge and the disk component even in the local Universe. For both cases, the sSFR decreases when an AGN is present being this effect higher for the bulges in competition with the effect of the bars. We identify the same behavior among different morphological types such as Sa/Sab and Sb/Sbc. Again, due to the short timescale traced by the H$\alpha$ line emission we cannot infer whether the AGN phase is cause, consequence or coeval with the star formation quenching/damping process. Besides, as discussed in Section~\ref{agn}, we cannot exclude that a fraction of these low-luminosity AGN could be powered by hot evolved stars in regions with basically null recent star formation.

\begin{figure}
\includegraphics[width=85mm]{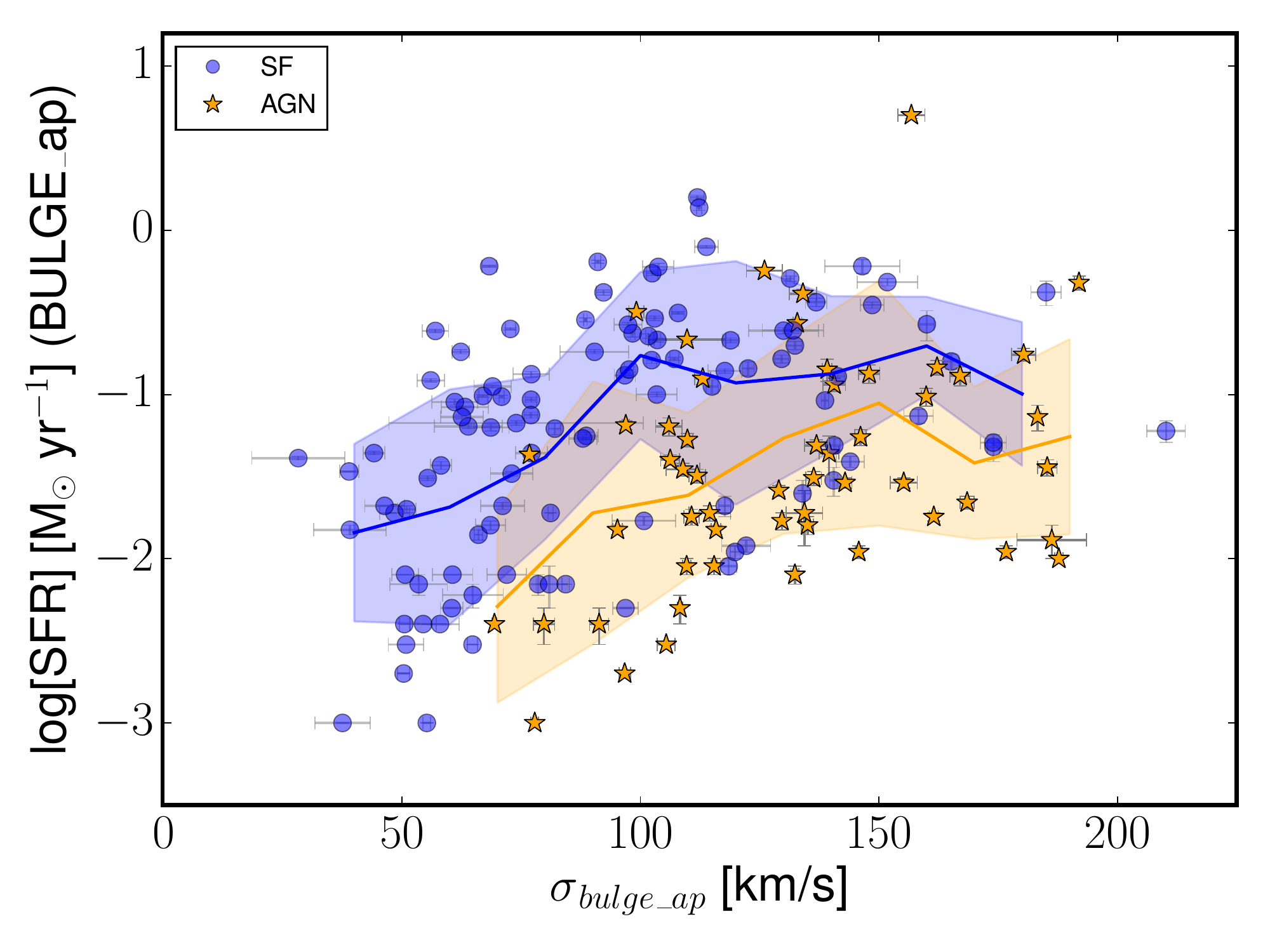} \\
\includegraphics[width=88mm]{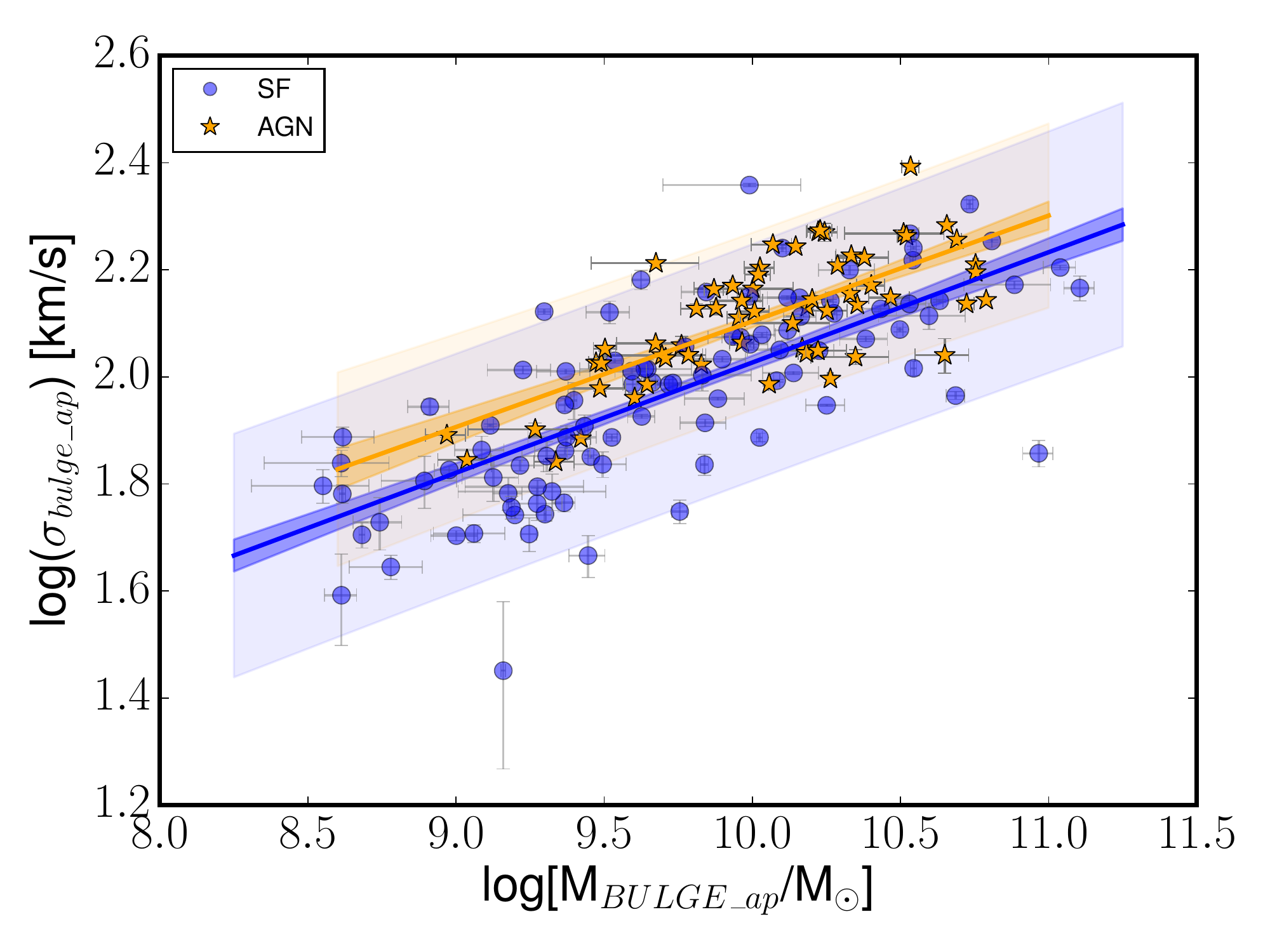} 
\caption{Top panel: SFR in the aperture of the bulge component versus line-of-sight stellar velocity dispersion. Galaxies are plotted using different colors and symbols, blue circles for star-forming objects and orange stars for type-2 AGN. Blue (orange) filled contours represent the 1$\sigma$ dispersion around the mean value showed in blue (orange) solid line for SF (type-2 AGN) galaxies. Bottom panel: Faber-Jackson relation for the bulges in our sample. The blue (orange) solid line shows the best-fitting for SF (AGN) galaxies. Dark shaded areas correspond to the error bands of the fittings when only errors associated to slope and intercept are taking into account. Light shaded areas mark the global uncertainty bands once an additional {\it s} parameter that takes into account the intrinsic variation of the points is also included. \label{fig:sfr_kinematics}}
\end{figure}

\begin{figure*}
\includegraphics[width=85mm]{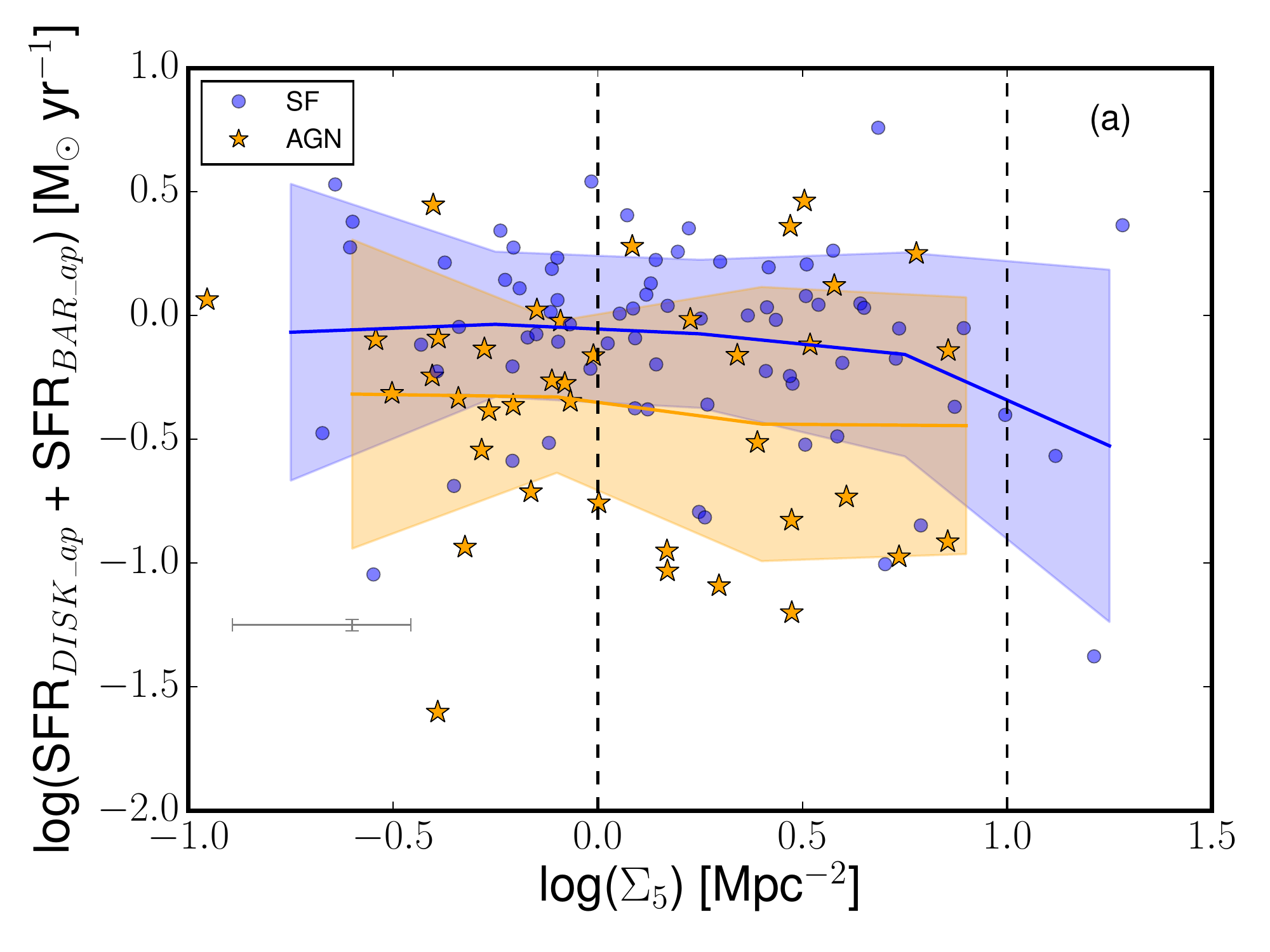} 
\includegraphics[width=85mm]{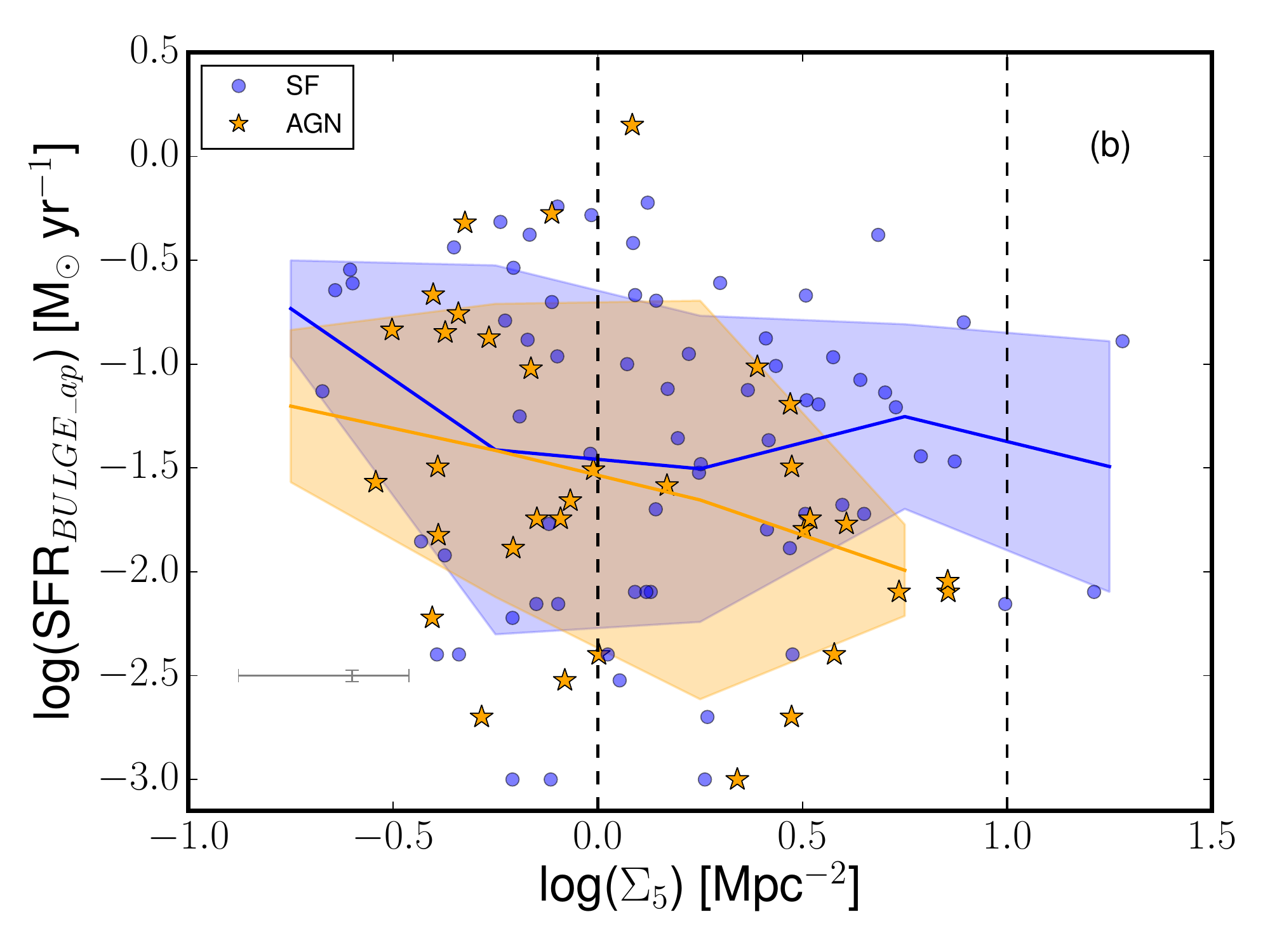} \\
\includegraphics[width=85mm]{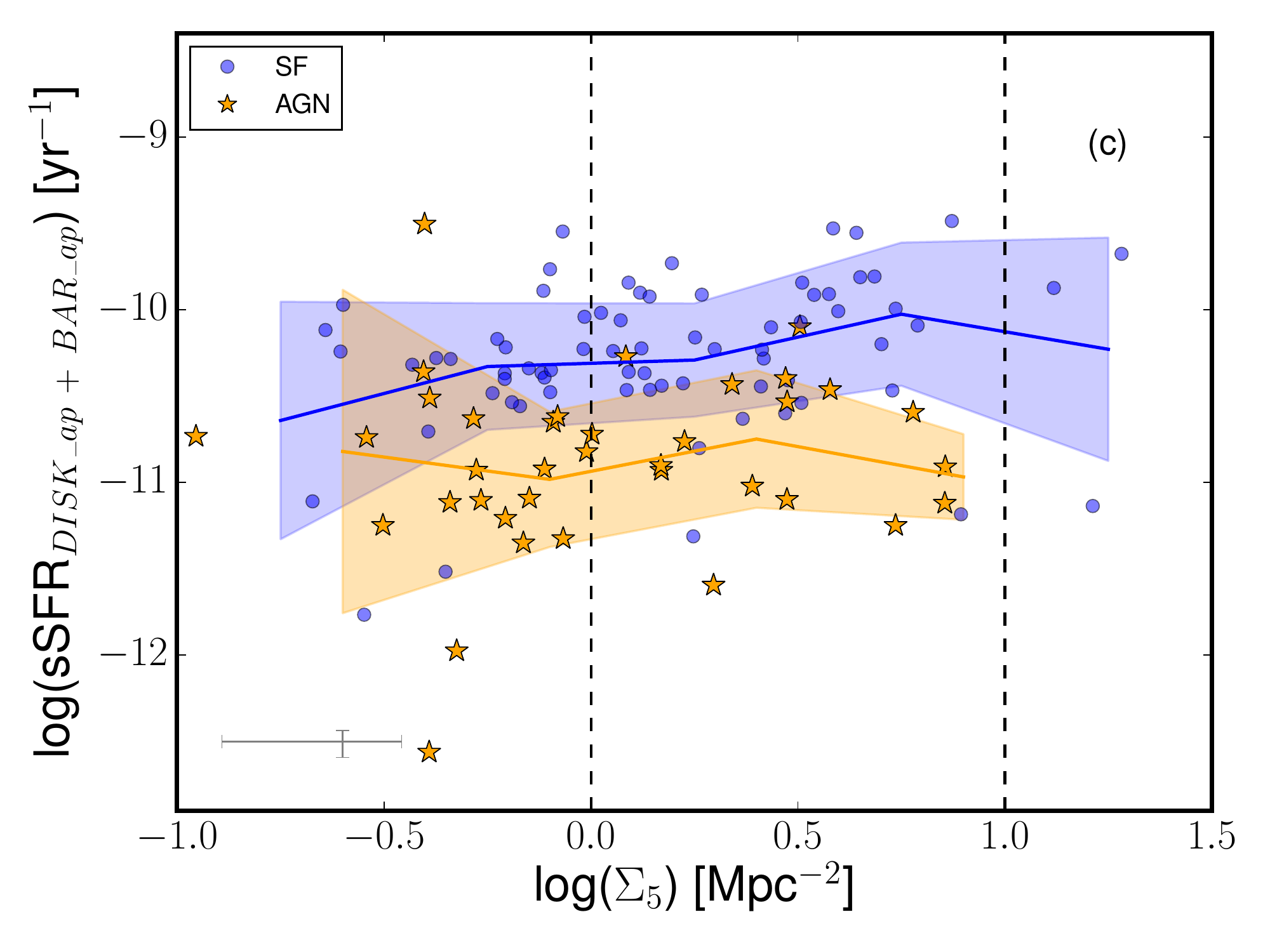} 
\includegraphics[width=85mm]{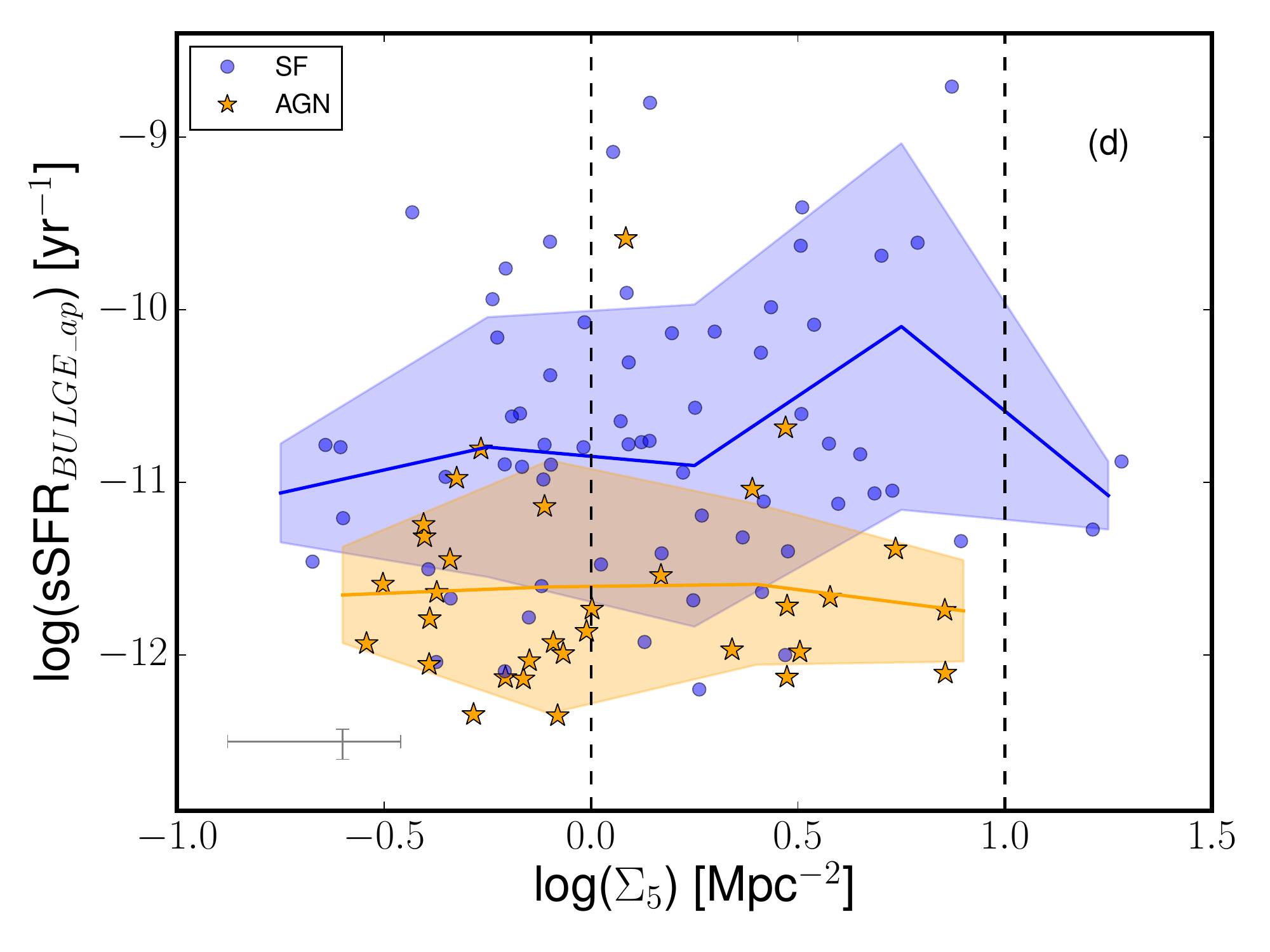} \\
\caption{Top left panel (a): Variation of the SFR in the aperture of the disk + bar component with the $\Sigma$$_{5}$ parameter. Blue (orange) points denote star-forming (type-2 AGN) galaxies. Blue (orange) filled contours represent the 1$\sigma$ dispersion around the mean value expressed as blue (orange) solid line for the star-forming (type-2 AGN) galaxies. Vertical dashed lines correspond to log($\Sigma$$_{5}$) $=$ 0.0 and log($\Sigma$$_{5}$) $=$ 1.0, i.\,e., the demarcation between low-medium and medium-high density environments, respectively. Top right panel (b): Same description as panel (a) but for the SFR in the bulge component. Bottom left panel (c): Variation of the sSFR in the disk + bar component as a function of the $\Sigma$$_{5}$ parameter. Color-coding and symbols are the same as in previous panels. Bottom right panel (d): sSFR in the bulge component as a function of the $\Sigma$$_{5}$ value. Same color-coding and symbols as in previous panels. \label{fig:sfr_sigma}}
\end{figure*}

%%%%%%%%%%%%%%%%%%%%%%%%%%%%%%%%%%%%%%%%%%%%%%%%%%%%%%%%%
%%%%%%%%%%%%%%%%%%%%%%%%%%%%%%%%%%%%%%%%%%%%%%%%%%%%%%%%%

%%%%%%%%%%%%%%%%%%%%%%%%%%%%%%%%%%%%%%%%%%%%%%%%%%%%%%%%%

\subsection{Relation with other parameters}\label{other_parameters}

As discussed previously in Sections~\ref{main_sequence} and \ref{specific_sfr}, stellar mass seems to be the main driver of the star formation, and, after it, AGN activity also plays an important role. Nevertheless, it is worth exploring the role of other (possibly secondary) parameters that are known to either trigger or quench star formation. In that regard, the following subsections aim to shed some light on the effect that stellar kinematics and the environment have on the star formation processes taking place in our sample.

\subsubsection{Stellar kinematics} \label{kinematics}

In this section we explore how stellar kinematics could regulate the star formation in the inner regions of our galaxies. With this aim in mind, we analyze the line-of-sight (LOS) stellar velocity dispersions for the bulge component. We have restricted the analysis of the LOS velocity dispersions to the bulge component due to the fact that the LOS velocity dispersion distribution for each component (bulge, bar, and disk) is quite distinct in the regions where they coexist. This case is specially important for the internal parts of the disks where the values could be affected by the bulge contamination as this component tend to be the more prominent there. Thus, measuring the stellar velocity dispersion of only the disk component presents intrinsic limitations. For the previous reason, we will focus here on the possible impact of the stellar velocity dispersion in bulges on their SFR. 

We employ the CALIFA stellar velocity dispersion maps created by \citet{Falcon_Barroso_2016} using V1200 grating data. In order to calculate the integrate velocity dispersions for the bulge, we first multiply the stellar velocity dispersion map by the luminosity-weight map of the bulge component in the {\it g}-band (previously derived as explained in Section \ref{2d_photometric}). Then, we divide it by the {\it g}-band luminosity taking into account only those pixels where the dispersion values are greater than zero. The method applied to obtain the luminosity-weight maps is the same as the one explained in Section \ref{2d_photometric}. The only difference is that here we used Voronoi bins instead of spaxels as each Voronoi bin provides its own velocity dispersion for the stars. Thus, the expression used to obtain the LOS stellar velocity dispersion for each bulge component is the following:

\begin{equation}
 \sigma_{bulge} = \frac{\sum_{i=1}^{N} F_{i} \cdot \sigma_{i}}{\sum_{i=1}^{N} F_{i}}
\label{dispersion_formula}
\end{equation}

where the {\it i} subscript refers to the Voronoi bin used in each case. The values of the $\sigma$$_{bulge}$ calculated in this Section are given in Table~\ref{table}. The methodology followed in this work is similar to other kinematical parameters based on 2D spectroscopic data used in the literature (see e.g. \cite{Emsellem_2011} for a similar recipe for $\lambda_{\mathrm{R}}$) and it is easily reproducible by other authors using data from different instruments. Moreover, it allows to go beyond the standard method as we apply the luminosity-weight maps for the bulge component that should restrict in a better way the calculation of the LOS velocity dispersions. 

In Figure~\ref{fig:sfr_kinematics} (top panel) we show the relation between the SFR in bulges and the stellar velocity dispersions computed as in Equation~\ref{dispersion_formula}. The sample is separated by spectral class (AGN and SF). We find that for the same LOS velocity dispersion, star-forming galaxies show higher bulge SFRs that those of AGN hosts. This, in principle, might be simply due to the correlation between stellar mass and stellar velocity dispersion found in ellipticals and bulges \citep{Faber_jackson_1976,Chilingarian_2008,Falcon_Barroso_2011}  and the noisy correlation between the former and the SFR (see bottom left panel in Figure~\ref{fig: main_sequence_component}). 

%% unir los siguientes cachos %%
Additionally, a higher $\sigma$$_{bulge}$ (even for the same stellar mass) could also contribute to dynamically heating the gas and to reduce the efficiency of star formation.  In order to test whether both effects (or only the stellar mass) are at play, we compare the $\sigma$$_{bulge}$ and stellar mass values of our bulges in Figure~\ref{fig:sfr_kinematics} (bottom panel). The blue (orange) solid line shows the best-fitting for SF (AGN) galaxies. We have employed the Markov chain Monte Carlo (MCMC) method to sample the probability density function of our model parameters. The {\it Pymc3} code \citep{pymc3} is used to implement the analysis. Slope and intercept of a line are computed considering uncertainties in both axes. Also, an additional {\it s} parameter that takes into account intrinsic variations of the individual points is included. The best-fitting for SF galaxies is $-$0.035 ($\pm$ 0.180) + 0.206 ($\pm$ 0.018) $\times$ log[M$_{BULGE}$/M$_{\odot}$] with a {\it s} $=$ 0.110 $\pm$ 0.009 while for the AGNs is 0.130 ($\pm$ 0.253) + 0.197 ($\pm$ 0.025) $\times$ log[M$_{BULGE}$/M$_{\odot}$] with a {\it s} $=$ 0.082 $\pm$ 0.008. A similar value for the slope in both cases is found while there is a slighter higher value for the intercept of AGNs. This indicates higher $\sigma$$_{bulge}$ values for the AGNs at stellar masses larger than 10$^{9.5}$ M$_{bulge}$/M$_{\odot}$ (the bulge stellar mass range where most of the SF and AGN coexist). Dark shaded area corresponds to the error bands of the fitting when only errors associated to slope and intercept are taking into account. Light shaded area marks the global uncertainty bands once the additional {\it s} is also included. If we fix the slope of the fits to both datasets to 1/4 then the mean difference in $\sigma$$_{bulge}$ between the two samples would be 0.03 dex. This Faber-Jackson relation shows that even for the same stellar mass, star-forming galaxies tend to have a lower $\sigma$$_{bulge}$, suggesting that a dynamically cooler stellar population in the bulges can more easily host star formation.

%%%%%%%%%%%%%%%%%%%%%%%%%%%%%%%%%%%%%%%%%%%%%%%%%%%%%%%%%

%%%%%%%%%%%%%%%%%%%%%%%%%%%%%%%%%%%%%%%%%%%%%%%%%%%%%%%%%
\subsubsection{Environment} \label{environment}

Environment is another parameter that can strongly affect the SFR and further stellar mass growth of galaxies and components within galaxies. It is also thought to be the cause of the well known morphology-density relation \citep{Dressler_1980}. The main three broad mechanisms proposed to play a role in this sense are mergers/interactions (sometimes referred as galaxy harassment), ram pressure and viscous stripping of cold gas and strangulation in the supply of warm/hot gas \citep[see][]{Kawata_2008}. As these mechanisms act differently in different regions of galaxies and on different timescales, the study of the distribution of the current SFR is key to determine whether or not they are contributing on specific objects and which one dominates in each case \citep{Boselli_2006}. Moreover, in the case of mergers and interactions, they might lead to either quenching or triggering of the star formation depending on the type of interaction (mass ratios, impact parameters) and on the region considered (nuclear regions, outer disks or even tidal tails). Thus, to investigate whether or not the environment is playing a significant role on the SFR or sSFR of the different structural components of our galaxies, we use the local density values from the projected comoving distance to the 5$^{th}$ nearest neighbor of the target galaxy. The projected galaxy density, $\Sigma$$_{5}$, in number of galaxies per Mpc$^{2}$ is calculated as:

\begin{equation}
 \Sigma_{5} = \frac{N}{\pi (d_{5}) ^{2}}
\label{sigma_formula}
\end{equation}

We have reliable measurements for a total of 140 objects while we lack $\Sigma$$_{5}$ measurements for 87 galaxies (see Table~\ref{table}). This is mainly because the area enclosing the nearest neighbor lies outside the footprint of the SDSS survey. This means that for these galaxies we cannot obtain a reliable measurement of the density, since we do not know whether there is another close galaxy outside the survey area.

In Figure~\ref{fig:sfr_sigma} (top panels), we present the variation of the H$\alpha$$-$based SFR in the disk and in the bulge components as a function of galaxy density, $\Sigma$$_{5}$. We appreciate a weak trend between both parameters. Galaxies tend to have lower values of their SFR in both components (bulges and disks) for higher values of the galaxy density. These $\Sigma$$_{5}$ values are associated with medium and high density environments although the latter case is not well-sampled due to a lack of galaxies in this position of the diagram. The previous trend is consistent with other works that used galaxy density to estimate environmental effects associated with SFR but using integrated values, e.g., \citet{Gomez_2003}, and with high resolution cosmological simulations that show a reduction of the SFR in high-density environments at z=0 \citep{Tonnesen_2014}.

In order to properly assess this effect, which is also related with the mass of the galaxies, the bottom panels of Figure~\ref{fig:sfr_sigma} represent the relationship between sSFR (sSFR(disk) and sSFR(bulge)) and galaxy density. The evidence for a decrease in the SFR and sSFR in bulges and disks with the presence of type-2 AGN has been already discussed in Sections~\ref{main_sequence} and~\ref{specific_sfr}. We will focus here in the case of the SFR and sSFR measured in the disks of star-forming galaxies (left panels), as the trends found for bulges are clearly more noisy, albeit having similar slopes. It is clear that disks in star-forming galaxies with intermediate-to-high values of $\Sigma$$_{5}$ show higher sSFR values. The analysis of the morphological types that are responsible for the increase in the sSFR at intermediate densities (groups) indicates that this is due to a larger number of Sd (or later) galaxies being found in groups than in the field for our galaxy sample. The number of galaxies when split by environment and type is not large enough to drive firm conclusions. Despite that fact, an enhancement in the disk star formation activity for galaxies located in groups could increase the number of these objects in the sample due to either a positive bias towards actively star-forming systems being included in CALIFA or by means of a morphological transformation towards later types.

%%%%%%%%%%%%%%%%%%%%%%%%%%%%%%%%%%%%%%%%%%%%%%%%%%%%%%%%%

\section{Conclusions} \label{conclusions}

The uniqueness of combining IFS data and a 2D multi-component photometric decomposition makes possible to disentangle the distribution of the extinction-corrected H$\alpha$-based SFR within different stellar structures in galaxies (bulges, bars, and disks). It also allows to determine how these morphological components would grow in stellar mass due to {\it in-situ} star formation. With this aim in mind, we have analyzed which mechanisms might either trigger or quench the star formation in a sample of 219 CALIFA nearby galaxies.

This work led to the following main conclusions:

 \begin{enumerate}

\item{There is an enhancement of the central SFR and sSFR due to the presence of bars for star-forming galaxies in agreement with the results found in previous works \citep{de_Jong_1984,Devereux_1987,Ellison_2011,Wang_2012,Florido_2015}. This finding supports the idea that gas might be funneled into the central part of the galaxies triggering the star formation processes. On the other hand, this effect is reduced when a type-2 AGN is present making the SFR values in barred and unbarred galaxies more similar between them in terms of SFR (Section~\ref{sfr_ratios}).}

\item{We examine the SFR-M$_{\star}$ plane focusing on the Star-Forming Main Sequence treating galaxies as entire systems and also analyzing this sequence for their basic stellar structures (bulges, bars, and disks). The results indicate that there is a turnover in the Main Sequence not only for integrated values but also for disks, i.\,e., in the correlation between the SFR(disk) and the M$_{\star}$(disk). This fact means that also the disks of massive galaxies have been more efficiently quenched than their lower-mass counterparts (Section~\ref{main_sequence}).}

\item{The correlation between sSFR in the stellar components of the galaxies (bulge, bar, and disk) and the total stellar mass is analyzed to identify which mechanism(s) might be damping the star formation in more massive systems. First, we observe a decline associated to the sSFR(bulge) that is present across the entire stellar mass range while in the case of the sSFR(disk + bar) the decrease becomes significantly for M$_{\star}$ $>$ 10$^{10.5}$ M$_{\odot}$. We also find that galaxies hosting a type-2 AGN tend to have lower values of their sSFR in both bulges and disks, separately. We previously reported this behavior for entire systems in \citet{catalan_torrecilla_2015}. This effect is more important for the case of the bulge component in comparison with the disk component, $-$0.89($-$0.52) dex lower in the median value of the sSFR bulge(disk + bar) and $+$0.32($+$0.41) dex more massive in terms of the median value of the total stellar mass. As type-2 AGN tend to be in the more massive systems in our sample, [10$^{10}$ - 10$^{11.5}$] M$_{\odot}$, we analyze if this trend is also present in terms of the morphological type. We restrict the analysis to the more abundant objects with these morphological type and stellar masses, i.\,e, Sa/Sab and Sb/Sbc objects. Bulges of Sa/Sab (Sb/Sbc) show a median sSFR that is 0.27 (0.84) dex below that of star-forming galaxies while the difference in the median value of the stellar mass is 0.08 (0.14) dex. For the case of the disks, Sa/Sab (Sb/Sbc) galaxies exhibit a difference in the median values of sSFR for star-forming and AGNs of 0.11 (0.23) dex while the difference in stellar masses is 0.04 (0.16) dex (Section~\ref{specific_sfr}).}

\item{The previous point supports the idea of negative feedback produced by type-2 AGN galaxies. We cannot exclude, however, that other possibilities might be at a play. On one hand, at least a fraction of the LLAGN that are classified as LINERs could be powered by hot evolved stars. In those cases, the low SFR and sSFR values derived would indicate that these galaxies define a lower photoionization envelope (i.e. a minimum EW$_{\mathrm{H}\alpha}$) associated to evolved (non-star-forming) stellar populations in very massive systems \citep{cid_fernandes_2011}. Thus, for these galaxies mass would be solely the parameter driving the level of current SFR in galaxies and in components within galaxies. On the other hand, AGN damping might be coeval with another mechanism(s) that is (are) regulating the star formation processes (Section~\ref{specific_sfr}).}

\item{The role that stellar kinematics could have in regulating the star formation processes is analyzed by means of the light-weighted LOS stellar velocity dispersion of the bulge component, $\sigma$$_{bulge}$. Type-2 AGN galaxies show higher values of the $\sigma$$_{bulge}$ than star-forming objects. This bimodality is also displayed in the Faber-Jackson relation where type-2 AGN galaxies present higher values of the $\sigma$$_{bulge}$ for the same stellar mass than star-forming objects (Section~\ref{kinematics}).}

\item{The effect that environment has on the star formation processes is studied using the projected galaxy density, $\Sigma$$_{5}$. We find that galaxies have lower values of the SFR in both bulges and disks when they are located in intermediate- and high-density environments (Section~\ref{environment}).}

\end{enumerate}

In brief, this study concludes that the parameter that is affecting more strongly the current SFR of a galaxy, even the SFR associated to their basic stellar structures, is the stellar mass. Star formation damping by type-2 AGN plays also a significant role for bulges but also for disks. Nevertheless, we do not discard the possibility that AGN might be coeval with other processes affecting the star formation processes in the galaxies. In addition to the stellar mass and the nuclear activity, it seems that kinematics and environment act as a secondary parameter in regulating the SFR, at least, in our sample of galaxies. 
We emphasize the importance of applying 2D multi-component photometry decomposition over IFS data to understand the role that different mechanisms play at quenching or triggering the star formation in the structural components that form galaxies.

\twocolumngrid

\acknowledgements
This study makes uses of the data provided by the Calar Alto Legacy Integral Field Area (CALIFA) survey (http://califa.caha.es). CALIFA is the first legacy survey being performed at Calar Alto. The CALIFA collaboration would like to thank the IAA-CSIC and MPIA-MPG as major partners of the observatory, and CAHA itself, for the unique access to telescope time and support in manpower and infrastructures. The CALIFA collaboration thanks also the CAHA staff for the dedication to this project. We would like to thank A. Arag\'on-Salamanca for useful comments and suggestions. C. C.-T. gratefully acknowledges the support of the Spanish {\it Ministerio de Educaci\'on, Cultura y Deporte} by means of the FPU Fellowship Program and the Postdoctoral Fellowship of the {\it Youth Employment Initiative (YEI) European Program}. The authors also thank the support from the {\it Plan Nacional de Investigaci\'on y Desarrollo} funding programs, AYA2012-30717 and AyA2013-46724P, of Spanish {\it Ministerio de Econom\'ia y Competitividad} (MINECO).

%\bibliographystyle{apj} % style aa.bst
%\bibliography{paperII_biblio} % your references Yourfile.bib

\newpage

\clearpage

\onecolumngrid

\LongTables

\begin{landscape}

\begin{longtable}{llccccccccccc}

\caption{ \label{table}} \\

  \hline\hline 
  \\[0.1pt]
  \multicolumn{1}{c}{ID} &
  \multicolumn{1}{c}{name} &
  \multicolumn{1}{c}{M$_{BULGE}$} &
  \multicolumn{1}{c}{M$_{BAR}$} &
  \multicolumn{1}{c}{M$_{DISK}$} &
  \multicolumn{1}{c}{SFR(H$\alpha$$_{BULGE}$)} &
  \multicolumn{1}{c}{SFR(H$\alpha$$_{BAR}$)} &
  \multicolumn{1}{c}{SFR(H$\alpha$$_{DISK}$)} &
  \multicolumn{1}{c}{HT} &
  \multicolumn{1}{c}{$\Sigma$$_{5}$} &
  \multicolumn{1}{c}{AGN} &
  \multicolumn{1}{c}{$\sigma$$_{bulge}$} \\
  
 &  & \centering [$\times$10$^{10}$ M$_{\odot}$] & \centering [$\times$10$^{10}$ M$_{\odot}$]  &  \centering [$\times$10$^{10}$ M$_{\odot}$] &  \centering [M$_{\odot}$ yr$^{-1}$]  & \centering [M$_{\odot}$ yr$^{-1}$] & \centering [M$_{\odot}$ yr$^{-1}$] &  &  \centering [Mpc$^{-2}$] & & [km s$^{-1}$] \\
 
 \centering (1) &  \multicolumn{1}{p{2.1cm}}{\centering (2)} &  \centering (3) &  \centering (4) &  \centering (5) &  \centering (6) &  \centering (7) &  \centering (8) &  \centering (9) &  \centering (10) &  \centering (11) &  \centering (12) & \\
\\[0.1pt]
\hline \\

\endfirsthead
\hline\hline 
\\[0.1 pt]
  \multicolumn{1}{c}{ID} &
  \multicolumn{1}{c}{name} &
  \multicolumn{1}{c}{M$_{BULGE}$} &
  \multicolumn{1}{c}{M$_{BAR}$} &
  \multicolumn{1}{c}{M$_{DISK}$} &
  \multicolumn{1}{c}{SFR(H$\alpha$$_{BULGE}$)} &
  \multicolumn{1}{c}{SFR(H$\alpha$$_{BAR}$)} &
  \multicolumn{1}{c}{SFR(H$\alpha$$_{DISK}$)} &
  \multicolumn{1}{c}{HT} &
  \multicolumn{1}{c}{$\Sigma$$_{5}$} &
  \multicolumn{1}{c}{AGN} &
  \multicolumn{1}{c}{$\sigma$$_{bulge}$} \\
   &  & \centering [$\times$10$^{10}$ M$_{\star}$/M$_{\odot}$] & \centering [$\times$10$^{10}$ M$_{\star}$/M$_{\odot}$]  &  \centering [$\times$10$^{10}$ M$_{\star}$/M$_{\odot}$] &  \centering [M$_{\odot}$ yr$^{-1}$]  & \centering [M$_{\odot}$ yr$^{-1}$] & \centering [M$_{\odot}$ yr$^{-1}$] &  &  \centering [Mpc$^{-2}$] & & [km s$^{-1}$] \\
    \centering (1) &   \multicolumn{1}{p{2.1cm}}{\centering (2)} &  \centering (3) &  \centering (4) &  \centering (5) &  \centering (6) &  \centering (7) &  \centering (8) &  \centering (9) &  \centering (10) &  \centering (11) &  \centering (12) & \\
   \\[0.1pt]
\hline \\

\endhead 
 2    &     UGC 00005  &  0.30$\pm$0.02  &  \ldots  &  5.54$\pm$0.30  &   0.040$\pm$0.003  &  \ldots  &  3.890$\pm$0.155  &  Sbc    &      \ldots   &    yes  &    106.24$\pm$1.98  \\   
 3    &     NGC 7819    &    0.16$\pm$0.03  &  0.22$\pm$0.04  &  0.85$\pm$0.17  &   0.605$\pm$0.008  &  0.268$\pm$0.009  &  0.609$\pm$0.021  &  Sc  &   \ldots   &    no    &    68.29$\pm$1.55  \\   
  5    &     IC 1528    &     0.20$\pm$0.00  &  \ldots  &  1.07$\pm$0.01  &   0.031$\pm$0.001  &  \ldots  &  1.017$\pm$0.023  &  Sbc    &      \ldots   &    no    &    55.40$\pm$1.60  \\   
  6    &     NGC 7824    &    5.41$\pm$0.12  &  \ldots  &  8.78$\pm$0.20  &   0.060$\pm$0.009  &  \ldots  &  0.201$\pm$0.017  &  Sab    &      \ldots   &    no    &    210.19$\pm$4.03  \\   
 7    &     UGC 00036  &  1.26$\pm$0.04  &  1.14$\pm$0.03  &  4.89$\pm$0.15  &   0.051$\pm$0.003  &  0.087$\pm$0.006  &  0.530$\pm$0.052  &  Sab    &      \ldots   &    no    &    173.99$\pm$2.70  \\   
 8    &     NGC 0001    &    1.68$\pm$0.20  &  \ldots  &  2.70$\pm$0.33  &   1.586$\pm$0.051  &  \ldots  &  1.021$\pm$0.042  &  Sbc    &      \ldots   &    no    &    111.93$\pm$1.42  \\   
 10    &    NGC 0036    &    0.65$\pm$0.08  &  0.92$\pm$0.11  &  5.58$\pm$0.65  &   0.411$\pm$0.027  &  0.215$\pm$0.012  &  1.210$\pm$0.105  &  Sb  &   \ldots   &    yes  &    134.06$\pm$2.90  \\   
 11    &    UGC 00139  &  0.01$\pm$0.00  &  \ldots  &  0.40$\pm$0.06  &   0.055$\pm$0.001  &  \ldots  &  0.534$\pm$0.013  &  Scd    &      \ldots   &    no  &  \ldots     \\     
  13    &    MCG-02-02-030  &  0.26$\pm$0.02  &  \ldots  &  1.75$\pm$0.13  &   0.043$\pm$0.003  &  \ldots  &  0.376$\pm$0.027  &  Sb  &   \ldots   &    yes  &    76.72$\pm$0.56  \\   
  14    &    UGC 00312  &  0.04$\pm$0.02  &  0.03$\pm$0.01  &  0.43$\pm$0.19  &   0.112$\pm$0.002  &  0.285$\pm$0.004  &  2.632$\pm$0.044  &  Sd  &   \ldots   &    no    &    68.99$\pm$3.88  \\   
  19    &    ESO 540-G003  &    0.09$\pm$0.00  &  0.20$\pm$0.00  &  0.46$\pm$0.01  &   0.000$\pm$0.000  &  0.070$\pm$0.005  &  0.177$\pm$0.018  &  Sb  &   \ldots   &    no  &  \ldots     \\     
  21    &    NGC 0165    &    0.31$\pm$0.06  &  0.34$\pm$0.06  &  2.78$\pm$0.51  &   0.383$\pm$0.006  &  0.044$\pm$0.002  &  1.024$\pm$0.058  &  Sb  &  1.22$\pm$0.00  &    no  &  \ldots     \\     
  23    &    NGC 0171    &    0.47$\pm$0.12  &  0.42$\pm$0.11  &  2.02$\pm$0.53  &   0.009$\pm$0.001  &  0.055$\pm$0.003  &  0.819$\pm$0.032  &  Sb  &   \ldots   &    yes  &    115.44$\pm$0.61  \\   
 25    &    NGC 0180    &    0.34$\pm$0.02  &  0.56$\pm$0.03  &  3.83$\pm$0.21  &   0.165$\pm$0.004  &  0.040$\pm$0.002  &  0.758$\pm$0.038  &  Sb  &   \ldots   &    no    &    107.11$\pm$0.91  \\  
 26    &    NGC 0192    &    1.88$\pm$0.04  &  \ldots  &  5.41$\pm$0.12  &   0.509$\pm$0.019  &  \ldots  &  0.825$\pm$0.065  &  Sab    &      \ldots   &    no    &    131.38$\pm$0.85  \\   
 28    &    NGC 0214    &    0.58$\pm$0.01  &  0.28$\pm$0.01  &  2.37$\pm$0.05  &   0.019$\pm$0.002  &  0.259$\pm$0.018  &  2.723$\pm$0.200  &  Sbc    &      \ldots   &    yes  &    114.53$\pm$4.37  \\   
 30    &    NGC 0237    &    0.28$\pm$0.01  &  \ldots  &  1.36$\pm$0.03  &   0.097$\pm$0.002  &  \ldots  &  1.621$\pm$0.035  &  Sc  &   \ldots   &    no    &    70.90$\pm$0.53  \\   
  31    &    NGC 0234    &    0.23$\pm$0.05  &  \ldots  &  4.00$\pm$0.84  &   0.251$\pm$0.004  &  \ldots  &  4.194$\pm$0.074  &  Sc  &   \ldots   &    no    &    72.72$\pm$1.07  \\   
  33    &    NGC 0257    &    1.21$\pm$0.03  &  \ldots  &  5.54$\pm$0.13  &   0.236$\pm$0.012  &  \ldots  &  3.882$\pm$0.235  &  Sc  &   \ldots   &    no    &    98.35$\pm$0.49  \\   
  34    &    NGC 0309    &    0.91$\pm$0.21  &  0.28$\pm$0.06  &  4.99$\pm$1.12  &   0.009$\pm$0.001  &  0.014$\pm$0.001  &  7.389$\pm$0.076  &  Scd    &      \ldots   &    no    &    118.50$\pm$1.71  \\   
 38    &    NGC 0447    &    2.93$\pm$0.46  &  0.97$\pm$0.15  &  5.41$\pm$0.85  &   0.115$\pm$0.011  &  0.056$\pm$0.004  &  0.000$\pm$0.000  &  Sa  &   \ldots   &    yes  &    140.60$\pm$2.39  \\   
 42    &    NGC 0477    &    0.34$\pm$0.01  &  \ldots  &  2.33$\pm$0.10  &   0.093$\pm$0.003  &  \ldots  &  1.739$\pm$0.082  &  Sbc    &      \ldots   &    no    &    77.05$\pm$1.17  \\   
  43    &    IC 1683    &     0.59$\pm$0.02  &  0.76$\pm$0.02  &  1.61$\pm$0.05  &   0.794$\pm$0.016  &  0.597$\pm$0.018  &  0.264$\pm$0.026  &  Sb  &   \ldots   &    no    &    113.82$\pm$2.44  \\   
 45    &    NGC 0496    &    0.14$\pm$$^{\dagger}$ &  \ldots  &  1.93$\pm$0.04  &   0.041$\pm$0.001  &  \ldots  &  2.696$\pm$0.084  &  Scd    &      \ldots   &    no    &    28.25$\pm$9.74  \\   
  49    &    UGC 00987  &  1.47$\pm$0.03  &  \ldots  &  2.28$\pm$0.05  &   0.125$\pm$0.010  &  \ldots  &  0.415$\pm$0.022  &  Sa  &   \ldots   &    yes  &    113.05$\pm$1.68  \\   
 50    &    NGC 0528    &    3.42$\pm$0.23  &  \ldots  &  2.46$\pm$0.16  &   0.027$\pm$0.005  &  \ldots  &  0.026$\pm$0.001  &  S0  &   \ldots   &    yes  &    246.98$\pm$2.61  \\   
 52    &    NGC 0551    &    0.39$\pm$0.03  &  0.13$\pm$0.01  &  3.37$\pm$0.24  &   0.005$\pm$$^{\dagger}$  &  0.036$\pm$0.001  &  1.243$\pm$0.043  &  Sbc    &      \ldots   &    no    &    96.85$\pm$2.64  \\   
  58    &    IC 0159    &     0.04$\pm$0.02  &  \ldots  &  0.38$\pm$0.18  &   0.090$\pm$0.001  &  \ldots  &  0.951$\pm$0.017  &  Sdm    &      \ldots   &    no  &  \ldots     \\     
  60    &    UGC A021    &    0.0010 $\pm$ 0.0003  &  0.01$\pm$$^{\dagger}$ &  0.10$\pm$0.03  &   0.004$\pm$$^{\dagger}$  &  0.008$\pm$0.001  &  0.253$\pm$0.016  &  Sdm    &      \ldots   &    no    &    \ldots  \\   
 65    &    NGC 0716 (*)   &    0.20$\pm$0.01  &  0.90$\pm$0.04  &  3.03$\pm$0.15  &   0.199$\pm$0.011  &  2.943$\pm$0.168  &  2.846$\pm$0.184  &  Sb  &   \ldots   &    no    &    132.38$\pm$1.52  \\   
 66    &    UGC 01368  &  0.70$\pm$0.12  &  \ldots  &  5.46$\pm$0.97  &   0.039$\pm$0.003  &  \ldots  &  1.129$\pm$0.162  &  Sab    &      \ldots   &    no    &    143.99$\pm$2.95  \\   
 69    &    NGC 0755    &   \ldots &  \ldots  &  0.18$\pm$$^{\dagger}$ &   \ldots  &  \ldots  &  0.326$\pm$0.007  &  Scd    &      \ldots   &    no  &  \ldots     \\     
  71    &    NGC 0768    &    4.46$\pm$0.91  &  \ldots  &  0.89$\pm$0.18  &   0.216$\pm$0.007  &  \ldots  &  2.799$\pm$0.161  &  Sc  &  0.40$\pm$0.14  &    yes  &    109.76$\pm$8.15  \\   
 73    &    NGC 0776    &    0.79$\pm$0.16  &  0.42$\pm$0.08  &  2.89$\pm$0.57  &   0.314$\pm$0.006  &  0.341$\pm$0.007  &  1.791$\pm$0.055  &  Sb  &   \ldots   &    no    &    107.88$\pm$1.28  \\   
 75    &    NGC 0787    &    3.49$\pm$0.65  &  \ldots  &  3.84$\pm$0.71  &   0.031$\pm$0.003  &  \ldots  &  0.339$\pm$0.017  &  Sa  &   \ldots   &    yes  &  \ldots     \\   
  80    &    NGC 0842    &    1.83$\pm$0.04  &  0.57$\pm$0.01  &  3.37$\pm$0.08  &   0.000$\pm$0.000  &  0.008$\pm$$^{\dagger}$  &  0.026$\pm$0.002  &  S0  &   \ldots   &    no    &    138.68$\pm$0.85  \\   
 81    &    UGC 01659  &  0.17$\pm$0.01  &  0.37$\pm$0.03  &  2.68$\pm$0.22  &   0.007$\pm$0.001  &  0.116$\pm$0.008  &  1.626$\pm$0.099  &  Sc  &   \ldots   &    yes  &  \ldots     \\    
  84    &    NGC 0873    &    0.25$\pm$0.01  &  \ldots  &  1.43$\pm$0.03  &   0.182$\pm$0.005  &  \ldots  &  4.067$\pm$0.082  &  Scd    &      \ldots   &    no    &    90.37$\pm$7.19  \\   
  85    &    NGC 0924    &    2.69$\pm$0.19  &  0.63$\pm$0.04  &  2.58$\pm$0.18  &   0.044$\pm$0.001  &  0.017$\pm$$^{\dagger}$  &  0.101$\pm$0.001  &  S0  &   \ldots   &    yes  &  \ldots     \\    
  86    &    UGC 01918  &  0.32$\pm$0.02  &  0.28$\pm$0.02  &  3.20$\pm$0.24  &   0.425$\pm$0.026  &  0.263$\pm$0.015  &  0.295$\pm$0.030  &  Sb  &   \ldots   &    no  &  \ldots     \\     
  89    &    NGC 0941 (*)   &  0.00086$\pm$0.00025  &  \ldots  &  0.14$\pm$0.04  &   0.004$\pm$$^{\dagger}$  &  \ldots  &  0.207$\pm$0.004  &  Scd    &      \ldots   &    no    &    54.43$\pm$2.71  \\   
  94    &    NGC 0976    &    3.15$\pm$0.07  &  0.63$\pm$0.01  &  3.69$\pm$0.08  &   0.144$\pm$0.014  &  0.737$\pm$0.049  &  2.507$\pm$0.474  &  Sbc    &      \ldots   &    no    &    122.57$\pm$1.24  \\   
 95    &    NGC 0991 (*)  &   0.00019$\pm$0.00004  &  0.02$\pm$0.01  &  0.35$\pm$0.08  &   0.000$\pm$0.000  &  0.031$\pm$0.001  &  0.069$\pm$0.001  &  Scd    &      \ldots   &    no    &    37.60$\pm$4.67  \\   
  97    &    UGC 02134  &  0.08$\pm$0.01  &  0.11$\pm$0.02  &  3.70$\pm$0.59  &   0.054$\pm$0.002  &  0.056$\pm$0.002  &  1.251$\pm$0.039  &  Sb  &   \ldots   &    no    &    87.94$\pm$2.91  \\   
  102  &    NGC 1070    &    0.74$\pm$0.05  &  \ldots  &  5.36$\pm$0.40  &   0.011$\pm$0.001  &  \ldots  &  1.153$\pm$0.053  &  Sb  &   \ldots   &    yes  &    145.77$\pm$0.72  \\   
 107  &    NGC 1094    &    1.53$\pm$0.03  &  \ldots  &  3.64$\pm$0.08  &   0.016$\pm$0.002  &  \ldots  &  2.903$\pm$0.164  &  Sb  &  3.20$\pm$1.41  &    yes  &    135.04$\pm$1.26  \\   
 108  &    NGC 1093    &    1.14$\pm$0.01  &  0.25$\pm$$^{\dagger}$ &  1.91$\pm$0.01  &   0.065$\pm$0.004  &  0.140$\pm$0.007  &  0.579$\pm$0.027  &  Sbc    &      \ldots   &    yes  &    96.92$\pm$0.73  \\   
  109  &    UGC 02311  &  0.86$\pm$0.07  &  0.72$\pm$0.06  &  3.44$\pm$0.28  &   0.214$\pm$0.007  &  0.288$\pm$0.011  &  0.911$\pm$0.062  &  Sbc    &     3.22$\pm$0.47  &    no    &    118.90$\pm$1.63  \\   
 113  &    UGC 02367  &  12.74$\pm$1.49    &     \ldots  &  5.41$\pm$0.63  &   0.605$\pm$0.060  &  \ldots  &  1.122$\pm$0.122  &  Sab    &      \ldots   &    no    &    146.51$\pm$7.86  \\   
 115  &    UGC 02403  &  2.67$\pm$0.05  &  0.84$\pm$0.02  &  1.70$\pm$0.03  &   0.717$\pm$0.020  &  0.744$\pm$0.034  &  0.221$\pm$0.011  &  Sb  &   \ldots   &    no  &  \ldots     \\     
  116  &    UGC 02405  &  0.88$\pm$0.16  &  \ldots  &  2.13$\pm$0.38  &   0.040$\pm$0.002  &  \ldots  &  1.988$\pm$0.115  &  Sbc    &      \ldots   &    no  &  \ldots     \\     
  117  &    UGC 02443  &  0.03$\pm$0.01  &  \ldots  &  0.41$\pm$0.11  &   0.001$\pm$$^{\dagger}$  &  \ldots  &  0.204$\pm$0.005  &  Scd    &      \ldots   &    no  &  \ldots     \\     
  121  &    NGC 1211    &    2.23$\pm$0.66  &  0.64$\pm$0.19  &  1.35$\pm$0.40  &   0.035$\pm$0.003  &  0.000$\pm$0.000  &  0.000$\pm$0.000  &  S0a    &      \ldots   &    yes  &    108.90$\pm$3.53  \\   
 122  &    MCG-01-09-006  &  3.94$\pm$1.28  &  \ldots  &  2.24$\pm$0.73  &   0.245$\pm$0.010  &  \ldots  &  2.394$\pm$0.176  &  Sbc    &     0.25$\pm$0.07  &    no    &    130.00$\pm$7.33  \\   
 123  &    IC 0307    &     3.41$\pm$0.24  &  \ldots  &  5.97$\pm$0.41  &   0.420$\pm$0.072  &  \ldots  &  0.000$\pm$0.000  &  Sab    &     0.68$\pm$0.13  &    no    &    185.05$\pm$3.18  \\   
 125  &    UGC 02690  &  0.11$\pm$$^{\dagger}$ &  \ldots  &  0.79$\pm$0.02  &   0.146$\pm$0.007  &  \ldots  &  3.481$\pm$0.136  &  Scd    &      \ldots   &    no  &  \ldots     \\     
  126  &    NGC 1324    &    4.28$\pm$0.11  &  \ldots  &  8.85$\pm$0.23  &   0.092$\pm$0.009  &  \ldots  &  1.999$\pm$0.131  &  Sb  &   \ldots   &    no    &    138.69$\pm$0.76  \\   
 127  &    NGC 1349    &    3.24$\pm$1.19  &  \ldots  &  4.03$\pm$1.48  &   0.036$\pm$0.004  &  \ldots  &  0.443$\pm$0.042  &  E6  &   \ldots   &    yes  &    185.28$\pm$2.03  \\   
 129  &    MCG-01-10-015  &  0.05$\pm$0.01  &  \ldots  &  0.91$\pm$0.12  &   0.007$\pm$$^{\dagger}$  &  \ldots  &  0.442$\pm$0.018  &  Sc  &   \ldots   &    no  &  \ldots     \\     
  130  &    MCG-01-10-019  &  0.23$\pm$0.02  &  \ldots  &  1.03$\pm$0.09  &   0.037$\pm$0.002  &  \ldots  &  0.610$\pm$0.035  &  Sbc    &     0.96$\pm$0.58  &    no    &    58.16$\pm$2.23  \\   
  134  &    NGC 1645    &    1.17$\pm$0.18  &  1.09$\pm$0.17  &  3.47$\pm$0.54  &   0.011$\pm$0.001  &  0.021$\pm$0.002  &  0.000$\pm$0.000  &  S0a    &      \ldots   &    yes  &    176.69$\pm$0.63  \\   
 137  &    NGC 1659    &    0.01$\pm$$^{\dagger}$ &  0.24$\pm$0.04  &  3.00$\pm$0.48  &   0.139$\pm$0.004  &  0.198$\pm$0.006  &  1.858$\pm$0.070  &  Sbc    &      \ldots   &    no    &    117.69$\pm$3.54  \\   
 139  &    NGC 1666    &    0.79$\pm$0.02  &  0.24$\pm$0.01  &  2.12$\pm$0.05  &   0.005$\pm$0.001  &  0.003$\pm$$^{\dagger}$  &  0.030$\pm$0.001  &  S0a    &      \ldots   &    no  &  \ldots     \\     
  140  &    NGC 1667    &    1.00$\pm$0.37  &  0.52$\pm$0.19  &  4.47$\pm$1.64  &   0.055$\pm$0.006  &  0.088$\pm$0.007  &  6.783$\pm$0.285  &  Sbc    &      \ldots   &    yes  &    146.13$\pm$1.17  \\   
 146  &    UGC 03253  &  0.33$\pm$0.01  &  0.60$\pm$0.02  &  1.99$\pm$0.08  &   0.060$\pm$0.004  &  0.053$\pm$0.004  &  0.000$\pm$0.000  &  Sb  &   \ldots   &    no  &  \ldots     &     \\
  147  &    NGC 2253    &    0.46$\pm$0.01  &  0.75$\pm$0.02  &  2.97$\pm$0.07  &   0.142$\pm$0.005  &  0.048$\pm$0.002  &  2.339$\pm$0.055  &  Sbc    &      \ldots   &    no    &    97.62$\pm$1.01  \\   
 149  &    NGC 2347    &    7.67$\pm$2.49  &  0.11$\pm$0.03  &  0.73$\pm$0.24  &   0.351$\pm$0.015  &  0.211$\pm$0.014  &  1.485$\pm$0.101  &  Sbc    &      \ldots   &    no    &    148.59$\pm$2.45  \\   
 151  &    NGC 2410    &    1.79$\pm$0.30  &  0.48$\pm$0.08  &  4.30$\pm$0.72  &   0.272$\pm$0.017  &  0.106$\pm$0.007  &  1.188$\pm$0.084  &  Sb  &   \ldots   &    yes  &    132.87$\pm$0.91  \\   
 152  &    UGC 03944 (*) &  0.01$\pm$$^{\dagger}$ &  0.09$\pm$$^{\dagger}$ &  0.71$\pm$0.02  &   0.001$\pm$$^{\dagger}$  &  0.403$\pm$0.013  &  0.631$\pm$0.016  &  Sbc    &     0.77$\pm$0.46  &    no    &    37.53$\pm$5.79  \\   
 154  &    UGC 03973  &  0.55$\pm$0.22  &  1.22$\pm$0.48  &  2.34$\pm$0.92  &   1.415$\pm$0.303  &  0.773$\pm$0.078  &  1.130$\pm$0.122  &  Sbc    &     1.21$\pm$0.27  &    yes  &   \ldots    \\
  155  &    UGC 03995  &  0.86$\pm$0.02  &  0.88$\pm$0.02  &  4.36$\pm$0.10  &   0.134$\pm$0.017  &  0.052$\pm$0.004  &  0.361$\pm$0.027  &  Sb  &  0.54$\pm$0.34  &    yes  &    147.94$\pm$2.24  \\   
 156  &    NGC 2449    &    2.26$\pm$0.05  &  0.59$\pm$0.01  &  3.98$\pm$0.09  &   0.031$\pm$0.003  &  0.022$\pm$0.002  &  0.667$\pm$0.081  &  Sab    &     0.98$\pm$0.63  &    yes  &    136.32$\pm$1.62  \\   
 164  &    NGC 2487    &    1.52$\pm$0.09  &  0.46$\pm$0.03  &  3.80$\pm$0.23  &   0.018$\pm$0.002  &  0.099$\pm$0.006  &  0.851$\pm$0.052  &  Sb  &  0.81$\pm$0.32  &    yes  &    110.71$\pm$1.63  \\   
 165  &    UGC 04132  &  4.85$\pm$0.35  &  0.21$\pm$0.02  &  3.47$\pm$0.25  &   0.419$\pm$0.015  &  0.175$\pm$0.008  &  5.564$\pm$0.341  &  Sbc    &     4.84$\pm$0.39  &    no    &    92.25$\pm$1.56  \\   
 167  &    UGC 04145  &  1.06$\pm$0.12  &  1.22$\pm$0.14  &  2.00$\pm$0.23  &   0.097$\pm$0.012  &  0.147$\pm$0.020  &  0.160$\pm$0.022  &  Sa  &  2.45$\pm$0.49  &    yes  &    159.91$\pm$1.00  \\   
 176  &    UGC 04195  &  0.13$\pm$0.07  &  0.21$\pm$0.12  &  1.59$\pm$0.90  &   0.064$\pm$0.002  &  0.031$\pm$0.001  &  0.498$\pm$0.035  &  Sb  &   \ldots   &    no  &  \ldots     &  \\   
  179  &    NGC 2530    &    0.08$\pm$0.02  &  0.07$\pm$0.02  &  0.84$\pm$0.24  &   0.064$\pm$0.002  &  0.066$\pm$0.002  &  1.040$\pm$0.032  &  Sd  &  3.46$\pm$2.30  &    no    &    63.90$\pm$7.10  \\   
 183  &    NGC 2540 (*)   &    0.17$\pm$0.04  &  0.45$\pm$0.11  &  2.66$\pm$0.64  &   0.291$\pm$0.009  &  0.945$\pm$0.025  &  0.938$\pm$0.045  &  Sbc    &     0.62$\pm$0.36  &    no    &    102.98$\pm$1.63  \\   
 187  &    UGC 04308  &  0.12$\pm$0.02  &  0.10$\pm$0.02  &  1.31$\pm$0.25  &   0.033$\pm$0.001  &  0.023$\pm$0.001  &  0.950$\pm$0.024  &  Sc  &  1.78$\pm$0.46  &    no    &    72.99$\pm$4.43  \\   
 190  &    UGC 04262  &  2.41$\pm$0.44  &  \ldots  &  2.05$\pm$0.37  &   0.021$\pm$0.003  &  \ldots  &  0.832$\pm$0.038  &  Sbc    &      \ldots   &    no    &    117.68$\pm$1.28  \\   
 192  &    NGC 2565    &    1.44$\pm$0.32  &  0.69$\pm$0.15  &  2.61$\pm$0.58  &   0.030$\pm$0.006  &  0.041$\pm$0.005  &  0.120$\pm$0.017  &  Sb  &  1.77$\pm$0.34  &    no    &    140.51$\pm$0.84  \\   
 194  &    NGC 2572    &    1.75$\pm$0.18  &  1.56$\pm$0.16  &  5.43$\pm$0.57  &   0.013$\pm$0.003  &  0.062$\pm$0.004  &  0.372$\pm$0.041  &  Sa  &  0.62$\pm$0.45  &    yes  &    186.24$\pm$7.31  \\   
 196  &    UGC 04375  &  0.16$\pm$0.05  &  \ldots  &  0.97$\pm$0.32  &   0.001$\pm$$^{\dagger}$  &  \ldots  &  0.153$\pm$0.005  &  Sbc    &     1.83$\pm$1.63  &    no    &    55.21$\pm$0.78  \\   
 208  &    UGC 04461 (*) &  0.02$\pm$$^{\dagger}$ &  \ldots  &  1.12$\pm$0.32  &   0.067$\pm$0.002  &  \ldots  &  1.610$\pm$0.044  &  Sbc    &     3.24$\pm$1.17  &    no    &    73.95$\pm$26.67  \\   
 209  &    NGC 2604 (*)   &   0.00023$\pm$0.00007 &  0.02$\pm$0.01  &  0.38$\pm$0.12  &   0.084$\pm$0.001  &  0.538$\pm$0.008  &  0.581$\pm$0.009  &  Sd  &  4.38$\pm$3.34  &    no    &    63.19$\pm$4.93  \\   
 219  &    NGC 2639    &    5.67$\pm$0.13  &  \ldots  &  8.60$\pm$0.20  &   0.146$\pm$0.015  &  \ldots  &  0.485$\pm$0.065  &  Sa  &  0.31$\pm$0.25  &    yes  &    162.19$\pm$0.38  \\   
 232  &    NGC 2730    &    0.06$\pm$0.02  &  \ldots  &  0.97$\pm$0.27  &   0.044$\pm$0.001  &  \ldots  &  1.811$\pm$0.023  &  Scd    &     1.57$\pm$0.80  &    no    &    44.14$\pm$2.29  \\   
 260  &    NGC 2805    &    0.21$\pm$0.11  &  \ldots  &  0.37$\pm$0.19  &   0.090$\pm$0.004  &  \ldots  &  0.107$\pm$0.006  &  Sc  &   \ldots   &    no    &    61.06$\pm$4.81  \\   
 275  &    NGC 2906    &    0.67$\pm$0.02  &  \ldots  &  2.21$\pm$0.05  &   0.003$\pm$$^{\dagger}$  &  \ldots  &  0.532$\pm$0.018  &  Sbc    &     0.83$\pm$0.65  &    yes  &    105.39$\pm$1.84  \\   
 277  &    NGC 2916    &    0.92$\pm$0.09  &  \ldots  &  2.63$\pm$0.25  &   0.015$\pm$0.001  &  \ldots  &  0.811$\pm$0.052  &  Sbc    &     0.41$\pm$0.32  &    yes  &    115.86$\pm$0.99  \\   
 278  &    UGC 05108 (*) &  0.42$\pm$0.03  &  1.64$\pm$0.12  &  5.06$\pm$0.36  &   0.484$\pm$0.025  &  0.477$\pm$0.020  &  0.000$\pm$0.000  &  Sb  &  0.58$\pm$0.32  &    no    &    151.76$\pm$6.35  \\   
 307  &    UGC 05359  &  1.31$\pm$0.13  &  0.11$\pm$0.01  &  3.01$\pm$0.30  &   0.012$\pm$0.001  &  0.006$\pm$0.001  &  1.631$\pm$0.143  &  Sb  &  0.42$\pm$0.20  &    no    &    122.16$\pm$5.12  \\   
 309  &    UGC 05396  &  0.06$\pm$0.01  &  0.13$\pm$0.02  &  1.63$\pm$0.30  &   0.007$\pm$0.001  &  0.157$\pm$0.009  &  0.627$\pm$0.046  &  Sbc    &     0.80$\pm$0.65  &    no    &    53.50$\pm$6.01  \\   
 311  &    NGC 3106    &    4.89$\pm$0.17  &  \ldots  &  6.04$\pm$0.21  &   0.175$\pm$0.016  &  \ldots  &  0.463$\pm$0.073  &  Sab    &     0.46$\pm$0.17  &    yes  &    180.34$\pm$2.53  \\   
 312  &    NGC 3057    &    \ldots  &  0.0044$\pm$0.0006  &  0.10$\pm$0.01  &   \ldots  &  0.012$\pm$$^{\dagger}$  &  0.101$\pm$0.002  &  Sdm    &      \ldots   &    no  &  \ldots     &     \\
  353  &    NGC 3381    &    0.01$\pm$$^{\dagger}$ &  0.04$\pm$0.01  &  0.32$\pm$0.12  &   0.019$\pm$$^{\dagger}$  &  0.050$\pm$0.001  &  0.251$\pm$0.003  &  Sd  &  3.21$\pm$0.01  &    no    &    48.44$\pm$3.14  \\   
 381  &    IC 0674    &     2.13$\pm$0.45  &  0.74$\pm$0.16  &  3.57$\pm$0.76  &   0.074$\pm$0.007  &  0.034$\pm$0.004  &  0.301$\pm$0.028  &  Sab    &     0.21$\pm$0.09  &    no    &    158.30$\pm$3.06  \\   
 386  &    UGC 06312  &  6.15$\pm$0.14  &  \ldots  &  3.90$\pm$0.09  &   0.142$\pm$0.022  &  \ldots  &  0.000$\pm$0.000  &  Sab    &     0.42$\pm$0.26  &    yes  &    139.14$\pm$1.03  \\   
 388  &    NGC 3614    &    0.68$\pm$0.31  &  \ldots  &  0.71$\pm$0.33  &   0.017$\pm$0.001  &  \ldots  &  0.306$\pm$0.007  &  Sbc    &     0.76$\pm$0.61  &    no    &    100.73$\pm$6.64  \\   
 414  &    NGC 3687    &    0.44$\pm$0.01  &  0.09$\pm$$^{\dagger}$ &  1.13$\pm$0.03  &   0.002$\pm$$^{\dagger}$  &  0.002$\pm$$^{\dagger}$  &  0.284$\pm$0.006  &  Sb  &  0.52$\pm$0.25  &    yes  &    96.69$\pm$1.21  \\   
 436  &    NGC 3811    &    0.23$\pm$0.05  &  0.38$\pm$0.08  &  1.68$\pm$0.34  &   0.162$\pm$0.007  &  0.063$\pm$0.004  &  1.331$\pm$0.055  &  Sbc    &     0.59$\pm$0.51  &    no    &    102.35$\pm$1.03  \\   
 437  &    NGC 3815    &    0.42$\pm$0.04  &  0.57$\pm$0.06  &  1.27$\pm$0.13  &   0.007$\pm$$^{\dagger}$  &  0.061$\pm$0.002  &  0.780$\pm$0.023  &  Sbc    &     0.71$\pm$0.34  &    no    &    84.33$\pm$0.75  \\   
 476  &    NGC 3994    &    0.98$\pm$0.09  &  \ldots  &  1.10$\pm$0.10  &   0.129$\pm$0.005  &  \ldots  &  2.318$\pm$0.094  &  Sbc    &     19.14$\pm$2.95  &   no    &    141.36$\pm$1.44  \\   
 479  &    NGC 4003    &    3.39$\pm$0.46  &  1.96$\pm$0.26  &  4.78$\pm$0.65  &   0.365$\pm$0.021  &  0.090$\pm$0.012  &  0.115$\pm$0.028  &  S0a    &     0.45$\pm$0.10  &    no    &    136.83$\pm$2.21  \\   
 486  &    UGC 07012  & 0.0017$\pm$0.0016  &  \ldots  &  0.13$\pm$0.12  &   0.034$\pm$$^{\dagger}$  &  \ldots  &  0.428$\pm$0.008  &  Scd    &     7.44$\pm$0.00  &    no    &    38.91$\pm$2.00  \\   
 489  &    NGC 4047    &    1.78$\pm$0.26  &  \ldots  &  3.29$\pm$0.49  &   0.285$\pm$0.007  &  \ldots  &  1.886$\pm$0.048  &  Sbc    &     0.25$\pm$0.07  &    no    &    88.46$\pm$0.36  \\   
 515  &    NGC 4185    &    0.18$\pm$0.08  &  0.17$\pm$0.07  &  3.67$\pm$1.49  &   0.004$\pm$0.001  &  0.014$\pm$0.001  &  1.308$\pm$0.028  &  Sbc    &     3.78$\pm$0.00  &    yes  &    79.77$\pm$2.24  \\   
 518  &    NGC 4210    &    0.09$\pm$0.01  &  0.18$\pm$0.03  &  1.70$\pm$0.26  &   0.001$\pm$$^{\dagger}$  &  0.003$\pm$$^{\dagger}$  &  0.688$\pm$0.016  &  Sb  &  2.19$\pm$0.78  &    yes  &    77.83$\pm$0.83  \\   
 528  &    IC 0776    &     0.01$\pm$$^{\dagger}$ &  \ldots  &  0.18$\pm$0.06  &   0.036$\pm$0.001  &  \ldots  &  0.142$\pm$0.002  &  Sdm    &     6.15$\pm$2.91  &    no    &     \ldots     \\     
  548  &    NGC 4470    &    \ldots  &  \ldots  &  0.88$\pm$0.02  &   \ldots  &  \ldots  &  0.887$\pm$0.011  &  Sc  &  5.45$\pm$3.73  &    no    &     \ldots     \\     
  569  &    NGC 4644    &    0.69$\pm$0.12  &  \ldots  &  1.97$\pm$0.34  &   0.062$\pm$0.005  &  \ldots  &  0.670$\pm$0.028  &  Sb  &  5.35$\pm$1.46  &    no    &    82.04$\pm$1.27  \\   
 580  &    NGC 4711    &    0.19$\pm$0.04  &  \ldots  &  1.74$\pm$0.37  &   0.004$\pm$$^{\dagger}$  &  \ldots  &  0.900$\pm$0.038  &  Sbc    &     0.46$\pm$0.29  &    no    &    57.96$\pm$4.03  \\   
 581  &    UGC 08004  &  0.12$\pm$0.02  &  \ldots  &  0.81$\pm$0.16  &   0.004$\pm$$^{\dagger}$  &  \ldots  &  0.772$\pm$0.026  &  Scd    &     1.06$\pm$0.37  &    no    &     \ldots     \\     
  603  &    NGC 4961    &    0.03$\pm$0.01  &  0.07$\pm$0.01  &  0.28$\pm$0.05  &   0.002$\pm$$^{\dagger}$  &  0.136$\pm$0.002  &  0.301$\pm$0.008  &  Scd    &     1.85$\pm$1.59  &    no    &    50.35$\pm$1.26  \\   
 606  &    UGC 08231  &  \ldots  &  0.01$\pm$$^{\dagger}$ &  0.10$\pm$0.02  &  \ldots  &  0.011$\pm$$^{\dagger}$  &  0.314$\pm$0.005  &  Sd  &  3.85$\pm$3.58  &    no    &     \ldots     \\     
  608  &    NGC 5000    &    0.33$\pm$0.05  &  0.50$\pm$0.08  &  2.30$\pm$0.38  &   0.246$\pm$0.010  &  0.262$\pm$0.015  &  1.389$\pm$0.083  &  Sbc    &     1.99$\pm$0.00  &    no    &    132.00$\pm$6.44  \\   
 611  &    NGC 5016 (*)   &   0.0038$\pm$0.0006  &  \ldots  &  1.59$\pm$0.24  &   0.014$\pm$0.001  &  \ldots  &  0.763$\pm$0.038  &  Sbc    &     0.37$\pm$0.12  &    no    &    66.03$\pm$0.87  \\   
 624  &    NGC 5157    &    2.16$\pm$0.24  &  1.27$\pm$0.14  &  8.24$\pm$0.92  &   0.022$\pm$0.002  &  0.024$\pm$0.002  &  0.426$\pm$0.036  &  Sab    &     0.86$\pm$0.03  &    yes  &    168.49$\pm$1.43  \\   
 630  &    NGC 5205    &    0.11$\pm$0.02  &  0.08$\pm$0.02  &  0.71$\pm$0.14  &   0.000$\pm$0.000  &  0.008$\pm$0.001  &  0.085$\pm$0.004  &  Sbc    &     1.48$\pm$0.81  &    yes  &    69.93$\pm$0.52  \\   
 651  &    NGC 5267    &    1.40$\pm$0.02  &  1.87$\pm$0.02  &  4.36$\pm$0.05  &   0.000$\pm$0.000  &  0.057$\pm$0.006  &  0.676$\pm$0.057  &  Sab    &     0.53$\pm$0.28  &    yes  &    175.17$\pm$1.11  \\   
 653  &    NGC 5289    &    0.90$\pm$0.08  &  \ldots  &  0.89$\pm$0.08  &   0.026$\pm$0.002  &  \ldots  &  0.112$\pm$0.003  &  Sab    &     1.48$\pm$0.10  &    yes  &    129.00$\pm$0.55  \\   
 659  &    NGC 5320    &    0.10$\pm$0.02  &  \ldots  &  1.36$\pm$0.25  &   0.004$\pm$$^{\dagger}$  &  \ldots  &  0.531$\pm$0.012  &  Sbc    &     2.99$\pm$0.22  &    no    &    50.49$\pm$1.05  \\   
 663  &    IC 0944    &     3.48$\pm$0.08  &  \ldots  &  13.61$\pm$0.31  &  0.159$\pm$0.017  &  \ldots  &  0.890$\pm$0.157  &  Sab    &     7.83$\pm$0.00  &    no    &    165.17$\pm$1.57  \\   
 665  &    UGC 08781  &  2.52$\pm$0.27  &  0.90$\pm$0.10  &  6.09$\pm$0.65  &   0.000$\pm$0.000  &  0.020$\pm$0.004  &  1.761$\pm$0.130  &  Sb  &  6.00$\pm$1.42  &    yes  &    148.13$\pm$2.51  \\   
 673  &    NGC 5379    &    0.15$\pm$0.02  &  \ldots  &  0.58$\pm$0.07  &   0.008$\pm$$^{\dagger}$  &  \ldots  &  0.042$\pm$0.003  &  Sab    &     16.30$\pm$11.85  &  no    &    60.58$\pm$4.24  \\   
 676  &    NGC 5378    &    0.50$\pm$0.18  &  0.35$\pm$0.13  &  1.26$\pm$0.47  &   0.009$\pm$0.001  &  0.011$\pm$0.001  &  0.111$\pm$0.013  &  Sb  &  7.16$\pm$1.72  &    yes  &    109.68$\pm$0.85  \\   
 684  &    NGC 5406    &    1.42$\pm$0.07  &  2.06$\pm$0.11  &  7.08$\pm$0.37  &   0.018$\pm$0.004  &  0.067$\pm$0.003  &  0.694$\pm$0.040  &  Sb  &  3.30$\pm$0.00  &    yes  &     \ldots     \\    
  690  &    NGC 5443    &    0.19$\pm$0.02  &  \ldots  &  1.88$\pm$0.20  &   0.008$\pm$0.001  &  \ldots  &  0.106$\pm$0.005  &  Sab    &     5.44$\pm$1.40  &    yes  &     \ldots     \\    
  707  &    NGC 5480    &    0.09$\pm$0.01  &  \ldots  &  1.21$\pm$0.13  &   0.098$\pm$0.002  &  \ldots  &  0.961$\pm$0.017  &  Scd    &     2.72$\pm$1.31  &    no    &    67.00$\pm$0.93  \\   
 714  &    UGC 09067  &  0.44$\pm$0.18  &  \ldots  &  2.93$\pm$1.20  &   0.100$\pm$0.003  &  \ldots  &  2.543$\pm$0.141  &  Sbc    &     1.18$\pm$0.00  &    no    &    103.42$\pm$4.25  \\   
 715  &    NGC 5520 (*)    &    0.13$\pm$$^{\dagger}$ &  0.30$\pm$0.01  &  0.40$\pm$0.01  &   0.019$\pm$$^{\dagger}$  &  0.579$\pm$0.017  &  0.496$\pm$0.014  &  Sbc    &     4.47$\pm$2.51  &    no    &    81.15$\pm$0.49  \\   
 719  &    NGC 5519    &    1.38$\pm$0.03  &  0.55$\pm$0.01  &  2.92$\pm$0.07  &   0.574$\pm$0.017  &  0.361$\pm$0.015  &  0.794$\pm$0.057  &  Sb  &  0.80$\pm$0.16  &    no    &     \ldots     \\     
  720  &    NGC 5522    &    1.56$\pm$0.13  &  1.01$\pm$0.09  &  3.27$\pm$0.28  &   0.075$\pm$0.004  &  0.266$\pm$0.014  &  0.735$\pm$0.041  &  Sb  &  2.33$\pm$0.00  &    no    &     \ldots     \\     
  724  &    NGC 5533    &    7.32$\pm$0.72  &  \ldots  &  4.56$\pm$0.45  &   0.531$\pm$0.048  &  \ldots  &  0.545$\pm$0.043  &  Sab    &     0.77$\pm$0.36  &    yes  &     \ldots     \\    
  732  &    NGC 5559    &    0.55$\pm$0.05  &  \ldots  &  3.00$\pm$0.28  &   0.043$\pm$0.003  &  \ldots  &  1.567$\pm$0.056  &  Sb  &  2.61$\pm$1.62  &    no    &     \ldots     \\     
  736  &    NGC 5587    &    0.27$\pm$0.01  &  \ldots  &  1.87$\pm$0.04  &   0.002$\pm$$^{\dagger}$  &  \ldots  &  0.149$\pm$0.005  &  Sa  &  2.97$\pm$2.57  &    yes  &     \ldots     \\    
  742  &    NGC 5610    &    1.20$\pm$0.13  &  1.57$\pm$0.16  &  2.24$\pm$0.23  &   0.199$\pm$0.008  &  1.124$\pm$0.052  &  0.420$\pm$0.017  &  Sb  &  0.77$\pm$0.04  &    no    &     \ldots     \\     
  743  &    NGC 5622    &    0.74$\pm$0.02  &  \ldots  &  1.45$\pm$0.03  &   0.006$\pm$$^{\dagger}$  &  \ldots  &  0.623$\pm$0.019  &  Sbc    &     0.62$\pm$0.49  &    no    &     \ldots     \\     
  748  &    NGC 5633 (*)   &  0.00037$\pm$0.00004  &  \ldots  &  1.77$\pm$0.20  &   0.003$\pm$$^{\dagger}$  &  \ldots  &  1.018$\pm$0.021  &  Sbc    &     1.13$\pm$0.55  &    no    &    64.79$\pm$1.06  \\   
 749  &    NGC 5630    &    \ldots  &  0.02$\pm$$^{\dagger}$ &  0.31$\pm$0.01  &   \ldots  &  0.129$\pm$0.002  &  0.793$\pm$0.010  &  Sdm    &     0.85$\pm$0.54  &    no    &     \ldots     \\     
  750  &    NGC 5635    &    13.04$\pm$0.24    &     \ldots  &  4.35$\pm$0.08  &   0.095$\pm$0.008  &  \ldots  &  0.194$\pm$0.017  &  Sa  &  0.69$\pm$0.06  &    yes  &     \ldots     \\    
  751  &    UGC 09291  &  0.01$\pm$$^{\dagger}$ &  0.04$\pm$0.02  &  0.61$\pm$0.27  &   0.001$\pm$$^{\dagger}$  &  0.002$\pm$$^{\dagger}$  &  0.257$\pm$0.006  &  Scd    &     0.62$\pm$0.19  &    no    &     \ldots     \\     
  753  &    NGC 5656    &    0.67$\pm$0.07  &  \ldots  &  3.15$\pm$0.33  &   0.008$\pm$0.001  &  \ldots  &  1.349$\pm$0.057  &  Sb  &  1.35$\pm$0.37  &    no    &     \ldots     \\     
  754  &    NGC 5657    &    0.43$\pm$0.02  &  0.66$\pm$0.02  &  1.20$\pm$0.05  &   0.215$\pm$0.006  &  0.359$\pm$0.009  &  0.451$\pm$0.023  &  Sbc    &     1.23$\pm$0.57  &    no    &    103.54$\pm$0.65  \\   
 755  &    NGC 5659    &    1.30$\pm$0.19  &  \ldots  &  2.27$\pm$0.32  &   0.013$\pm$0.001  &  \ldots  &  0.570$\pm$0.018  &  Sb  &  2.95$\pm$0.00  &    no    &     \ldots     \\     
  756  &    NGC 5665    &    0.04$\pm$$^{\dagger}$ &  \ldots  &  1.00$\pm$0.02  &   0.109$\pm$0.001  &  \ldots  &  1.712$\pm$0.021  &  Sc  &  0.80$\pm$0.37  &    no    &     \ldots     \\     
  758  &    NGC 5682    &    \ldots  &  0.01$\pm$$^{\dagger}$ &  0.19$\pm$0.05  &   \ldots  &  0.021$\pm$$^{\dagger}$  &  0.250$\pm$0.003  &  Scd    &     13.13$\pm$11.20  &  no    &     \ldots     \\     
  764  &    NGC 5720    &    0.92$\pm$0.16  &  0.69$\pm$0.12  &  5.58$\pm$0.98  &   0.000$\pm$0.000  &  0.010$\pm$0.003  &  1.149$\pm$0.101  &  Sbc    &     0.11$\pm$0.04  &    yes  &    138.74$\pm$1.08  \\   
 768  &    NGC 5732    &    0.28$\pm$0.04  &  \ldots  &  0.66$\pm$0.09  &   0.021$\pm$$^{\dagger}$  &  \ldots  &  0.644$\pm$0.019  &  Sbc    &     3.96$\pm$1.66  &    no    &    46.35$\pm$4.15  \\   
 769  &    UGC 09476  &  0.11$\pm$0.03  &  \ldots  &  1.41$\pm$0.38  &   0.020$\pm$0.001  &  \ldots  &  1.678$\pm$0.020  &  Sbc    &     1.39$\pm$0.88  &    no    &    50.97$\pm$1.92  \\   
 771  &    NGC 5735    &    0.11$\pm$0.02  &  0.20$\pm$0.04  &  1.10$\pm$0.21  &   0.006$\pm$0.001  &  0.064$\pm$0.002  &  0.506$\pm$0.014  &  Sbc    &     0.39$\pm$0.17  &    yes  &     \ldots     \\    
  772  &    UGC 09492  &  3.62$\pm$0.02  &  1.32$\pm$0.01  &  7.78$\pm$0.04  &   0.032$\pm$0.003  &  0.025$\pm$0.001  &  0.000$\pm$0.000  &  Sab    &     0.41$\pm$0.06  &    yes  &     \ldots     \\    
  775  &    UGC 09542  &  0.69$\pm$0.02  &  \ldots  &  1.84$\pm$0.06  &   0.016$\pm$0.001  &  \ldots  &  1.080$\pm$0.054  &  Sc  &  2.59$\pm$1.77  &    no    &    68.54$\pm$3.13  \\   
 777  &    NGC 5772    &    2.31$\pm$0.45  &  \ldots  &  4.38$\pm$0.85  &   0.027$\pm$0.002  &  \ldots  &  0.798$\pm$0.041  &  Sab    &     0.29$\pm$0.01  &    yes  &     \ldots     \\    
  778  &    NGC 5784    &    4.53$\pm$0.10  &  \ldots  &  10.95$\pm$0.25  &  0.479$\pm$0.046  &  \ldots  &  0.116$\pm$0.013  &  S0  &  0.47$\pm$0.18  &    yes  &    191.94$\pm$0.95  \\   
 779  &    UGC 09598  &  0.13$\pm$$^{\dagger}$ &  \ldots  &  3.02$\pm$0.07  &   0.004$\pm$0.001  &  \ldots  &  0.595$\pm$0.027  &  Sbc    &     0.40$\pm$0.13  &    no    &     \ldots     \\     
  782  &    UGC 09629  &  6.43$\pm$0.01  &  \ldots  &  5.25$\pm$0.01  &   0.000$\pm$0.000  &  \ldots  &  0.090$\pm$0.006  &  E7  &  0.28$\pm$0.04  &    no    &    179.38$\pm$0.92  \\   
 784  &    NGC 5829    &    1.96$\pm$1.33  &  \ldots  &  3.02$\pm$2.05  &   0.076$\pm$0.005  &  \ldots  &  1.095$\pm$0.097  &  Sc  &  1.48$\pm$0.71  &    no    &     \ldots     \\     
  787  &    NGC 5876    &    1.67$\pm$0.15  &  0.86$\pm$0.08  &  2.33$\pm$0.21  &   0.000$\pm$0.000  &  0.007$\pm$$^{\dagger}$  &  0.074$\pm$0.004  &  S0a    &     1.98$\pm$1.36  &    yes  &    186.70$\pm$1.02  \\   
 789  &    NGC 5888    &    1.94$\pm$0.09  &  1.67$\pm$0.07  &  11.32$\pm$0.51  &  0.018$\pm$0.001  &  0.036$\pm$0.005  &  1.017$\pm$0.088  &  Sb  &  0.71$\pm$0.02  &    yes  &    161.50$\pm$0.94  \\   
 790  &    UGC 09777  &  0.01$\pm$$^{\dagger}$ &  0.28$\pm$0.04  &  1.57$\pm$0.20  &   0.202$\pm$0.004  &  0.245$\pm$0.005  &  0.390$\pm$0.015  &  Sbc    &     1.39$\pm$1.06  &    no    &     \ldots     \\     
  792  &    UGC 09842  &  0.62$\pm$0.13  &  1.16$\pm$0.24  &  2.67$\pm$0.55  &   0.522$\pm$0.018  &  1.017$\pm$0.027  &  2.463$\pm$0.141  &  Sbc    &     0.96$\pm$0.52  &    no    &     \ldots     \\     
  803  &    NGC 5957    &    0.22$\pm$0.03  &  0.18$\pm$0.03  &  0.74$\pm$0.11  &   0.004$\pm$$^{\dagger}$  &  0.009$\pm$$^{\dagger}$  &  0.166$\pm$0.004  &  Sb  &  1.01$\pm$0.14  &    yes  &    69.38$\pm$0.51  \\   
 804  &    NGC 5971    &    1.66$\pm$0.42  &  \ldots  &  0.22$\pm$0.05  &   0.032$\pm$0.003  &  \ldots  &  0.063$\pm$0.004  &  Sb  &  2.98$\pm$1.80  &    yes  &    111.86$\pm$1.73  \\   
 807  &    IC 4566    &     1.02$\pm$0.12  &  0.66$\pm$0.08  &  5.21$\pm$0.62  &   0.008$\pm$0.001  &  0.009$\pm$0.001  &  0.713$\pm$0.067  &  Sb  &  7.18$\pm$5.67  &    yes  &    132.38$\pm$0.86  \\   
 810  &    NGC 5980    &    1.38$\pm$0.29  &  \ldots  &  4.44$\pm$0.95  &   0.227$\pm$0.007  &  \ldots  &  3.387$\pm$0.166  &  Sbc    &     0.23$\pm$0.12  &    no    &    101.69$\pm$0.60  \\   
 813  &    NGC 6004    &    0.23$\pm$0.01  &  0.25$\pm$0.01  &  4.17$\pm$0.20  &   0.056$\pm$0.003  &  0.092$\pm$0.002  &  1.197$\pm$0.028  &  Sbc    &     0.64$\pm$0.38  &    no    &    88.64$\pm$2.46  \\   
 817  &    IC 1151    &     0.05$\pm$$^{\dagger}$ &  \ldots  &  0.29$\pm$0.01  &   0.008$\pm$$^{\dagger}$  &  \ldots  &  0.422$\pm$0.007  &  Scd    &     1.23$\pm$1.01  &    no    &    50.67$\pm$2.78  \\   
 820  &    NGC 6032    &    0.24$\pm$0.06  &  0.55$\pm$0.14  &  1.11$\pm$0.29  &   0.133$\pm$0.005  &  0.181$\pm$0.024  &  0.231$\pm$0.067  &  Sbc    &     2.57$\pm$0.66  &    no    &    77.11$\pm$3.81  \\   
 821  &    NGC 6060    &    0.98$\pm$0.15  &  \ldots  &  6.01$\pm$0.89  &   0.112$\pm$0.011  &  \ldots  &  2.252$\pm$0.138  &  Sb  &  1.67$\pm$0.24  &    no    &    114.99$\pm$1.26  \\   
 823  &    NGC 6063    &    0.18$\pm$0.01  &  \ldots  &  0.80$\pm$0.05  &   0.003$\pm$$^{\dagger}$  &  \ldots  &  0.533$\pm$0.010  &  Sbc    &      \ldots   &    no    &    50.85$\pm$3.71  \\   
 824  &    IC 1199    &     0.52$\pm$0.05  &  \ldots  &  2.95$\pm$0.27  &   0.131$\pm$0.004  &  \ldots  &  0.816$\pm$0.041  &  Sb  &  0.67$\pm$0.38  &    no    &    96.72$\pm$2.21  \\   
 830  &    UGC 10337  &  0.75$\pm$0.02  &  \ldots  &  9.11$\pm$0.21  &   0.019$\pm$0.007  &  \ldots  &  1.303$\pm$0.169  &  Sb  &   \ldots   &    yes  &    134.37$\pm$3.81  \\   
 831  &    NGC 6132    &    9.25$\pm$1.08  &  \ldots  &  0.97$\pm$0.11  &   0.008$\pm$$^{\dagger}$  &  \ldots  &  1.215$\pm$0.049  &  Sbc    &     1.31$\pm$0.39  &    no    &    71.96$\pm$4.08  \\   
 833  &    NGC 6154    &    1.59$\pm$0.18  &  0.79$\pm$0.09  &  3.86$\pm$0.44  &   0.044$\pm$0.012  &  0.064$\pm$0.008  &  1.212$\pm$0.067  &  Sab    &      \ldots   &    yes  &    139.56$\pm$2.05  \\   
 836  &    NGC 6155    &    0.05$\pm$$^{\dagger}$ &  \ldots  &  1.31$\pm$0.12  &   0.008$\pm$$^{\dagger}$  &  \ldots  &  0.935$\pm$0.015  &  Sc  &   \ldots   &    no    &     \ldots     \\     
  838  &    UGC 10388  &  1.08$\pm$0.12  &  0.23$\pm$0.02  &  2.70$\pm$0.30  &   0.011$\pm$0.001  &  0.000$\pm$0.000  &  0.103$\pm$0.010  &  Sa  &   \ldots   &    no    &    119.84$\pm$1.06  \\   
 842  &    NGC 6186    &    0.77$\pm$0.17  &  0.73$\pm$0.17  &  1.89$\pm$0.43  &   0.643$\pm$0.013  &  0.555$\pm$0.016  &  0.174$\pm$0.008  &  Sb  &   \ldots   &    no    &    91.03$\pm$0.60  \\   
 844  &    NGC 6278    &    1.69$\pm$0.16  &  0.38$\pm$0.04  &  3.17$\pm$0.30  &   0.010$\pm$$^{\dagger}$  &  0.003$\pm$$^{\dagger}$  &  0.012$\pm$$^{\dagger}$  &  S0a    &      \ldots   &    yes  &    187.72$\pm$0.85  \\   
 849  &    NGC 6301    &    0.27$\pm$0.01  &  \ldots  &  5.79$\pm$0.16  &   0.007$\pm$0.002  &  \ldots  &  6.640$\pm$0.222  &  Sbc    &      \ldots   &    no    &    80.86$\pm$4.04  \\   
 850  &    NGC 6314    &    10.94$\pm$1.36    &     \ldots  &  5.09$\pm$0.63  &   0.268$\pm$0.056  &  \ldots  &  0.000$\pm$0.000  &  Sab    &      \ldots   &    no    &    160.03$\pm$1.56  \\   
 852  &    UGC 10796  &  0.04$\pm$0.02  &  0.06$\pm$0.02  &  0.10$\pm$0.04  &   0.073$\pm$0.002  &  0.045$\pm$0.002  &  0.054$\pm$0.002  &  Scd    &     5.03$\pm$3.50  &    no    &    62.57$\pm$4.50  \\   
 854  &    UGC 10811  &  0.47$\pm$0.19  &  1.04$\pm$0.41  &  4.56$\pm$1.81  &   0.000$\pm$0.000  &  0.035$\pm$0.006  &  0.930$\pm$0.089  &  Sb  &  1.68$\pm$0.67  &    yes  &    163.05$\pm$1.83  \\   
 856  &    IC 1256 (*)   &     0.04$\pm$0.01  &  \ldots  &  1.76$\pm$0.48  &   0.044$\pm$0.002  &  \ldots  &  0.828$\pm$0.033  &  Sb  &   \ldots   &    no    &    77.15$\pm$3.35  \\   
 857  &    NGC 6394    &    0.31$\pm$0.01  &  1.49$\pm$0.03  &  4.25$\pm$0.10  &   0.064$\pm$0.008  &  1.187$\pm$0.088  &  1.107$\pm$0.101  &  Sbc    &     2.95$\pm$0.31  &    yes  &    105.97$\pm$2.78  \\   
 862  &    NGC 6478    &    1.46$\pm$0.18  &  \ldots  &  8.08$\pm$1.01  &   0.165$\pm$0.009  &  \ldots  &  3.801$\pm$0.301  &  Sc  &   \ldots   &    no    &    129.60$\pm$1.47  \\   
 863  &    NGC 6497    &    2.63$\pm$0.06  &  1.26$\pm$0.03  &  4.87$\pm$0.11  &   0.008$\pm$0.001  &  0.008$\pm$0.001  &  0.253$\pm$0.021  &  Sab    &      \ldots   &    yes  &     \ldots     \\    
  865  &    UGC 11228  &  3.31$\pm$0.23  &  0.40$\pm$0.03  &  4.33$\pm$0.30  &   0.073$\pm$0.013  &  0.009$\pm$0.002  &  0.000$\pm$0.000  &  S0  &   \ldots   &    yes  &    183.26$\pm$0.74  \\   
 866  &    UGC 11262  &  0.13$\pm$0.03  &  \ldots  &  0.63$\pm$0.14  &   0.006$\pm$0.001  &  \ldots  &  0.247$\pm$0.014  &  Sc  &   \ldots   &    no    &    64.85$\pm$6.37  \\   
 867  &    NGC 6762    &    0.31$\pm$0.07  &  \ldots  &  1.00$\pm$0.22  &   0.015$\pm$0.001  &  \ldots  &  0.013$\pm$0.003  &  Sab    &      \ldots   &    yes  &    95.18$\pm$0.98  \\   
 868  &    MCG-02-51-004  &  0.61$\pm$0.09  &  \ldots  &  4.27$\pm$0.62  &   0.053$\pm$0.005  &  \ldots  &  2.173$\pm$0.126  &  Sb  &   \ldots   &    yes  &    109.84$\pm$1.14  \\   
 869  &    NGC 6941    &    1.31$\pm$0.32  &  0.76$\pm$0.19  &  6.57$\pm$1.61  &   0.049$\pm$0.005  &  0.014$\pm$0.002  &  1.408$\pm$0.070  &  Sb  &   \ldots   &    no    &    140.64$\pm$1.22  \\   
 871  &    NGC 6978    &    5.28$\pm$0.31  &  \ldots  &  6.97$\pm$0.41  &   0.049$\pm$0.004  &  \ldots  &  0.562$\pm$0.054  &  Sb  &   \ldots   &    yes  &    136.91$\pm$2.52  \\   
 872  &    UGC 11649  &  0.32$\pm$0.03  &  0.35$\pm$0.03  &  1.89$\pm$0.16  &   0.000$\pm$0.000  &  0.000$\pm$0.000  &  0.230$\pm$0.012  &  Sab    &      \ldots   &    yes  &    112.77$\pm$0.91  \\   
 876  &    NGC 7047    &    0.40$\pm$0.01  &  \ldots  &  4.75$\pm$0.11  &   0.004$\pm$0.001  &  \ldots  &  1.340$\pm$0.072  &  Sbc    &      \ldots   &    yes  &    91.33$\pm$1.98  \\   
 879  &    UGC 11740  &  0.20$\pm$0.04  &  0.99$\pm$0.17  &  1.01$\pm$0.18  &   0.021$\pm$0.001  &  0.144$\pm$0.009  &  0.685$\pm$0.037  &  Sbc    &      \ldots   &    no    &    71.09$\pm$4.59  \\   
 886  &    NGC 7311    &    3.49$\pm$0.08  &  \ldots  &  5.96$\pm$0.14  &   0.048$\pm$0.009  &  \ldots  &  1.569$\pm$0.131  &  Sa  &   \ldots   &    no    &    174.01$\pm$1.24  \\   
 887  &    NGC 7321    &    1.05$\pm$0.10  &  1.17$\pm$0.12  &  6.08$\pm$0.60  &   0.029$\pm$0.002  &  0.200$\pm$0.008  &  2.723$\pm$0.144  &  Sbc    &      \ldots   &    yes  &    155.20$\pm$2.85  \\   
 890  &    UGC 12185  &  2.71$\pm$0.35  &  0.64$\pm$0.08  &  1.29$\pm$0.17  &   0.025$\pm$0.005  &  0.205$\pm$0.014  &  0.309$\pm$0.031  &  Sb  &   \ldots   &    no    &    133.98$\pm$1.63  \\   
 891  &    UGC 12224  &  0.04$\pm$0.01  &  \ldots  &  0.81$\pm$0.10  &   0.015$\pm$0.001  &  \ldots  &  0.656$\pm$0.024  &  Sc  &   \ldots   &    no    &    39.07$\pm$7.57  \\   
 894  &    UGC 12274  &  2.14$\pm$0.05  &  \ldots  &  8.04$\pm$0.18  &   0.029$\pm$0.002  &  \ldots  &  0.114$\pm$0.005  &  Sa  &   \ldots   &    yes  &    142.90$\pm$1.41  \\   
 896  &    NGC 7466    &    1.37$\pm$0.46  &  \ldots  &  2.88$\pm$0.96  &   0.567$\pm$0.024  &  \ldots  &  1.846$\pm$0.104  &  Sbc    &      \ldots   &    yes  &    126.00$\pm$3.77  \\   
 898  &    NGC 7489 (*)    &    0.31$\pm$0.06  &  \ldots  &  2.14$\pm$0.43  &   0.063$\pm$0.003  &  \ldots  &  3.655$\pm$0.160  &  Sbc    &      \ldots   &    no    &    68.62$\pm$3.80  \\   
 899  &    NGC 7536    &    0.64$\pm$0.05  &  \ldots  &  1.48$\pm$0.12  &   0.108$\pm$0.004  &  \ldots  &  1.829$\pm$0.048  &  Sc  &  3.76$\pm$0.59  &    no    &     \ldots     \\    
  901  &    NGC 7549    &    0.39$\pm$0.05  &  0.36$\pm$0.04  &  0.95$\pm$0.12  &   0.548$\pm$0.015  &  0.723$\pm$0.025  &  0.627$\pm$0.019  &  Sbc    &      \ldots   &    no    &    102.46$\pm$1.20  \\   
 904  &    NGC 7591    &    1.24$\pm$0.27  &  4.52$\pm$1.00  &  2.27$\pm$0.50  &   1.375$\pm$0.053  &  0.415$\pm$0.013  &  2.259$\pm$0.128  &  Sbc    &      \ldots   &    no    &    112.31$\pm$0.55  \\   
 906  &    IC 5309    &     0.57$\pm$0.03  &  \ldots  &  1.25$\pm$0.07  &   0.122$\pm$0.003  &  \ldots  &  0.658$\pm$0.021  &  Sc  &   \ldots   &    no    &    56.02$\pm$2.86  \\   
 914  &    NGC 7631    &    1.06$\pm$0.02  &  \ldots  &  1.47$\pm$0.03  &   0.075$\pm$0.003  &  \ldots  &  0.460$\pm$0.017  &  Sb  &   \ldots   &    no    &    77.04$\pm$1.43  \\   
 915  &    NGC 7653    &    1.83$\pm$0.04  &  \ldots  &  1.60$\pm$0.04  &   0.318$\pm$0.008  &  \ldots  &  1.740$\pm$0.037  &  Sb  &   \ldots   &    yes  &    99.14$\pm$1.54  \\   
 916  &    NGC 7671    &    0.98$\pm$0.48  &  0.33$\pm$0.16  &  3.27$\pm$1.60  &   0.000$\pm$0.000  &  0.000$\pm$0.000  &  0.063$\pm$0.001  &  S0  &   \ldots   &    no    &    228.13$\pm$1.50  \\   
 920  &    NGC 7691    &    0.04$\pm$$^{\dagger}$ &  0.13$\pm$$^{\dagger}$ &  1.40$\pm$0.03  &   0.005$\pm$$^{\dagger}$  &  0.016$\pm$0.001  &  0.655$\pm$0.090  &  Sbc    &      \ldots   &    no    &    60.43$\pm$2.37  \\   
 924  &    NGC 7716    &    0.51$\pm$0.01  &  0.14$\pm$$^{\dagger}$ &  1.23$\pm$0.03  &   0.005$\pm$0.001  &  0.037$\pm$0.001  &  0.354$\pm$0.013  &  Sb  &   \ldots   &    yes  &    108.28$\pm$0.52  \\   
 925  &    NGC 7722    &    2.39$\pm$0.48  &  \ldots  &  6.76$\pm$1.37  &   0.130$\pm$0.017  &  \ldots  &  0.000$\pm$0.000  &  Sab    &      \ldots   &    yes  &    167.02$\pm$2.12  \\   
 927  &    NGC 7738    &    5.67$\pm$0.13  &  4.01$\pm$0.09  &  1.98$\pm$0.05  &   5.035$\pm$0.230  &  0.744$\pm$0.055  &  0.867$\pm$0.074  &  Sb  &   \ldots   &    yes  &    156.79$\pm$2.82  \\   
 929  &    UGC 12810  &  0.54$\pm$0.13  &  0.48$\pm$0.12  &  3.83$\pm$0.95  &   0.266$\pm$0.009  &  0.182$\pm$0.009  &  1.934$\pm$0.138  &  Sbc    &      \ldots   &    no    &    97.39$\pm$2.93  \\   
 930  &    UGC 12816  &  0.19$\pm$0.08  &  \ldots  &  0.11$\pm$0.05  &   0.182$\pm$0.005  &  \ldots  &  0.436$\pm$0.017  &  Sc  &   \ldots   &    no    &    62.32$\pm$1.77  \\   
 931  &    NGC 7782    &    2.01$\pm$0.65  &  \ldots  &  10.48$\pm$3.40  &  0.025$\pm$0.003  &  \ldots  &  3.558$\pm$0.118  &  Sb  &   \ldots   &    no    &     \ldots     \\     
  933  &    NGC 7787    &    3.50$\pm$0.08  &  \ldots  &  0.70$\pm$0.02  &   0.599$\pm$0.030  &  \ldots  &  0.418$\pm$0.032  &  Sab    &     1.32$\pm$0.49  &    no    &    103.75$\pm$3.26  \\   
 935  &    UGC 12864  &  0.15$\pm$$^{\dagger}$ &  0.12$\pm$$^{\dagger}$ &  0.67$\pm$0.02  &   0.244$\pm$0.005  &  0.146$\pm$0.004  &  0.277$\pm$0.011  &  Sc  &   \ldots   &    no    &    57.00$\pm$2.77  \\   
\hline 

\end{longtable}

Col. (1): ID CALIFA identifier. Col. (2): Galaxy name. Cols. (3) $-$ (5): Stellar mass of the bulge, bar, and disk component, respectively. Stellar masses have been derived as explained in Section~\ref{masses}. Cols. (6) $-$ (8): Extinction-corrected H$\alpha$ SFR for the bulge, bar, and disk component, respectively. The value provided for the SFR in the disk component has been corrected by aperture effects. Col. (9): Morphological type. Col. (10): Projected galaxy density ($\Sigma$$_{5}$). Col. (11): Type-2 AGN candidate using the BPT diagram to obtain the nuclear activity classification. Col. (12): Line-of-sight dispersion for the bulge component ($\sigma$$_{bulge}$).

\footnotesize

(*) These galaxies have a nuclear point source component instead of a bulge component.

$^{\dagger}$ These quantities have errors that are smaller than the last digit of the corresponding magnitude measured. Instead of adding more precision to the quantities quoted in this column, and for the sake of clarity in the formatting of the table, we have not included their errors here. 

\clearpage

\end{landscape}

\end{document}